\documentclass[11pt]{article}

\usepackage[margin=1in]{geometry}
\usepackage{setspace}
\usepackage{amsmath, amssymb, amsthm}
\usepackage{graphicx}
\usepackage{booktabs}
\usepackage{multirow}
\usepackage{subfigure}
\usepackage{soul}
\usepackage[ruled, vlined]{algorithm2e}

\usepackage{xcolor}
\definecolor{blue-violet}{rgb}{0.54, 0.17, 0.89}
\definecolor{myblue}{rgb}{0.2,0,0.9}

\usepackage{varioref}
\usepackage{xr-hyper}
\usepackage{hyperref}
\hypersetup{
    colorlinks,
    linkcolor=myblue,
    citecolor=blue-violet,
    urlcolor=myblue
}
\usepackage{xurl}

\usepackage{natbib}

\def\({\left(}
\def\){\right)}
\def\[{\left[}
\def\]{\right]}

\title{\Large MarketGANs: Multivariate financial time-series data augmentation using generative adversarial networks}

\usepackage{authblk}

\newsavebox\affbox
\author[1]{Jeonggyu Huh}
\author[2]{Seungwon Jeong}
\author[3]{Hyun-Gyoon Kim\footnote{Corresponding author, E-mail address: hyungoonkim@ajou.ac.kr, limbh@skku.edu}}
\author[3]{Hyeng Keun Koo}
\author[4]{Byung Hwa Lim$^*$}

\affil[1]{Department of Mathematics, Sungkyunkwan University, Republic of Korea.}
\affil[2]{Global-Learning \& Academic research institution for Master's $\cdot$ PhD students, and Postdocs, Chonnam National University, Republic of Korea.}
\affil[3]{Department of Financial Engineering, Ajou University, Republic of Korea.}
\affil[4]{Department of FinTech, SKK Business School, Sungkyunkwan University, Republic of Korea.}

\date{}
\begin{document}
\maketitle

\begin{abstract}
This paper introduces MarketGAN, a factor-based generative framework for high-dimensional asset return generation under severe data scarcity. We embed an explicit asset-pricing factor structure as an economic inductive bias and generate returns as a single joint vector, thereby preserving cross-sectional dependence and tail co-movement alongside inter-temporal dynamics. MarketGAN employs generative adversarial learning with a temporal convolutional network (TCN) backbone, which models stochastic, time-varying factor loadings and volatilities and captures long-range temporal dependence. Using daily returns of large U.S. equities, we find that MarketGAN more closely matches empirical stylized facts of asset returns, including heavy-tailed marginal distributions, volatility clustering, leverage effects, and, most notably, high-dimensional cross-sectional correlation structures and tail co-movement across assets, than conventional factor-model-based bootstrap approaches. In portfolio applications, covariance estimates derived from MarketGAN-generated samples outperform those derived from other methods when factor information is at least weakly informative, demonstrating tangible economic value.
\end{abstract}

\bigskip

\noindent
\textbf{Keywords:} Multivariate time-series, Data augmentation, Factor models, Generative adversarial networks

\externaldocument[A-]{Online_Appendix}

\section{Introduction}

Modern empirical finance increasingly confronts a fundamental tension between high dimensionality and limited data availability. Asset pricing, portfolio choice, and risk management problems routinely involve dozens or hundreds of assets, yet the effective length of available financial time series remains short relative to the dimensionality of the objects of interest. Even classical mean–variance portfolio optimization requires estimating a large number of parameters. For example, the mean vector and covariance matrix of 100 assets alone entail more than 5,000 parameters, and extensions incorporating higher-order co-moments quickly become infeasible, as described in \cite{martellini2010improved}.\footnote{If one considers higher-order co-moments to improve upon traditional methodologies \cite{martellini2010improved}, the number of required parameters rapidly becomes infeasible.
	Specifically, incorporating third-order co-moments requires estimating over 170,000 parameters, and including fourth-order moments increases this number to more than 4 million.
	Given that 20 years of daily data yield roughly 5,000 observations, and even fewer observations are available from less-noisy monthly data, these estimation problems are highly underdetermined. 
	Furthermore, even if one collects sufficient historical data over the past several decades, structural shifts in macroeconomic conditions and changes in individual firm-specific characteristics can significantly alter asset correlations over time, which makes the historical sample covariance matrix unreliable.} With only a few thousand daily observations over several decades—and even fewer reliable observations at lower frequencies—such estimation problems are inherently underdetermined. Moreover, structural changes in macroeconomic conditions and firm characteristics further undermine the reliability of purely historical estimates of dependence.

These challenges are amplified in the context of financial machine learning. Deep learning and reinforcement learning methods offer flexible nonlinear representations but are inherently data intensive \citep{goodfellow2016deep}. When trained on finite historical samples, such models are prone to overfitting and poor out-of-sample generalization. This issue has been formally documented in the context of deep empirical risk minimization and deep stochastic optimization, where neural-network-based models trained on fixed datasets exhibit strong in-sample performance but deteriorate sharply out of sample unless additional simulated or synthetic data are supplied \citep{Reppen2023deep, Reppen2023deepstochasticoptimization}. These findings suggest that, in high-dimensional financial settings, synthetic data play a structural role in enabling reliable learning and decision-making.

Motivated by this observation, a growing literature employs generative models, particularly generative adversarial networks (GANs), to augment financial datasets and improve model performance. Early studies show that GAN-generated data can replicate stylized features of asset returns and enhance forecasting or trading performance, primarily in single-asset or low-dimensional settings \citep{takahashi2019modeling, wiese2020quant, buehler2020data}. However, most existing approaches treat synthetic data as local augmentation aimed at reproducing marginal distributions, and rarely address joint return distributions, cross-sectional dependence, or portfolio-relevant risk structures. As a result, their evaluation is often confined to predictive accuracy rather than economically meaningful distributional properties.

Recent work emphasizes that the usefulness of synthetic data depends critically on whether it extends learning beyond historical realizations. Viewing financial markets as complex adaptive systems, \citet{wei2025enhancing} and \citet{Horvath2026marketgenerators} argue that merely replicating past distributions is insufficient to prepare models for regime shifts and rare but economically consequential events. At the same time, a growing literature on missing data and sparsity in high-dimensional asset pricing panels shows that even sophisticated regularization and imputation methods struggle to recover economically relevant dependence structures—particularly low-variance directions and cross-sectional correlations central to portfolio choice \citep{bryzgalova2025missing, bryzgalova2025forest, freyberger2025missingassetpricing, chen2024missing}. These findings point to a deeper limitation that the joint distribution of asset returns is only weakly identified from historical observations alone in high dimensions.

This paper addresses this gap by proposing MarketGAN, a factor-based generative framework for high-dimensional asset return generation. The central objective of MarketGAN is not to improve point prediction accuracy, but to learn and generate the joint distribution of asset returns in a manner that preserves economically meaningful dependence structures. The approach embeds a traditional asset-pricing factor model, such as the CAPM and Fama–French factor models, as an economic inductive bias, while using generative learning to model stochastic, time-varying factor loadings and idiosyncratic risks. By generating returns as a single joint vector, MarketGAN directly targets cross-sectional dependence across assets, which underpins diversification and portfolio risk.

While we illustrate our framework using financial portfolios, the problem we address is more general. Many decision-making environments are characterized by high-dimensional uncertainty and severe data scarcity, in which historical observations are insufficient to reliably estimate the joint distribution of relevant variables. Examples include demand forecasting across large product assortments in supply chains, risk aggregation in energy and insurance systems, and scenario generation for stress testing and capacity planning. In such settings, decisions are typically nonlinear functions of the underlying uncertainty, and small errors in dependence modeling can be amplified through optimization and control. Consequently, the central challenge is not point prediction, but learning and extending a joint distribution that is suitable as an input to downstream decision problems. From this perspective, financial portfolios serve as a canonical and transparent test-bed rather than a special case. Portfolio optimization provides a well-understood nonlinear decision mapping in which dependence structures, such as cross-sectional correlation and tail co-movement, play a decisive role. The proposed framework is therefore best viewed as a data-driven approach to decision-making under weak identification, where economic or structural constraints are used to regularize distributional learning and to generate synthetic scenarios that improve decision quality. This interpretation highlights the relevance of our approach beyond finance, positioning MarketGAN as a general framework for high-dimensional decision problems under data scarcity.

When generating synthetic multivariate asset return data, two aspects are of central importance. The first is the cross-sectional dependence across assets, and the other is inter-temporal dependence over time. To address inter-temporal dependence, we employ temporal convolutional networks (TCNs), which capture long-range causal dependence through stacked dilated convolutions \citep{bai2018empirical}.  Importantly, in our framework TCNs are used not to generate returns directly, but to model the time-varying evolution of factor model coefficients, thereby linking temporal dynamics to risk exposures.\footnote{To address inter-temporal dependence, we incorporate temporal architectures explicitly designed to model sequential dynamics, including recurrent neural networks \citep{elman1990finding}, self-attention-based transformers \citep{vaswani2017attention}, and temporal convolutional networks (TCNs) \citep{bai2018empirical}. In this study, we employ TCNs due to their structural advantages in modeling long-range causal dependence with efficient parallel computation. TCNs have been successfully applied to sequential data such as audio generation \citep{oord2016wavenet} and financial time series modeling \citep{wiese2020quant}. 
} The factor structure organizes cross-sectional dependence, while stochastic, time-varying coefficients allow dependence patterns to evolve with market conditions. We implement this framework using adversarial distribution learning as a flexible engine for approximating complex high-dimensional distributions.\footnote{Applying generative neural networks to financial data raises a fundamental challenge of data scarcity. Generative models themselves are data intensive, and learning high-dimensional joint distributions without additional structure can lead to underfitting or overfitting when historical samples are limited. To break this circular dilemma, MarketGAN incorporates domain knowledge in the form of asset-pricing factor models. The factor structure provides the overall organization of cross-sectional dependence across assets, while its coefficients which are intercepts, factor loadings, and idiosyncratic volatilities are modeled as stochastic, time-varying objects generated by a generative network. This design can be interpreted as a time-varying coefficient factor model similar to \citet{fabozzi1978beta}, \citet{chamberlain1982multivariate} and \citet{ang2007capm}, in which factor exposures evolve dynamically rather than remaining fixed or deterministically specified.} To ensure stable training, we adopt the Wasserstein GAN with gradient penalty \citep{arjovsky2017wasserstein, gulrajani2017improved}. Importantly, adversarial learning plays an instrumental role in providing distributional flexibility, while the factor structure supplies economic discipline.

We empirically evaluate MarketGAN using daily excess returns of 98 large U.S. equities from the S\&P 100 universe. Across one-, three-, and five-factor specifications, MarketGAN generates synthetic returns that (i) closely match empirical marginal distributions and higher-order moments, (ii) reproduce canonical intertemporal stylized facts such as volatility clustering and leverage effects, and—most importantly—(iii) substantially improve the fidelity of cross-sectional dependence, including dependence in extreme returns, relative to transparent factor-model-based bootstrap benchmarks. Quantitative distributional metrics commonly used in the generative modeling \citep{heusel2017gans, bonneel2015sliced} confirm these gains.

We further assess economic value through mean–variance portfolio optimization using MarketGAN-generated synthetic data. The results show that when future factor information is even weakly informative, MarketGAN achieves superior portfolio performance relative to factor-model-based bootstrap methods, while remaining competitive with strong covariance-based benchmarks. These gains do not arise from improved factor forecasting or marginal distributional accuracy alone, but from MarketGAN’s ability to generate coherent high-dimensional return distributions that preserve economically meaningful dependence structures. In sum, by embedding factor-model discipline into a generative architecture and by demonstrating gains in both statistical fidelity and downstream portfolio performance, MarketGAN provides a practical framework for decision-making in high-dimensional financial settings where historical data alone are insufficient.


\vspace{1cm}

\noindent{\bf Related Literature}

This paper is related to the growing literature that applies machine learning and deep learning methods to asset pricing and return modeling.
\citet{gu2020empirical} provide a comprehensive evaluation of a wide range of machine learning models using firm characteristics and macroeconomic variables, showing that flexible nonlinear methods can outperform traditional linear regressions in predicting equity risk premia and in portfolio applications.
Building on this insight, \citet{gu2021autoencoder} integrate deep learning with factor models by allowing factor exposures to depend nonlinearly on conditioning information. Their approach shares with ours a structural motivation grounded in factor models. While these approaches enhance predictive performance and factor representation, factor loadings are treated as deterministic functions of conditioning information, and the focus remains on prediction or estimation rather than on generating joint return distributions for data augmentation. Related work also applies adversarial learning to asset pricing problems. \citet{chen2024deep} propose an adversarially trained stochastic discount factor estimator combining macroeconomic state variables with a GAN-style objective. In contrast, our focus is not on SDF estimation but on learning and generating joint return distributions relevant for high-dimensional decision problems.

A second strand of the literature studies generative neural networks for financial time-series data, with GAN-based models playing a central role. Early contributions focus primarily on univariate or low-dimensional settings and demonstrate that generative models can reproduce canonical stylized facts such as heavy tails and volatility clustering \citep{takahashi2019modeling, wiese2020quant}. Subsequent work extends these approaches to multivariate settings \citep{yoon2019time, liao2020conditional, rizzato2023generative, cont2022tail}, but empirical validation is typically confined to small cross sections or to marginal distributions. As a result, these models often fall short of capturing the high-dimensional joint return distributions and cross-sectional dependence structures that are central to portfolio risk management.

More recent approaches begin to address higher-dimensional generation by incorporating structural or economic inductive biases \citep{wang2024factor, duan2022factorvae, tepelyan2023generative, cho2025diffolio}. In particular, \citet{cho2025diffolio} propose a hierarchical attention-based diffusion model that integrates asset-specific and systematic covariates to produce calibrated conditional predictive distributions for multivariate returns, with a primary focus on probabilistic forecasting and portfolio allocation.

In contrast, MarketGAN is designed as a distributional learning framework that embeds economic structure directly into the data generating process through a factor-model-based architecture. Rather than emphasizing conditional prediction at short horizons, MarketGAN aims to learn and extend the joint distribution of asset returns itself, preserving complex cross-sectional dependence and tail co-movement across a large universe of assets. While many existing models rely on latent factors without clear economic interpretation or generate assets independently and concatenate them post hoc, MarketGAN directly targets high-dimensional joint dependence in a manner that is explicitly grounded in asset-pricing theory.

Our work is also closely related, at a conceptual level, to a recent literature on missing data and sparsity in high-dimensional asset pricing panels.
\citet{bryzgalova2025missing, bryzgalova2025forest} show that missing characteristics and noise in large cross sections distort the estimation of economically relevant directions in return space, particularly low-variance components important for portfolio choice.
Similarly, \citet{freyberger2025missingassetpricing} document the instability of cross-sectional dependence in sparse panels, and \citet{chen2024missing} show that small covariance estimation errors induced by missing data can translate into large portfolio losses. While this literature focuses on imputation, regularization, and dimensionality reduction within observed panels, it highlights a deeper limitation that in high-dimensional settings, the joint distribution of asset returns is only weakly identified from historical observations alone. Our approach takes a complementary perspective by treating the problem as one of distributional representation rather than data reconstruction, and by directly learning and generating joint return distributions subject to an economically motivated structure.

The remainder of this paper is organized as follows. Section~\ref{sec: preliminaries} describes the components of our proposed model, including GANs, TCNs, and their integration.
The architecture of our proposed model and its perspective as a stochastic coefficient factor model are then described in Section~\ref{sec: marketgan}.
Empirical experiments are presented in Section~\ref{sec: experiments} where we evaluate the fidelity of the generated data, assess its ability to reproduce stylized facts. Section \ref{sec: portfolio optim} demonstrates the generated synthetic data's  utility in portfolio optimization and Section~\ref{sec: conclusion} concludes.

\section{Methodological Background} \label{sec: preliminaries}

\subsection{Generative Adversarial Networks (GANs)}

GANs provide a flexible framework for learning complex data distributions by formulating a two-player game between a generator and a discriminator  \citep{goodfellow2014generative}. The generator $G$ aims to produce samples that resemble the true data distribution $p_{data}(x)$, while the discriminator $D$ aims to distinguish real from generated samples. Through adversarial training, the generator learns to approximate the underlying data distribution without requiring an explicit likelihood specification. Formally, the generator $G$ takes as input a random noise vector $z\sim p_z(z)$ drawn from a simple prior distribution (e.g., Gaussian) and outputs synthetic samples $G(z)$ that resemble the real data. Then, these two neural networks compete against each other in a minimax game described by the following objective function:
\begin{equation*}
	\min_G \max_D V(D, G)=\mathbb{E}_{x\sim p_{data}(x)} \left[ \log D \left(x \right) \right] + \mathbb{E}_{z\sim p_z(z)} \left[ \log \left( 1-D\left(G(z)\right) \right) \right].
\end{equation*}
When the discriminator is optimal, the minimization of this objective corresponds to minimizing the Jensen-Shannon divergence between the real and generated distributions:
\begin{equation*}
	JSD(p_{data}, q_{G}) = \frac{1}{2} D_{KL} \left( p_{data} \Vert M  \right) + \frac{1}{2} D_{KL} \left( q_{G} \Vert M \right),
\end{equation*}
where $q_G$ denotes the model distribution induced by the generator, and $M=\frac{1}{2} \left(p_{data}+q_{G}\right)$ and $D_{KL}$ is the Kullback-Leibler divergence.
In essence, the generator is trained to approximate the real data distribution $p_{data}$.

In the context of financial data, GANs are attractive because they operate at the distributional level, rather than focusing on point predictions. This feature makes GANs well-suited for applications such as scenario generation, stress testing, and portfolio analysis, where the joint distribution of returns is more relevant than conditional means. However, standard GANs are known to be challenging to train and sensitive to data limitations, particularly in high-dimensional and noisy environments such as financial markets. 

To improve training stability, several variants of GANs have been proposed. Among them, the Wasserstein GAN with gradient penalty (WGAN-GP) replaces the original Jensen-Shannon divergence with the Wasserstein distance and enforces Lipschitz continuity through a gradient penalty \citep{arjovsky2017wasserstein, gulrajani2017improved}.\footnote{The formal objective function of WGAN-GP is provided in Online Appendix \ref{appendix: formal introduction of gan and tcn}.} We adopt the WGAN-GP objective to ensure stable training in continuous-valued financial data.

\subsection{Temporal Convolutional Networks (TCNs)} \label{sec: tcn}

Modeling temporal dependence is essential for financial time-series data. TCNs provide an effective alternative to recurrent architectures by employing causal and dilated convolutions along the time dimension \citep{bai2018empirical}. Through stacked convolutional layers with increasing dilation, TCNs capture long-range temporal dependencies while maintaining computational efficiency and stable gradients. Thanks to this advantage, in the context of deep learning-based sequence modeling, TCNs \citep{bai2018empirical} have been shown to achieve competitive performance compared to other widely-used recurrent or self-attention-based models. 
TCNs apply one-dimensional convolution along the temporal axis and effectively extract features from long-range input sequences through dilated convolutions. In addition, they can offer efficient computation owing to their low computational complexity and ability to perform parallel operations.

The receptive field size (RFS) of a TCN, i.e., the number of input time steps that influence a single output time step, grows exponentially with depth. Specifically, the RFS is given by
$
\mathrm{RFS} = 1 + 2(k - 1) \cdot \frac{D^L - 1}{D - 1},
$
where $k$ is the kernel size, $D$ is the dilation factor, and $L$ is the number of residual blocks.
This allows TCNs to model long-term temporal dependencies effectively, even with a relatively shallow architecture.

Compared to recurrent neural networks, TCNs offer several practical advantages in financial applications. They enable parallel computation over time, avoid vanishing or exploding gradients associated with long sequences, and provide a flexible receptive field that can be adjusted to the desired temporal horizon. As a result, TCNs have been successfully applied to various sequential data problems, including audio generation and financial time series modeling \citep{wiese2020quant}. 

\subsection{GAN with TCN backbone} \label{sec: gan with tcn backbone}

While GANs excel at capturing cross-sectional dependence, they do not natively account for temporal structure. A natural approach to modeling financial time series is therefore to combine GANs with architectures designed for temporal representation learning. In such hybrid models, temporal networks such as RNNs, transformers, or TCNs are used to encode historical information, while the GAN framework learns the distribution of future observations conditional on this temporal state. 

This separation of roles is particularly appealing in finance. The temporal network extracts latent features that summarize market dynamics over time, whereas the GAN focuses on learning the conditional distribution of returns. By integrating temporal modeling into the generative process, GAN-based models can generate synthetic time-series data that reflect both inter-temporal dependence and cross-sectional structure. 

Another important aspect of GANs with a TCN backbone is the inherent stochastic nature of the generator. Since the input to the generator is a sequence of random variables, the output synthetic sequence is also a random sequence.
Even under fixed conditions, repeated evaluations yield different synthetic sequences.
Hence, the generator models the conditional probability distribution of the data, which performs density estimation rather than deterministic point estimation. In addition, due to the receptive field structure of TCNs, the output at time $t$, denoted by $\hat{x}_t$, depends not only on $z_t$ but also on a range of past latent variables $\{z_{t-k}\}_{k=0}^{RFS-1}$. 
This dependence on a finitely extended temporal window implies that the process is non-Markovian, as future states are not independent of the distant past given the present. Thus, the GAN with a TCN backbone defines a multi-dimensional discrete-time non-Markovian stochastic process. This formulation enables the generator to capture intricate temporal dependencies and structural features of sequential financial data while allowing for generating diverse and statistically consistent synthetic samples. 

Building on the generative and temporal modeling frameworks reviewed in this section,
the next section introduces MarketGAN, a factor-based generative model tailored to
high-dimensional asset returns.

\section{MarketGAN} \label{sec: marketgan}

Building on the generative and temporal modeling frameworks in Section \ref{sec: preliminaries}, {we introduce a factor-based generative model tailored to high-dimensional asset return generation under severe data scarcity.} The central objective of MarketGAN is not to improve point predcition accuracy, but to learn joint distribution of asset returns in a manner that preserves economically meaningful cross-sectional dependence, inter-temporal dynamics, and portfolio relevant risk characteristics.

\subsection{MarketGAN Architecture}

The primary challenge addressed by MarketGAN is the severe scarcity of informative financial time-series data in high-dimensional settings. While generative models provide a natural mechanism for data augmentation, they themselves require structural regularization to remain effective in noisy and data-poor financial environments. To this end, MarketGAN incorporates domain knowledge in the form of an asset-pricing factor model, which serves as an economic inductive bias that displines the generative learning problem.

We model the excess return of $N$ assets at time $t+1$, dented by $\boldsymbol{r}_{t+1}\in\mathbb{R}^N$, using the following factor-model-based formulation:
\begin{equation} \label{eqn: marketgan}
	\boldsymbol{r}_{t+1} = \boldsymbol{\alpha}_t + \boldsymbol{\beta}_t \cdot \boldsymbol{F}_{t+1}+\boldsymbol{\sigma}_t \odot \boldsymbol{\epsilon}_{t+1},
\end{equation}
where $\boldsymbol{F}_{t+1}\in \mathbb{R}^K$ denotes observable risk factors such as CAPM or Fama-French factors, $\boldsymbol{\alpha}_t\in \mathbb{R}^N$ is the vector of intercepts, $\boldsymbol{\beta}_t\in \mathbb{R}^{N\times K}$ is the matrix of factor loadings, $\boldsymbol{\sigma}_t\in \mathbb{R}^N$ scales the idiosyncratic component, and $\boldsymbol{\epsilon}_{t+1}\in \mathbb{R}^N$ represents standardized residuals. The operator $\odot$ indicates element-wise multiplication.

The defining feature of MarketGAN is that the coefficients $\boldsymbol{\alpha}_t$, $\boldsymbol{\beta}_t$, and $\boldsymbol{\sigma}_t$ are not treated as deterministic parameters, but are instead generated as stochastic, time-varing objects by conditional GANs with a TCN backbone, as described in Section \ref{sec: gan with tcn backbone}.
This design integrates three complementary components. First, the TCN backbone extracts latent temporal features from historical data, allowing the model to capture long-range inter-temporal dependence. Second, adversarial training enables flexible learning of high-dimensional joint distributions across assets. Third, the factor structure imposes economically meaningful constraints that regularize the learning problem and improve robustness under data scarcity. 

Formally, the generative networks take as input sequences of latent noise vectors and observed covariates over a rolling temporal window determined by the RFS of the TCN. Specifically, for a normal random variable $\boldsymbol{z}_u\in\mathbb{R}^{d_z}$ and a covariate $\boldsymbol{y}_u\in \mathbb{R}^{d_y}$, let $\boldsymbol{z}_{t-(\mathrm{RFS}-1):t} := \{\boldsymbol{z}_u \}_{u=t-(\mathrm{RFS}-1)}^t$ denote a sequence of latent noise vectors, and $\boldsymbol{y}_{t-(\mathrm{RFS}-1):t} := \{\boldsymbol{y}_u \}_{u=t-(\mathrm{RFS}-1)}^t$ denote a sequence of covariates over the period from $t-(\mathrm{RFS}-1)$ to $t$. 
The networks are trained to extract informative features from these past sequences, thereby capturing both cross-sectional and inter-temporal dependencies conditioned on the covariates. 
In particular, the interweaving of the randomness from the $d_z$-dimensional noise vectors and the $d_y$-dimensional covariate inputs allows the model to approximate the joint conditional distribution of the $N$-dimensional asset returns. 

\begin{figure}[!t]
    \centering
    \includegraphics[width=\textwidth]{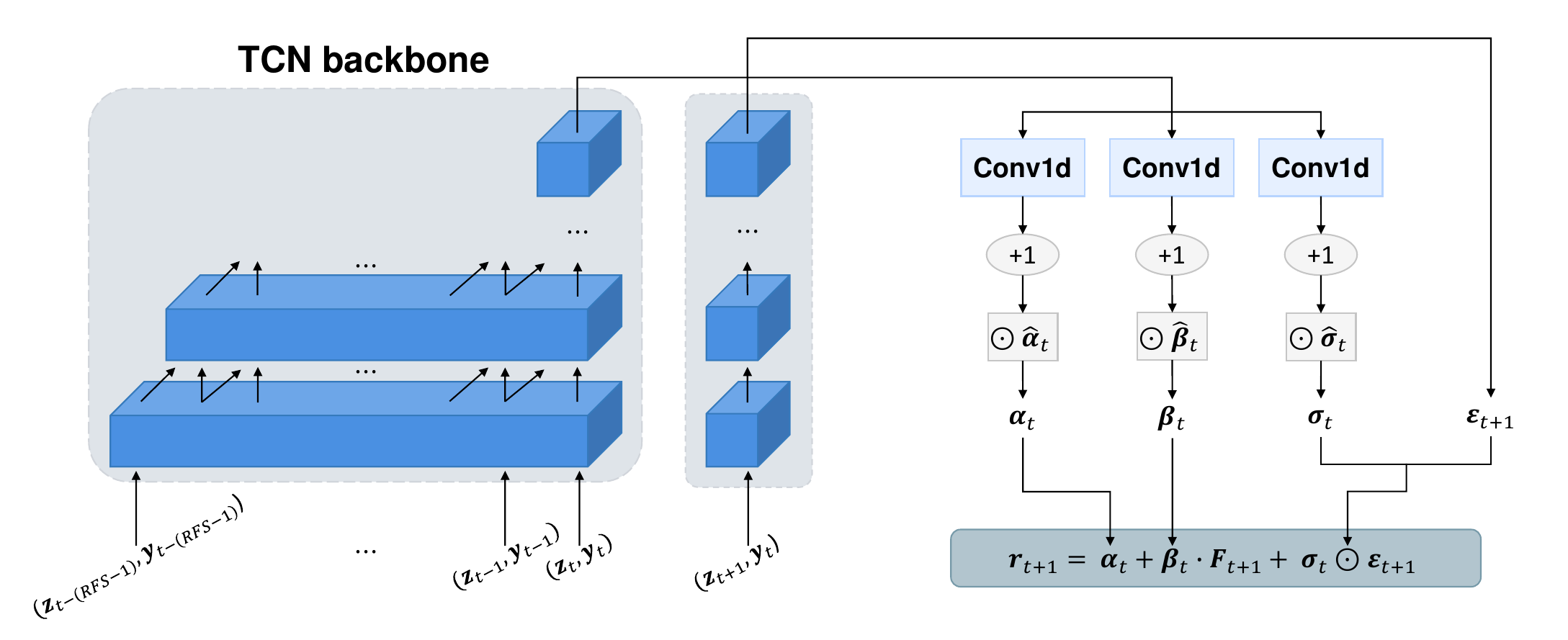}
    \caption{Architecture of MarketGAN.}
    \label{fig: marketgan architecture}
\end{figure}

To enhance training stability and anchor the generative process to economically plausible scales, MarketGAN incorporates ex-post coefficient estimates obtained from standard regression. 
\begin{equation} \label{eqn: alpha alphahat correction}
	\begin{gathered}
		\boldsymbol{\alpha}_t = \hat{\boldsymbol{\alpha}}_t \odot \left( 1 + f_\alpha \left(\boldsymbol{z}_{t-(\textrm{RFS}-1):t}, \boldsymbol{y}_{t-(\textrm{RFS}-1):t} \right) \right), \\
		\boldsymbol{\beta}_t = \hat{\boldsymbol{\beta}}_t \odot \left( 1 + f_\beta \left(\boldsymbol{z}_{t-(\textrm{RFS}-1):t}, \boldsymbol{y}_{t-(\textrm{RFS}-1):t} \right) \right), \\
		\boldsymbol{\sigma}_t = \hat{\boldsymbol{\sigma}}_t \odot \left( 1 + f_\sigma \left(\boldsymbol{z}_{t-(\textrm{RFS}-1):t}, \boldsymbol{y}_{t-(\textrm{RFS}-1):t} \right) \right), 
	\end{gathered}
\end{equation}
where $\hat{\boldsymbol{\alpha}}_t$, $\hat{\boldsymbol{\beta}}_t$, and $\hat{\boldsymbol{\sigma}}_t$ denote ex-post regression estimates, and $f_\alpha$, $f_\beta$, and $f_\sigma$ are TCN-based modules that generate stochastic correction terms. The multiplicative specification preserves the scale and sign structure of the baseline coefficients while allowing flexible stochastic deviations. Moreover, the three networks share all TCN layers except for their final $1 \times 1$ convolutional output layers, reflecting the idea that the coefficients are driven by common latent market dynamics. This parameter sharing scheme reduces computational complexity and improves estimation efficiency. The overall architecture of MarketGAN is illustrated in Figure \ref{fig: marketgan architecture}.

The idiosyncratic residual component $\boldsymbol{\epsilon}_{t+1}$ is produced via a conditional GAN with a simplified TCN structure. To ensure that residuals are independent of past uncertainty, the generator employs $1\times 1$ convolutions for all layers, resulting in a RFS of one, and takes $(\boldsymbol{z_{t+1}}, \boldsymbol{y_t})$ as input. This design allows the residual to depend only on contemporaneous latent noise and covariates and to be isolated from past uncertainty. Meanwhile, each noise sample $\boldsymbol{z}_{t+1}$—the innovation at time $t+1$—is subsequently reused to generate the factor model coefficients for the following period (e.g., $\boldsymbol{\alpha}_{t+1}, \boldsymbol{\beta}_{t+1}, \boldsymbol{\sigma}_{t+1}$). This reuse of latent noise across periods elaborates the model's temporal dependency and induces a nonlinear moving average type dependence structure. Consequently, MarketGAN captures rich inter-temporal dynamics without imposing a Markovian restriction.

The discriminator mirrors the generator's temporal structure and operates on a sequence of asset returns and covariates. For a given input window, it produces a single scalar score that reflects the discrepancy between real and generated return sequences, consistent with the WGAN-based training objective.\footnote{The discriminator employs a TCN with a final linear layer mapping to a scalar value, and takes as input a sequence $(\boldsymbol{x}_{t+1:t+T_L}, \boldsymbol{y}_{t+1:t+T_L})$, where $\boldsymbol{x}_u$ denotes either real or generated excess returns at time $u$.
	In other words, the discriminator outputs a single scalar value for a given sequence of length $T_L$.
	In standard GAN formulations, this scalar passes through a sigmoid activation function, and, in WGANs, it directly serves as a critic score for measuring the Wasserstein distance.}

In the next subsection, we interpret MarkeGAN as both a data generator and a stochastic coefficient factor model, and discuss how these perspectives clarify its role in financial applications. 

\subsection{MarketGAN as a Data Generator and a Stochastic Coefficient Factor Model}

MarketGAN is designed primarily as a data generation framework that produces synthetic realizations of high-dimensional asset returns. Rather than estimating model coefficients once and treating them as fixed, MarketGAN generates the coefficients of the factor model through neural networks and uses them to construct return realizations. As each inference, the model produces a distinct $N$-dimensional vector of excess returns, allowing the generation of an effectively unlimited number of synthetic samples. These samples are designed to preserve the statistical properties of the high-dimensional real data distribution learned during training.

Because MarketGAN is implemented using a conditional GAN, it naturally supports conditional data generation. By conditioning on observed covariates, such as macroeconomic variables or market state indicators, the model approximates time-varying conditional return distributions. This capability enables MarketGAN to generate realistic return paths that adapt to evolving economic conditions, making it suitable for applications such as scenario analysis, stress testing, and portfolio evaluation under alternative economic states. From this perspective, MarketGAN functions as a flexible return generator operating at the distributional level rather than as a point forecasting model.

Beyond its role as a return data generator, MarketGAN can also be interpreted as a stochastic coefficient factor model. While generative neural networks are often used primarily for data augmentation, their fundamental capability lies in high-dimensional density estimation. {In this framework, training MarketGAN corresponds to optimizing the deterministic weights of the generative neural networks that parameterize the stochastic coefficient processes. Specifically, instead of providing fixed point-estimates for the model parameters, MarketGAN implicitly learns the high-dimensional joint distribution of these coefficients—$\boldsymbol{\alpha}_t, \boldsymbol{\beta}_t,$ and $\boldsymbol{\sigma}_t$—across the $N$ assets via adversarial training.}

The use of conditional GANs allows the coefficient distributions to vary over time as functions of observed covariates, capturing changes in market conditions and risk exposures. Moreover, since MarketGAN employs a TCN backbone to model temporal dependence, the resulting coefficient processes are non-Markovian as described in Section \ref{sec: gan with tcn backbone}. Current realizations depend on a finite but extended history of latent noise and covariates, rather than solely on the most recent state. From this viewpoint, fitting MarketGAN to historical data can be seen as calibrating nonlinear, non-Markovian stochastic processes that govern the dynamics of factor model coefficients. 

This interpretation connects MarketGAN to the broader literature on stochastic coefficient and time-varying factor models \citep{fabozzi1978beta, chamberlain1982multivariate, ang2007capm, dangl2012predictive, jostova2005bayesian, boloorforoosh2020beta, armstrong2013factor}. Classical approaches in this literature allow factor loadings to evolve over time, typically through linear state-space or parametric specifications. MarketGAN generalizes these models by allowing for nonlinear dynamics, high-dimensional cross-sectional dependence, and flexible conditional distributions, while retaining the economically disciplined structure of a factor model.

\section{Empirical Experiments}  \label{sec: experiments}

This section evaluates the empirical performance of MarketGAN as a generative model for high-dimensional asset returns. The objective of the empirical analysis is not to assess predictive accuracy in the conventional sense, but to examine whether synthetic data generated by MarketGAN replicate key statistical and economic properties of real financial data. In particular, we focus on the model's ability to preserve cross-sectional dependence, inter-temporal dynamics, and portfolio-relevant risk characteristics under realistic data constraints.

\subsection{Data and Experimental Setup}\label{sec: experimental setup}

We conduct our empirical analysis using daily excess returns of stocks in the S\&P 100 index. The sample is constructed using constituents observed as of December 31, 2023, and includes all assets that appeared at least once during the in-sample period. The selection yields a balanced universe of 98 assets, allowing us to focus on high-dimensional cross-sectional dependence while avoiding survivorship bias from restricting attention to continuously traded stocks. The sample is divided into an in-sample period from January 1, 1991, to December 31, 2015, and an out-of-sample period from January 1, 2016, to December 31, 2023. The in-sample period is used to train and validate the generative models, while the out-of-sample period is reserved exclusively for evaluation. Excess returns are computed using the one-month U.S. Treasury bill rate as the risk-free rate. As covariates, we employ macroeconomic predictors from \citet{welch2008comprehensive} and \citet{gu2020empirical} which consist of the following eight variables, such as dividend-price ratio (dp), earnings-price ratio (ep), book-to-market ratio (bm), T-bill rate (tbl), term spread (tms), default spread (dfy), net equity {\color{magenta} issue} (ntis), and stock variance (svar).
Since these macroeconomic variables are available at a monthly frequency, missing values are forward-filled with the most recent observations.

Because not all assets are observed continuously throughout the in-sample period, missing returns and coefficients are imputed using a conventional regression-based factor model. For each asset with missing observations, factor model coefficients, $\hat{\boldsymbol{\alpha}}_i$ and $\hat{\boldsymbol{\beta}}_i$, are estimated using the full in-sample data. Missing values are then filled by identifying dates with macroeconomic conditions most similar to those of the missing observations, measured by Euclidean distance in the covariate space, and averaging the corresponding estimated coefficients. Specifically, the missing returns is filled with $\hat{\boldsymbol{\alpha}}_i + \hat{\boldsymbol{\beta}}_i\cdot \boldsymbol{F}$ where 
$\boldsymbol{F}$ denotes the factor values on that day.

We propose three types of MarketGAN models, each based on widely used factor models. The first is based on the CAPM and includes only the market factor. The second and third are based on the Fama-French three-factor and five-factor models, respectively. We refer to these factor models as FF-1, FF-3, and FF-5, and denote the corresponding generative models by MarketGAN1, MarketGAN3, and MarketGAN5, respectively. Then, MarketGAN1 is based on the single-factor CAPM and uses only the excess return on the market portfolio as the risk factor to capture time-varying market exposure and idiosyncratic risk dynamics while maintaining a parsimonious structure. MarketGAN3 extends this framework by incorporating the market factor, the size factor (SMB), and the value factor (HML). This specification allows the generative model to capture cross-sectional variation in returns associated with firm size and value characteristics, as well as their time-varying dynamics. MarketGAN5 further enriches the factor structure by adopting the market, size, and value factors with profitability (RMW) and investment (CMA) factors. By including these additional sources of systematic risk, MarketGAN5 provides the most flexible specification and allows us to assess whether a richer economic structure improves the quality of synthetic return generation.\footnote{For detailed descriptions of these factor portfolios, we refer to \cite{fama1993common, fama2015five}.} Across all three variants, the TCN backbone and adversarial training procedure remain identical. This design ensures that differences in empirical performance can be attributed primarily to the underlying factor specification rather than to changes in network architecture or training methodology.

Each FF-$k$, $k=1, 3, 5,$ model is of the form
\begin{equation} \label{eqn: ff model}
	\boldsymbol{r}_{t+1} = \hat{\boldsymbol{\alpha}}_t + \hat{\boldsymbol{\beta}}_t \cdot \boldsymbol{F}_{t+1}+\hat{\boldsymbol{\sigma}}_t \odot \boldsymbol{\epsilon}_{t+1},
\end{equation}
where the coefficients $\hat{\boldsymbol{\alpha}}_t$, $\hat{\boldsymbol{\beta}_t}$, and $\hat{\boldsymbol{\sigma}_t}$ are deterministically estimated at each time $t$, and $\boldsymbol{F}_{t+1}$ is the $k$-dimensional factor.
Specifically, the coefficients $\hat{\boldsymbol{\alpha}}_t$ and $\hat{\boldsymbol{\beta}}_t$ are estimated by rolling window regressions over the past one year of data including time $t$, and $\hat{\boldsymbol{\sigma}}_t$ is given by the standard deviation of the regression residuals within the same window. This conventional factor model serves two purposes. First, it provides a transparent benchmark against which the performance of MarketGAN can be evaluated. Second, rolling-window estimates of the factor model coefficients are used as ex-post baseline coefficients in the multiplicative correction structure of MarketGAN, defined in \eqref{eqn: alpha alphahat correction}. Specifically, intercepts and factor loadings are estimated using one-year rolling regressions, and idiosyncratic volatility is computed as the standard deviation of residuals within the same window.

The generators for the time-varying intercepts $\boldsymbol{\alpha}_t$, factor loadings $\boldsymbol{\beta}_t$, and volatilities $\boldsymbol{\sigma}_t$ share a common TCN backbone, reflecting the assumption that these coefficients are driven by shared latent market dynamics. The backbone employs a kernel size of two ($k = 2$), a dilation factor of two ($D = 2$), and six residual blocks ($L = 6$), resulting in a receptive field that spans 127 time steps (RFS=127). Each coefficient has its own output layer, while all intermediate layers are shared. The idiosyncratic residual component is generated by a separate conditional GAN with a simplified temporal structure, ensuring that residual shocks are contemporaneous while temporal dependence is captured primarily through the coefficients.\footnote{Specifically, these networks share all layers except the final output layer $\phi_O$: the input mapping $\phi_I$ and the residual blocks $g_\ell$ for $\ell = 1, 2, \dots, L$ are shared across the three coefficient networks, while each coefficient has its own output mapping layer, denoted by $\phi_O^\alpha$, $\phi_O^\beta$, and $\phi_O^\sigma$, respectively.
	On the other hand, the generator for $\boldsymbol{\epsilon}_{t+1}$ uses a kernel size of $k = 1$, a dilation factor of $D = 1$, and $L = 6$ residual blocks, which leads to an RFS of 1.
	The latent dimension of normal random variables $\boldsymbol{z}$ is set to $d_z = 10$.
	As a result, the generator takes a total of $(127 + 1) \times d_z = 1280$ independent standard normal random inputs $\boldsymbol{z}_{t-(\mathrm{RFS}-1):t+1}$ to generate a single sample of $N$-dimensional returns at each time step.}

All models are trained for 300 epochs. The in-sample period is split into training and validation subsets in a 7:1 ratio, with the most recent portion reserved for validation. Model selection is based on the similarity between the correlation matrix of generated returns and that of real returns, measured using the Frobenuis norm. To mitigate the loss of recent information due to this split, we perform an additional fine-tuning stage on the validation set with a one-tenth of the original learning rate. Consistent with standard practice in GAN training, the discriminator is updated more frequently than the generator to promote training stability. Other implementation details and hyperparameters used in the three MarketGAN models are presented in Online Appendix \ref{appendix: hyperparam}.

As benchmark methods, we consider factor-model-based bootstrap procedures corresponding to the FF-1, FF-3 and FF-5 models. These approaches generate synthetic returns by combining rolling-window estimates fo factor model coefficients with independently drawn Gaussian residuals. Specifically, these use the following formulation
\begin{equation}
	\hat{\boldsymbol{r}}_{t+1} = \hat{\boldsymbol{\alpha}}_t + \hat{\boldsymbol{\beta}}_t \cdot \boldsymbol{F}_{t+1} + \hat{\boldsymbol{\sigma}}_t \odot \boldsymbol{\epsilon}_{t+1}, \quad \boldsymbol{\epsilon}_{t+1}\sim\mathcal{N}(0, I),
\end{equation}
where $\boldsymbol{F}_{t+1}$ denotes the factor realizations at time $t+1$, defined by the respective FF-$k$ model. 
The coefficients $\hat{\boldsymbol{\alpha}}_t$, $\hat{\boldsymbol{\beta}}_t$, and $\hat{\boldsymbol{\sigma}}_t$ are estimated ex post using a rolling window of one-year data ending at time $t$, and are also used in \eqref{eqn: alpha alphahat correction} as ex-post coefficients.
Because these bootstrap methods share the same factor structure but lack a generative neural network component, they provide transparent baselines that isolate the incremental contribution of MarketGAN. Comparing MarketGAN to these benchmarks allows us to assess whether adversarial learning adds value beyond traditional factor-based resampling.

\subsection{Marginal Return Distributions} 
\label{sec:return_distribution}

We begin the empirical evaluation by examining whether MarketGAN preserves the marginal distributional properties of asset returns. Reproducing marginal distributions constitutes a necessary, though not sufficient, condition for realistic synthetic data generation, and has been widely used as a baseline diagnostic on financial generative models (e.g., \citealp{takahashi2019modeling, wiese2020quant, cont2022tail}). In financial markets, asset returns are known to exhibit non-Gaussian features such as heavy tails and excess kurtosis, which play a central role in risk assessment and portfolio management. We therefore assess the extent to which MarketGAN-generated returns replicate these stylized marginal characteristics relative to factor-model-based bootstrap benchmarks.

{We begin by examining market-level behavior, constructing an equal-weighted portfolio of the $N$
	assets and comparing its excess return paths across real data, MarketGAN-generated data, and factor-model-based bootstrap benchmarks.} We plot the realized equal-weighted excess returns over the out-of-sample period together with the mean and one-standard-deviation bands computed from 100 synthetic sample paths. Across all factor specifications, both MarketGAN and bootstrap methods closely track the realized aggregate returns. This result highlights the importance of the factor structure in capturing aggregate market dynamics and suggests that even simple factor-based approaches can effectively reproduce market-level behavior.\footnote{See Figure \ref{fig: ew return figure} in the Online Appendix. The relatively narrow dispersion bands indicate that idiosyncratic variations largely cancel out at the portfolio level. We also examine marginal distribution at the individual asset level, using GOOG as a representative example.  Figure \ref{fig: goog_return_comparison} compares the realized excess returns of GOOG with synthetic returns generated by MarketGAN and the corresponding bootstrap methods.} Similar to equal-weighted portfolio, all models capture the mean behavior reasonably well, but MarketGAN-generated returns exhibit substantially narrower standard deviation bands than those produced by bootstrap methods, indicating a closer fit to the empirical distribution of individual asset returns. In contrast, bootstrap-based synthetic paths exhibit greater variability, reflecting their reliance on independently drawn residuals rather than on a learned distributional structure.

To complement these visual comparisons, we quantify marginal reconstruction accuracy using root-mean-square error (RMSE) and mean absolute error (MAE), computed by comparing 100 synthetic sample paths with realized returns and averaging across assets and time. Table \ref{tbl:rmse_mae} reports these metrics for MarketGAN and bootstrap methods under different factor specifications. MarketGAN consistently outperforms the corresponding bootstrap benchmarks. It demonstrates superior marginal explanatory power. Moreover, performance improves for both approaches as the number of factors increases, consistent with the factor models (e.g., \citealp{fama1993common, fama2015five}). Notably, MarketGAN with three factors achieves lower error metrics than the five-factor bootstrap, which underscores the added value of adversarial distribution learning beyond factor richness alone. 

\begin{table}[!htb]
	\centering
	\begin{tabular}{lcccccc}
		\toprule
		\multicolumn{1}{c}{\multirow{2}{*}{Metric}} & \multicolumn{3}{c}{MarketGAN} & \multicolumn{3}{c}{Factor-model-based bootstrap} \\ 
		\cmidrule(lr){2-4} \cmidrule(lr){5-7}
		& 1 & 3 & 5 & 1 & 3 & 5 \\ 
		\midrule
		RMSE & 0.01743 & 0.01570 & \textbf{0.01541} & 0.01775 & 0.01648 & 0.01610 \\
		MAE  & 0.01390 & 0.01271 & \textbf{0.01256} & 0.01480 & 0.01375 & 0.01345 \\
		\bottomrule
	\end{tabular}
	\caption{\small Comparisons of RMSE and MAE of synthetic sample paths for models and factor specifications. \textbf{Bold} indicates the best.}
	\label{tbl:rmse_mae}
\end{table}

\begin{figure}[!p]
	\centering
	\subfigure[One-factor models]{
		\includegraphics[scale=0.6]{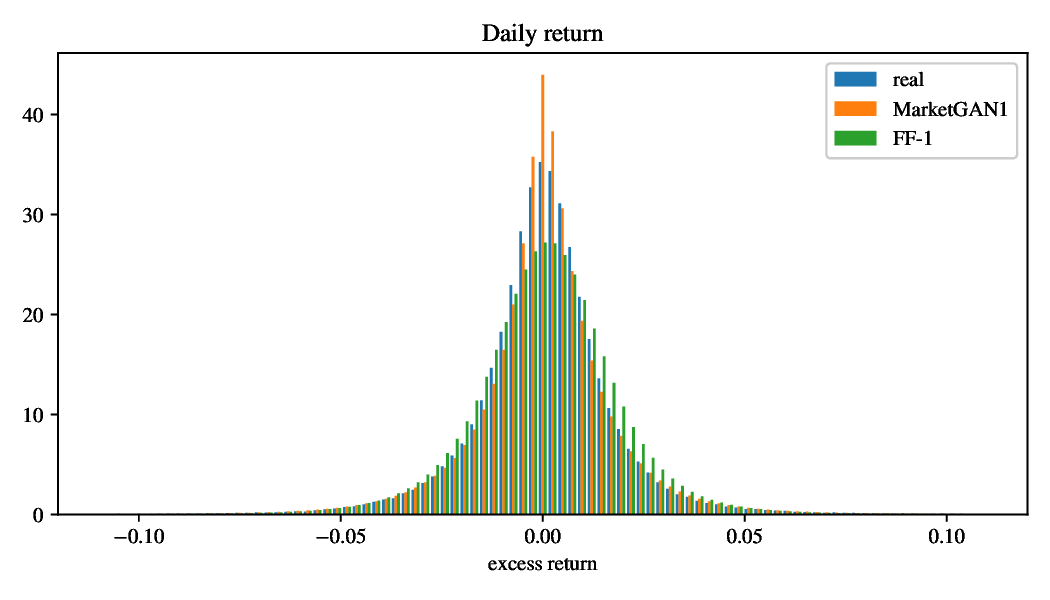}}
	\subfigure[Three-factor models]{
		\includegraphics[scale=0.6]{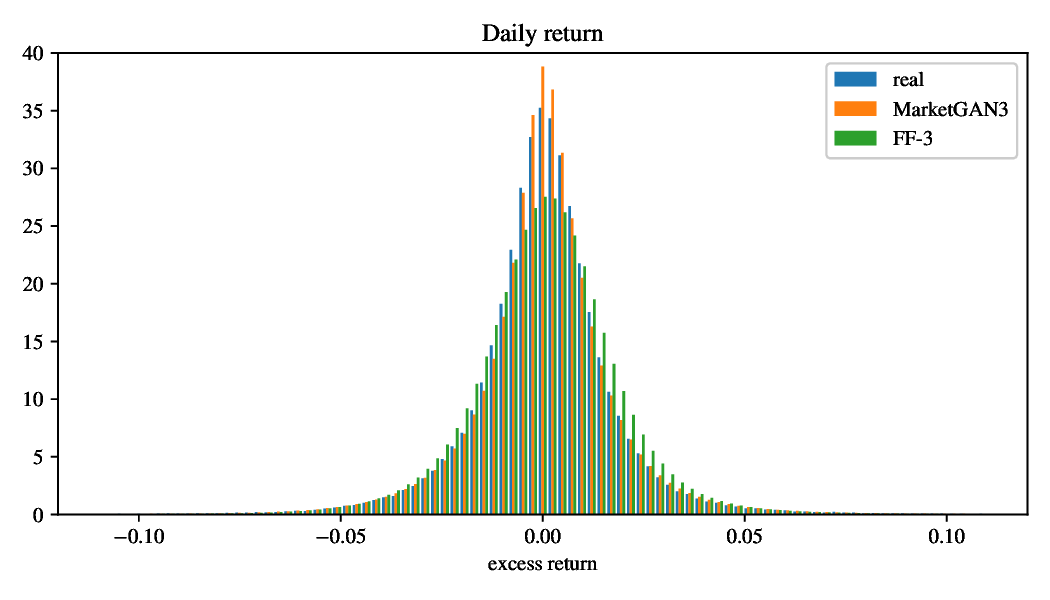}}
	\subfigure[Five-factor models]{
		\includegraphics[scale=0.6]{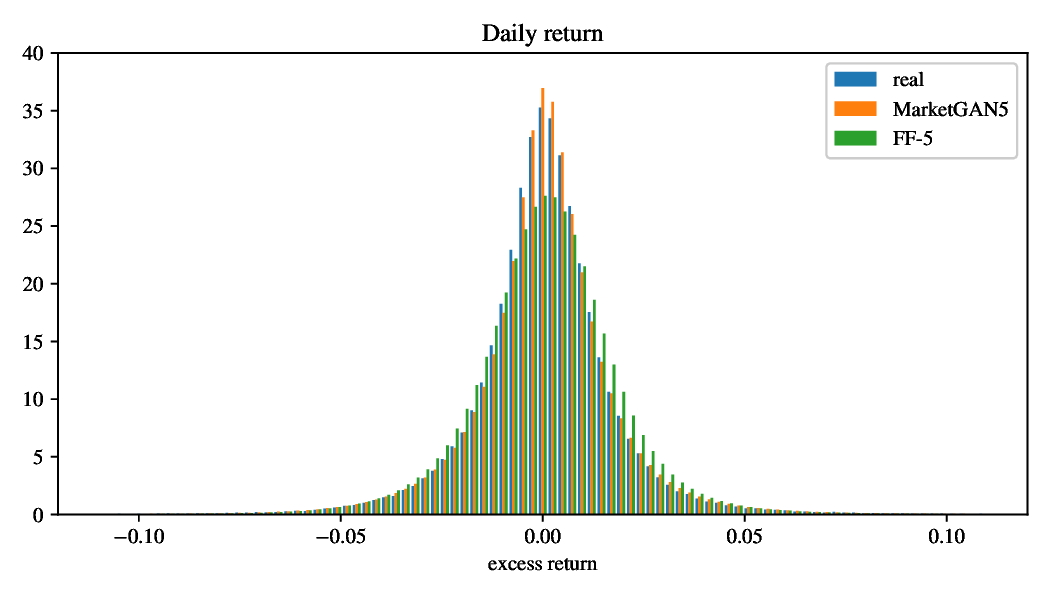}}
	\caption{Marginal distributions of daily excess returns. FF-1, FF-3, and FF-5 denote the respective factor-model-based bootstrap methods.
	}
	\label{fig: return histogram}
\end{figure}

We further assess the shape of marginal return distributions by comparing histograms of daily excess returns aggregated across assets. Figure \ref{fig: return histogram} contrasts the empirical distributions with those generated by MarketGAN and bootstrap methods under one-, three-, and five-factor specifications. Consistent with prior evidence on financial returns \citep{mandelbrot1963variation, cont2001empirical}, real data display pronounced leptokurtosis and heavy tails. MarketGAN-generated distributions closely resemble these features, whereas bootstrap-based distributions are visibly less peaked and exhibit thinner tails across all factor specifications. As the number of factors increases, MarketGAN tends to align more closely with the empirical distribution, whereas bootstrap methods do not exhibit a comparable improvement. We observe similar results for return distributions aggregated at weekly and monthly frequencies, where MarketGANs consistently demonstrate better alignment with real data than their bootstrap counterparts. These additional results are presented in Online Appendix \ref{appendix: weekly and monthly histogram}.

\begin{figure}[!h]
	\centering
	\subfigure[One-factor models]{
		\includegraphics[width=\textwidth]{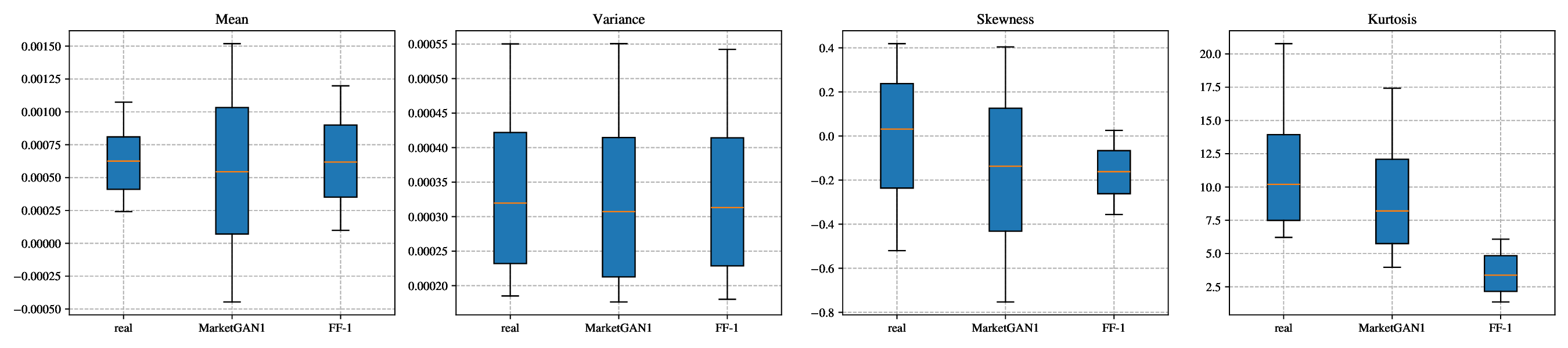}}
	\subfigure[Three-factor models]{
		\includegraphics[width=\textwidth]{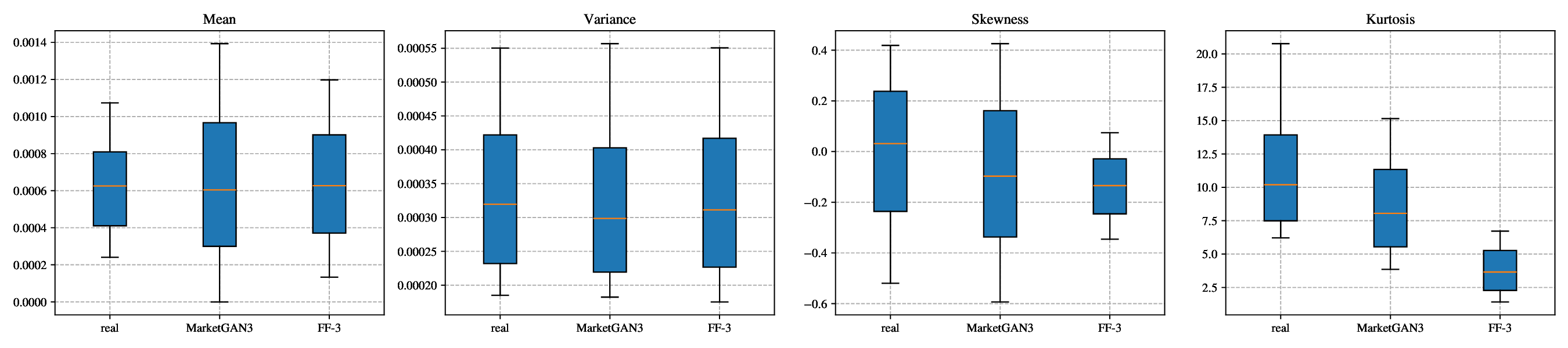}}
	\subfigure[Five-factor models]{
		\includegraphics[width=\textwidth]{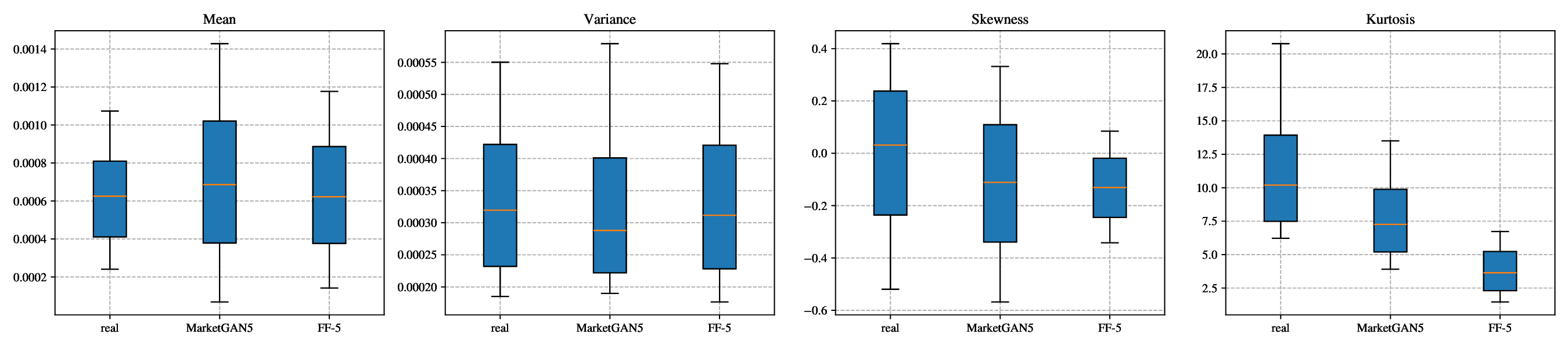}}
	\caption{Boxplots of the first to fourth moments (mean, variance, skewness, kurtosis) of asset returns generated by MarketGAN and factor-model-based bootstrap methods.
		Each boxplot summarizes the 10th, 25th, 50th, 75th, and 90th percentiles across assets.}
	\label{fig: Moments comparison}
\end{figure}

To provide a more systematic assessment, we compare the first four moments (mean, variance, skewness, and kurtosis) of marginal return distributions across assets. For each asset, these moments are computed using realized returns for the real data and using synthetic returns averaged over 100 generated paths for each model. Figure \ref{fig: Moments comparison} summarizes the cross-sectional distribution of these moments using boxplots. For the sample means, MarketGAN models may appear less aligned with the real data than the bootstrap methods. However, a closer inspection reveals that the discrepancies are not that significant, typically ranging between $2 \times 10^{-4}$ to $3 \times 10^{-4}$. The bootstrap methods’ slightly closer alignment can be attributed to their construction based on factor models with symmetric noise components. Conversely, MarketGAN generates synthetic returns by learning the distribution of stochastic coefficient processes via adversarial training without explicitly imposing symmetry restrictions. Nevertheless, MarketGAN models still produce nearly unbiased synthetic returns. Regarding the variance, both methods show comparable performance. However, MarketGAN exhibits a clear advantage in reproducing higher-order moments. MarketGAN captures skewness and kurtosis more accurately, reflecting its ability to model asymmetry and tail risk in return distribution. 


Despite these improvements, some limitations remain. For example, about 10\% of the real data exhibit extremely high kurtosis values exceeding 20 and a substantial deviation from the median kurtosis value of approximately 10, which MarketGAN does not fully reproduce. This suggests that while the model captures heavy-tailed behavior overall, it does not perfectly replicate the most extreme marginal tail events. These findings highlight both the strengths and current limitations of the proposed generative framework and motivate the subsequent analysis of inter-temporal and cross-sectional dependence, where MarketGAN’s advantages are more pronounced.

\subsection{Inter-temporal Dependencies}
\label{sec:inter-temporal_dependence}

We next evaluate whether MarketGAN reproduces inter-temporal dependence structures observed in financial return data. While marginal distributions capture static features of returns, realistic synthetic data must also reflect dynamic properties that characterize financial time series. We focus on three widely documented stylized facts: linear unpredictability, volatility clustering, and leverage effects. These features capture the absence of linear predictability in returns alongside the presence of persistent and asymmetric volatility dynamics. 

Linear unpredictability refers to the empirical observation that asset returns exhibit negligible autocorrelation at short horizons. This property is commonly assessed through the autocorrelation function (ACF), which measures the correlation between returns at different lags and typically decays rapidly toward zero. Specifically, $\mathrm{ACF}(k) = \mathrm{Corr}(r_{t+k}, r_{t})$. In contrast, volatility clustering is characterized by positive, persistent autocorrelation in squared returns, indicating that periods of high volatility tend to be followed by further periods of high volatility. It implies that $\mathrm{VC}(k) = \mathrm{Corr}(r_{t+k}^2, r_{t}^2)$ starts with a significant positive value and diminishes to zero. A related phenomenon is the leverage effect, whereby negative returns are associated with higher subsequent volatility, leading to a negative correlation between returns and future squared returns. In particular, $\mathrm{Lev}(k) = \mathrm{Corr}(r_{t+k}^2, r_{t})$ has a negative value for small lag and reverts to zero. The leverage effect is associated with increased volatility following negative returns, commonly observed in markets and closely related to negative skewness.

\begin{figure}[!htbp]
	\centering
	\subfigure[ACF of MarketGAN5]{
		\includegraphics[scale=0.48]{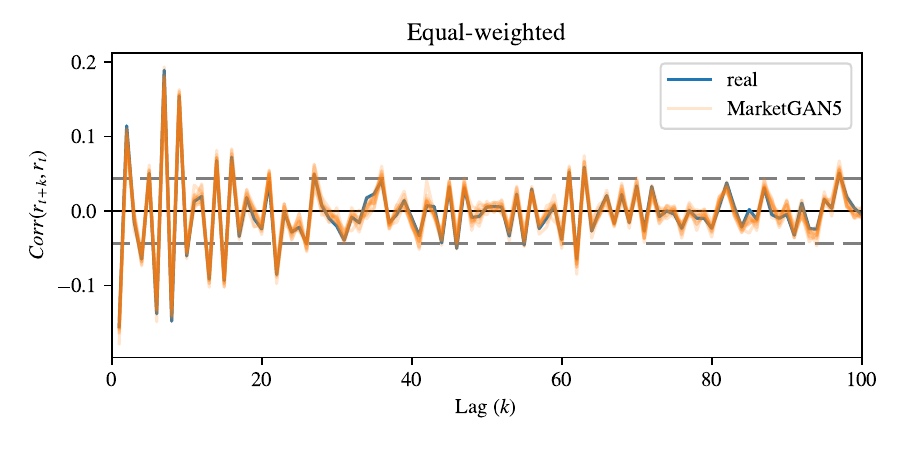}}
	\subfigure[ACF of FF-5 bootstrap]{
		\includegraphics[scale=0.48]{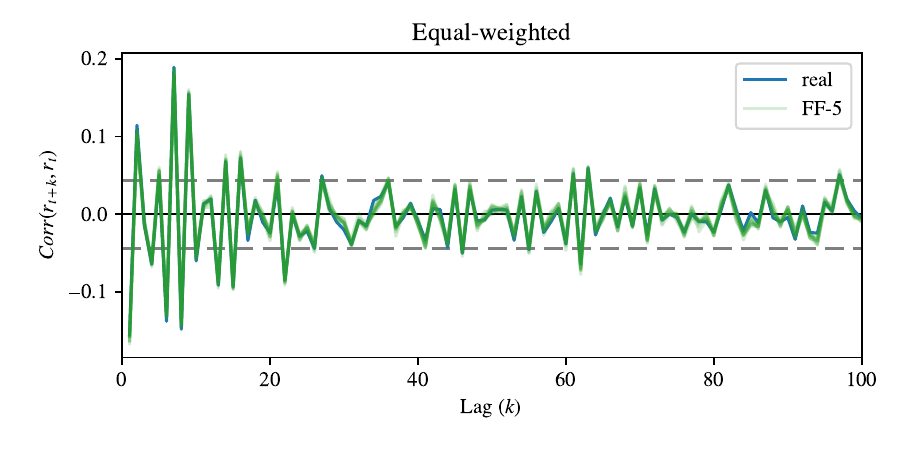}}
	\\
	\subfigure[VC of MarketGAN5]{
		\includegraphics[scale=0.48]{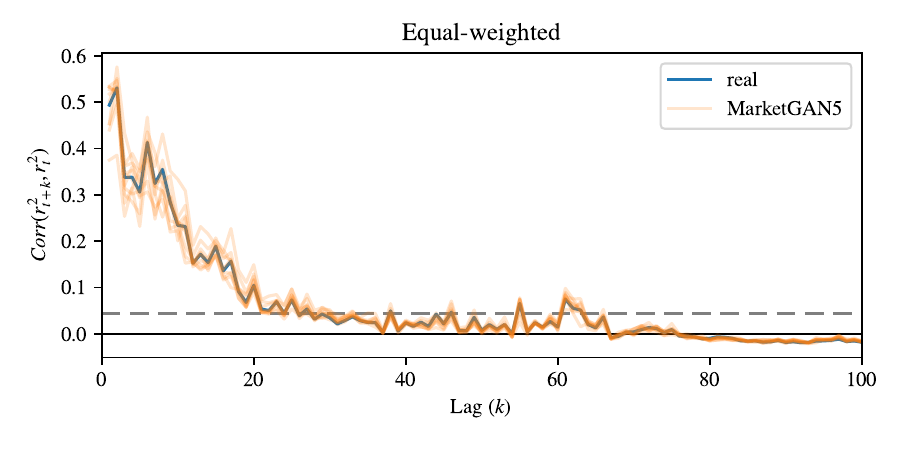}}
	\subfigure[VC of FF-5 bootstrap]{
		\includegraphics[scale=0.48]{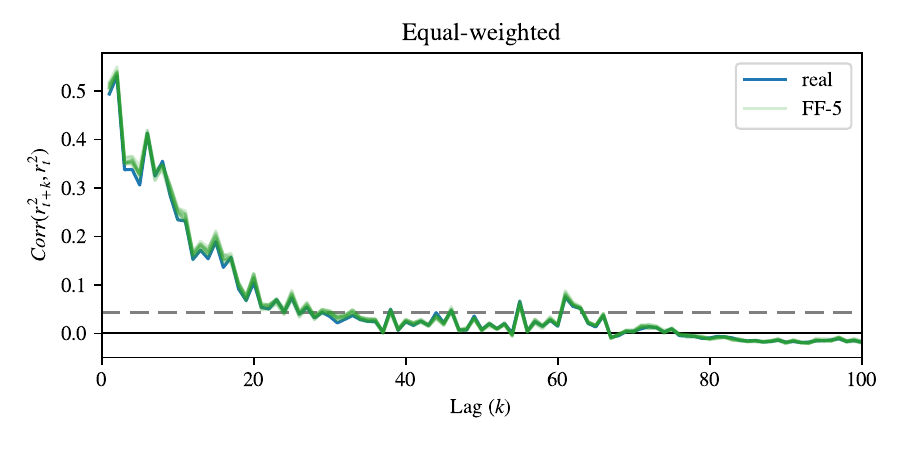}}
	\\
	\subfigure[Lev of MarketGAN5]{
		\includegraphics[scale=0.48]{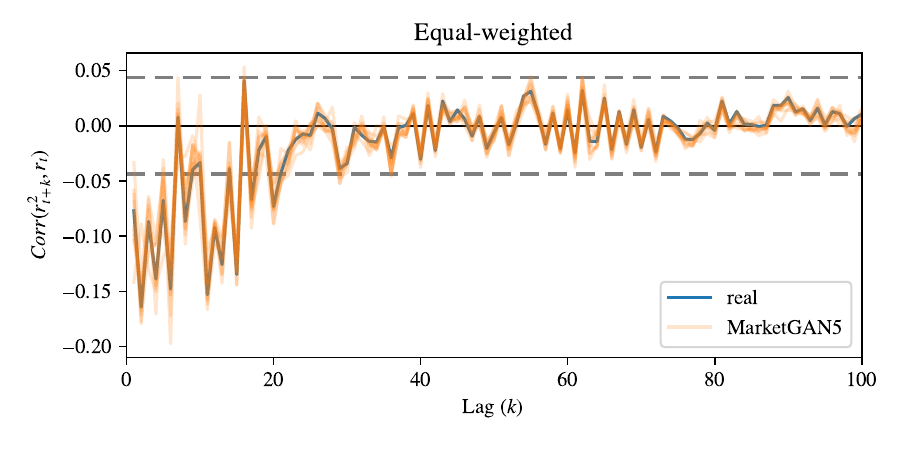}}
	\subfigure[Lev of FF-5 bootstrap]{
		\includegraphics[scale=0.48]{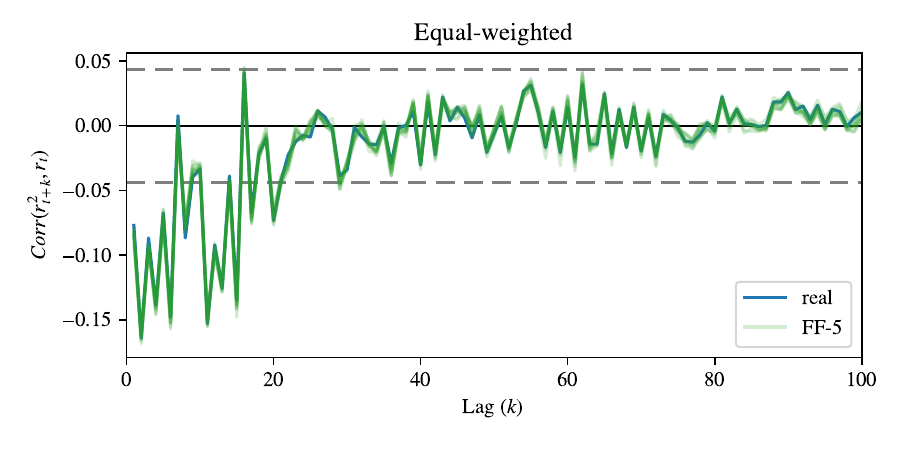}}
	\caption{ACF, VC, and Lev of real and synthetic equal-weighted excess return samples generated by the MarketGAN5 and FF-5 bootstrap.}
	\label{fig: autocorr_ew}
\end{figure}

We begin by examining these inter-temporal characteristics at the market level using equal-weighted returns. Figure \ref{fig: autocorr_ew} displays the ACF, volatility clustering (VC), and leverage effect (Lev) computed from synthetic sample paths generated by MarketGAN and from factor-model-based bootstrap methods, alongside the corresponding statistics from real data. For visual clarity, we randomly select 10 out of the 100 generated sample paths for plotting and focus on the five-factor specification in the main text, with results for one- and three-factor models reported separately in Online Appendix \ref{appendix: acf corrmatrix figures}. Across all three measures, MarketGAN-generated returns closely replicate the patterns observed in real data. The autocorrelation of returns remains close to zero across lags, while squared returns exhibit significant positive autocorrelation at short horizons that gradually decay. Similarly, the leverage effect is evident in the negative correlation between current returns and future volatility. {The factor-model-based bootstrap methods exhibit broadly similar aggregate behavior, reflecting the role of the factor structure in capturing market-wide dynamics}. This similarity is consistent with the accurate reconstruction of aggregate returns observed in the marginal distribution analysis. At the market level, therefore, both approaches succeed in reproducing key inter-temporal stylized facts.

We also examine the individual asset dynamics, again using GOOG as an illustrative example.\footnote{The ACF, VC, and Lev functions computed from real GOOG returns with those obtained from synthetic returns generated by five-factor MarketGAN and bootstrap methods are compared for one-, three-, and five-factor models, as provided in Online Appendix \ref{appendix: acf corrmatrix figures}.} Compared to the equal-weighted portfolio, individual asset returns exhibit greater variability in their autocorrelation patterns. Nevertheless, MarketGAN-generated series reproduce the qualitative features observed in the real data. Linear unpredictability is preserved, volatility clustering remains pronounced at short lags, and the leverage effect is clearly evident. Bootstrap methods yield similar qualitative patterns, though with slightly greater dispersion across synthetic paths.

Overall, the results indicate that MarketGAN effectively captures inter-temporal dependence structures characteristic of financial return series. Importantly, this performance is achieved without imposing explicit parametric assumptions on volatility dynamics, such as GARCH-type specifications. Instead, temporal dependence emerges endogenously through the TCN backbone that governs the evolution of stochastic factor loadings and volatilities. At the same time, the similarity between MarketGAN and bootstrap methods in some inter-temporal metrics suggests that factor structure alone accounts for a substantial portion of temporal dynamics at both the market and individual asset levels. These findings highlight a key distinction between temporal and cross-sectional aspects of generative fidelity. While factor-based approaches are often sufficient to reproduce aggregate and individual time-series dynamics, capturing cross-sectional dependence across many assets requires explicitly modeling the joint distribution. We examine this dimension in the next subsection, where we assess MarketGAN's ability to reproduce cross-sectional correlations and co-movements in asset returns. 

\subsection{Cross-sectional Dependence}
\label{sec:cross_dependence}

We now turn to cross-sectional dependence, which constitutes a central requirement for synthetic data intended for portfolio analysis and risk management. While marginal distributions and inter-temporal dynamics assess univariate and time-series fidelity, realistic multivariate return generation hinges on accurately reproducing dependence structures across assets. In particular, cross-sectional correlations and co-movements during extreme market events play a decisive role in diversification benefits and tail risk.

We assess the ability of MarketGAN to reproduce cross-sectional dependence along two complementary dimensions, which are cross-correlations of returns and cross-correlations of extreme returns. The first captures average co-movement across assets under normal market conditions, while the second focuses on dependence in the tail, which is known to intensify during market stress. These measures provide a comprehensive view of cross-asset dependence relevant for portfolio applications. 

\begin{figure}[!htbp]
	\centering
	
	\subfigure[Real data]{
		\includegraphics[scale=0.5]{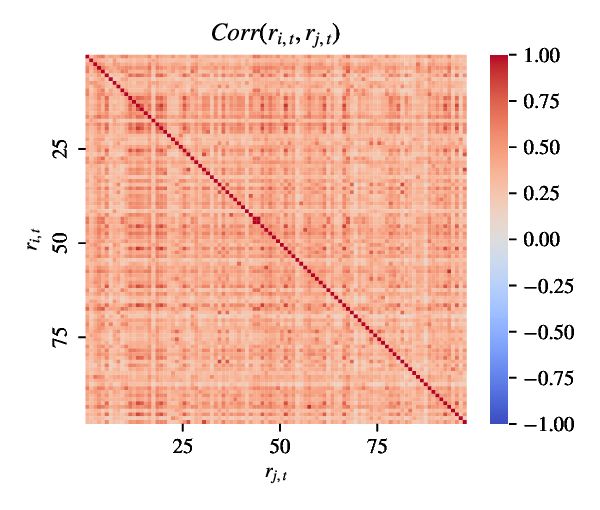}}
	\hfill
	\subfigure[MarketGAN1]{
		\includegraphics[scale=0.5]{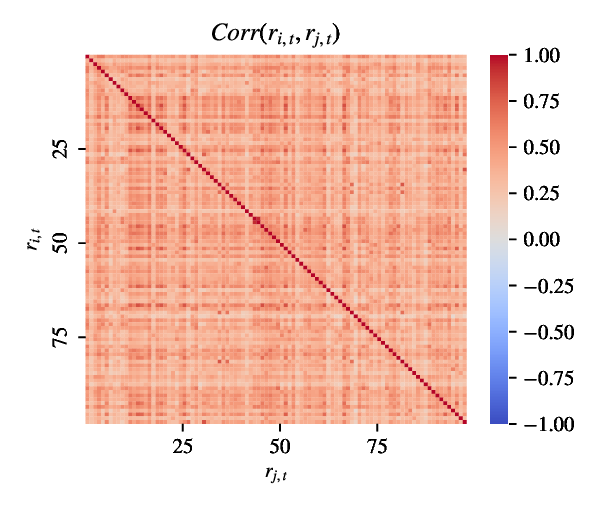}}
	\hfill
	\subfigure[FF-1 bootstrap]{
		\includegraphics[scale=0.5]{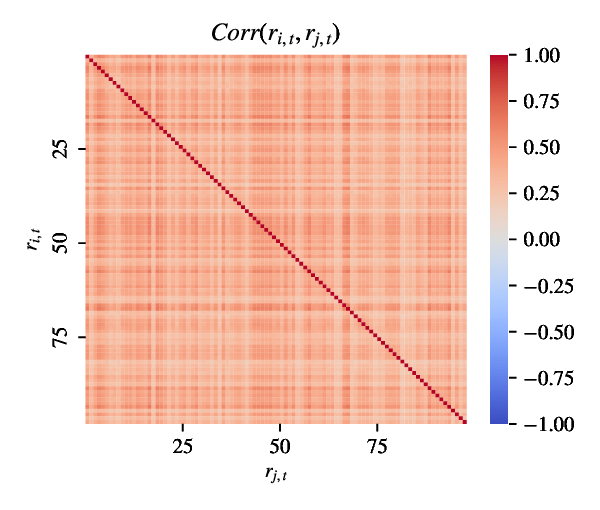}}
	\hfill
	\caption{Cross-correlation of real and synthetic excess return samples generated by MarketGANs and factor-model-based bootstrap methods.}
	\label{fig:crosscorr_onefactor}
\end{figure}

We begin by examining cross-correlations of returns defined by $\text{XCorr} = \text{Corr}(r_{i,t}, r_{j,t}),$ where $r_{i}$ and $r_j$ represent excess returns for assets $i$ and $j$, respectively. The matrix for real data represents the realized correlation of historical excess returns. For MarketGAN and bootstrap methods, we generate 100 independent synthetic paths covering the full out-of-sample period, compute the correlation matrix for each individual path and report the average of these 100 matrices. Figure \ref{fig:crosscorr_onefactor} shows the sample-estimated correlation matrices computed over the entire out-of-sample period for one-factor models. While the FF-1 bootstrap struggles to reproduce the rich correlation structure among nearly 100 assets using a single factor, MarketGAN captures a substantially more realistic dependence pattern. This result highlights the ability of adversarial learning to model complex joint distributions beyond what is achievable through factor structure alone,\footnote{Across all factor specifications, MarketGAN-generated correlation matrices more closely resemble those observed in real data than their bootstrap counterparts, as shown in Figure \ref{fig:crosscorr_comparison} in Online Appendix. This improvement is particularly pronounced in the one-factor specification.}

As the number of factors increases, both MarketGAN and bootstrap methods show improved performance in reproducing cross-sectional correlations. However, MarketGAN consistently exhibits closer alignment with the empirical correlation structure. In particular, the difference between MarketGAN and bootstrap methods narrows for the three- and five-factor models but does not disappear, suggesting that generative modeling complements, rather than replaces, economic factor structure in capturing cross-sectional dependence. This advantage can be attributed to the inherent ability of GANs to model complex joint distributions among multiple components, just as GANs applied to image data effectively capture pixel-level correlation structures. 

We next examine dependence in extreme returns defined by
$\mathrm{XCorr_E} = P(r_{i,t} < -q_i \mid r_{j,t} < -q_j),$ where $q_i$ and $q_j$ denote the 95\% quantile levels for the returns of assets $i$ and $j$, respectively. It is measured by the conditional probability that one asset experiences a large negative return, given that another asset simultaneously experiences an extreme loss. This measure captures the tendency for correlations to increase sharply during market downturns, a well-documented empirical phenomenon with direct implications for tail risk. {Extreme cross-sectional correlations display patterns similar to those observed for general cross-correlations, with MarketGAN models exhibiting superior performance and fidelity improving as the number of factors increases.} For brevity, detailed figures for extreme cross-correlations are provided in Online Appendix \ref{appendix: acf corrmatrix figures}.

These findings underscore a key distinction between factor-based resampling and generative modeling. Factor-model-based bootstrap methods rely on independently drawn residuals and therefore struggle to capture complex dependence structures, particularly in the tails. In contrast, MarketGAN generates returns as a single joint vector, allowing the model to learn and reproduce high-dimensional dependence patterns across assets. This joint-generation design is critical for accurately representing correlation structures that matter for portfolio diversification and risk assessment.

\subsection{Quantitative Evaluation Metrics}
\label{sec:quantitative_evaluation_metrics}
Having primarily focused on qualitative evaluations thus far, we now turn to quantitative metrics to rigorously assess MarketGAN's performance.  While the visual and descriptive results in Sections \ref{sec:return_distribution}-\ref{sec:cross_dependence} highlight how MarketGAN reproduces marginal distributions, inter-temporal dynamics, and cross-sectional dependence, quantitative measures provide a compact and comparable assessment across models and factor specifications. We consider four distance-based metrics that capture different aspects of distributional and temporal similarity. They are the Fréchet Inception Distance (FID), the Sliced Wasserstein Distance (SWD), the Mahalanobis Distance (MD), and the Dynamic Time Warping (DTW).\footnote{FID, SWD, and MD are widely utilized metrics for evaluating generative neural networks. 
	FID approximates the Wasserstein-2 distance by using only the first two moments, mean and covariance.
	This amounts to estimating the Wasserstein-2 distance assuming multivariate Gaussian distributions  \citep{heusel2017gans}.
	SWD approximates the Wasserstein-1 distance by projecting high-dimensional distributions onto multiple one-dimensional subspaces and averaging their respective Wasserstein distances \citep{bonneel2015sliced}.
	MD measures the distance between a point and a distribution considering the covariance structure.
	It thus becomes particularly sensitive to anomalies and useful for anomaly detection \citep{de2000mahalanobis}. 
	DTW computes an optimal alignment between two time-series, accommodating nonlinear temporal distortions and capturing phase-shifted patterns \citep{muller2007dynamic}.} In addition, we compute quantitative scores for the inter-temporal and cross-sectional characteristics examined previously, including measures based on ACF, VC, Lev, $\mathrm{XCorr}$, and $\mathrm{XCorr_E}$. For all metrics, lower values indicate closer alignment between synthetic and real data. These scores are defined as the $L^2$ distance (Frobenius norm) between estimates from real and synthetic data, with lags up to 100 considered for inter-temporal metrics.

All metrics are computed by averaging over 100 synthetic sample paths generated by each model. Table \ref{tbl:metric_full} reports the results for MarketGAN and factor-model-based bootstrap methods under one-, three-, and five-factor specifications. Consistent with the qualitative evidence, MarketGAN outperforms bootstrap methods across nearly all metrics. In particular, MarketGAN achieves substantially lower values in FID and SWD, indicating that the distributions induced by MarketGAN more closely match the empirical return distribution in terms of Wasserstein-based distances. Improvements are also evident in MD and DTW, suggesting that MarketGAN-generated samples exhibit fewer distributional anomalies and better temporal alignment with real return series.

\begin{table}[!htb]
	\centering
	\begin{tabular}{lccccccccc}
		\toprule
		\multicolumn{1}{c}{\multirow{2}{*}{Metrics}} &  & \multicolumn{3}{c}{MarketGAN} & \multicolumn{3}{c}{Factor-model-based bootstrap} \\ 
		\cmidrule(lr){3-5} \cmidrule(lr){6-8} 
		& {Factors} & 1  & 3 & 5 & 1 & 3 & 5 \\ 
		\midrule
		\multicolumn{2}{l}{FID} & 0.00189 & 0.00148 & \textbf{0.00146} & 0.00362 & 0.00189 & 0.00149 \\ \midrule
		\multicolumn{2}{l}{SWD} & 0.00116 & \textbf{0.00086} & 0.00092 & 0.00147 & 0.00123 & 0.00117 \\ \midrule
		\multicolumn{2}{l}{MD}  & 11.23570 & 10.69934 & \textbf{10.46770} & 15.44894 & 14.26671 & 13.86999 \\ \midrule
		\multicolumn{2}{l}{DTW} & 9.17709 & 8.43424 & \textbf{8.32902} & 9.19236 & 8.57615 & 8.41928 \\ \midrule
		\multicolumn{2}{l}{ACF} & 0.02438 & 0.02378 & 0.02356 & 0.02417 & 0.02326 & \textbf{0.02301} \\ \midrule
		\multicolumn{2}{l}{VC}  & 0.04108 & \textbf{0.03837} & 0.03952 & 0.05202 & 0.05330 & 0.05283 \\ \midrule
		\multicolumn{2}{l}{Lev} & 0.02609 & 0.02593 & \textbf{0.02570} & 0.02650 & 0.02613 & 0.02603 \\ \midrule
		\multicolumn{2}{l}{XCorr} & 5.54324 & \textbf{4.51026} & 4.87561 & 8.90126 & 5.93517 & 4.89255 \\ \midrule
		\multicolumn{2}{l}{$\mathrm{XCorr_E}$} & 5.76577 & 5.16640 & \textbf{4.97701} & 6.36357 & 5.49201 & 5.27223 \\ 
		\bottomrule
	\end{tabular}
	\caption{Comparison of evaluation metrics across the MarketGANs and factor-model-based bootstrap methods under different factor specifications. Lower values indicate better performance, and \textbf{bold} represents the best.}
	\label{tbl:metric_full}
\end{table}

Turning to metrics associated with inter-temporal dependence, MarketGAN and bootstrap methods perform similarly with respect to linear autocorrelation and leverage effects. It implies the strong role of factor structure in capturing these dynamics. However, MarketGAN exhibits a marked advantage in reproducing volatility clustering, consistent with the results in Section \ref{sec:inter-temporal_dependence} and the model’s explicit temporal representation through a TCN backbone.

The most pronounced differences arise in cross-sectional metrics. MarketGAN substantially outperforms bootstrap methods in both $\mathrm{XCorr}$ and $\mathrm{XCorr_E}$, particularly under lower-dimensional factor specifications. These results quantitatively confirm the visual evidence presented in Section \ref{sec:cross_dependence}. In particular, it generates asset returns as a single joint vector, allowing MarketGAN to capture complex dependence structures that are difficult to reproduce with independently sampled residuals. Notably, MarketGAN with three or five factors achieves comparable or superior performance relative to bootstrap methods with richer factor structures, underscoring the complementary role of adversarial learning and economic factor specification.

\section{Applications: Portfolio Optimization} \label{sec: portfolio optim}

The preceding section demonstrates that MarketGAN generates synthetic asset returns that closely replicate key statistical properties of real financial data. {In this section, we assess the economic value of MarketGAN-generated data through a standard portfolio optimization exercise—specifically, the construction of a daily-rebalanced long-only mean-variance tangency portfolio.\footnote{Since MarketGAN is designed to generate returns at a daily frequency, adopting a daily rebalancing scheme allows us to directly evaluate the economic value of the synthetic joint distribution.} }Portfolio choice provides a natural and demanding downstream test for generative models, as optimal allocations are highly sensitive to estimation errors in return distributions, especially in the presence of cross-sectional correlation and tail risk. Models that fail to capture joint dependence structures may appear adequate when evaluated marginally, yet perform poorly when used for portfolio construction. Our objective is not to propose a new portfolio strategy, but to evaluate whether synthetic data generated by MarketGAN can serve as reliable inputs for standard portfolio optimization problems. To this end, we compare portfolios constructed using MarketGAN-generated returns with those based on synthetic data from factor-model-based bootstrap methods and, where appropriate, real historical data. By holding the portfolio optimization procedure fixed, we isolate the effect of the data generation mechanism on economic outcomes. In words, we focus on whether improvements in statistical fidelity documented in Section \ref{sec: experiments} lead to economically meaningful gains in portfolio performance.

\subsection{Portfolio Construction and Experimental Design}

We consider daily-rebalaned long-only mean-variance tangency portfolios based on covariance matrices estimated from $N$-dimensional synthetic return samples. {The portfolio is based on the same universe of $N=98$ S\&P 100 constituents detailed in Section \ref{sec: experimental setup}, with the construction procedure implemented as follows. At each rebalancing date $t$, we first obtain a realized future factor vector $\boldsymbol{F}_{t+1}$ using the forecasting or simulation methods described below. MarketGAN then generates a large set of synthetic excess returns for time $t+1$ by combining generated stochastic coefficients with realizations of the factor vector. Specifically, for a given factor realization $\boldsymbol{F}_{t+1}$, MarketGAN exploits the stochasticity of its generated coefficients ($\boldsymbol{\alpha}_t, \boldsymbol{\beta}_t, \boldsymbol{\sigma}_t$) and idiosyncratic residuals ($\boldsymbol{\epsilon}_{t+1}$) to produce 10,000 distinct synthetic return vectors. The expected return vector $\hat{\mu}_{t+1}$ and the covariance matrix $\hat{\Sigma}_{t+1}$ are computed as the empirical mean and sample covariance of this generated set, respectively.} Because the covariance matrix is estimated dynamically at each date using these synthetic returns, portfolios are rebalanced daily to align with the data generation process.

A key practical consideration is the availability of future factor realizations $\boldsymbol{F}_{t+1}$ as given in \eqref{eqn: marketgan}. To address this issue, we consider two approaches. First, we forecast future factor values using simple time-series methods, including a one-year rolling average and a vector autoregression (VAR). Second, we conduct a perturbed forecast simulation in which controlled levels of noise are added to realized factor values to generate unbiased but noisy predictors with predetermined predictive accuracy. These two approaches allow us to separate the effect of factor forecasting error from the effect of the synthetic data generation mechanism itself.

To isolate the contribution of MarketGAN, we compare portfolio performance against several benchmark portfolios. These include the market portfolio and mean–variance portfolios constructed using alternative covariance estimation methods. Specifically, we consider covariance matrices estimated from historical sample returns, the Ledoit–Wolf shrinkage estimator \citep{ledoit2004well}, and factor-model-based covariance estimates using FF-1, FF-3, and FF-5 models. All benchmark covariance matrices are estimated using rolling windows of three years of daily return data. 

Portfolio performance is evaluated using multiple criteria, including annualized excess return, annualized volatility, Sharpe ratio, maximum drawdown, and turnover. Turnover is computed as given in \cite{gu2020empirical}, $$\mathrm{Turnover} = \frac{1}{T} \sum_{t=1}^T \left( \sum_i \left| w_{i, t+1}-\frac{w_{i, t} (1+r_{i, t+1})}{1+\sum_j w_{j, t}r_{j, t+1} }\right| \right).$$  It captures the extent of portfolio rebalancing required by each approach. While transaction costs are not explicitly incorporated into the optimization problem, we later assess the robustness of performance to plausible levels of trading costs.

MarketGAN models are trained using a rolling window procedure designed to reflect evolving market conditions. Initially, we train the models using data from the in-sample period spanning January 1, 1991, to December 31, 2015, splitting this interval into training and validation subsets with a 7:1 ratio as described in Section \ref{sec: experimental setup}. The trained model parameters are then updated quarterly by rolling the estimation window forward. Each rolling update maintains the same 7:1 training-to-validation ratio and uses the previously trained model parameters as initialization. Given the pre-trained initialization, we train MarketGANs for only 50 epochs per rolling window and apply an early stopping criterion based on the Frobenius norm between correlation matrices computed from synthetic and real validation data. To mitigate the exclusion of the most recent observations from direct training, we perform a brief fine-tuning step using the validation data with a reduced learning rate.\footnote{Training stops if no improvement is observed over 20 consecutive epochs, and the best-performing parameters are retained. However, this splitting strategy has a limitation: the most recent data in each window are used only for validation and are not included in the training set, so they are not directly fed into the model.
	Since the purpose of rolling training is to adapt to evolving market conditions, excluding the most recent data from direct training may reduce the model’s responsiveness. To address this issue, we fine-tune the trained models on the validation data for an additional 10 epochs, with the learning rate reduced to one-tenth of the original.}

\subsection{Results}

We now present portfolio optimization results and explicitly compare MarketGAN-based portfolios with a comprehensive set of benchmark portfolios. The objective is to assess whether synthetic data generated by MarketGAN provides incremental economic value relative to both traditional covariance estimators and factor-model-based bootstrap methods. 

We begin by summarizing the performance of benchmark portfolios under daily rebalancing. Table \ref{tbl:benchmark_portfolios_daily} reports annualized excess returns, volatility, Sharpe ratios, maximum drawdown, and turnover for the benchmark strategies. Portfolios constructed using factor-model-based covariance estimators achieve particularly strong performance, with Sharpe ratios close to or slightly above one. This strong performance reflects the favorable market conditions during the out-of-sample period and confirms that well-specified factor models already provide a competitive baseline for portfolio construction.

\begin{table}[!htb]
	\centering
	\begin{tabular}{lcccccc}
		\toprule
		Model & Factors & SR & Return & Std & MDD & Turnover \\
		\midrule
		Market Portfolio & - & 0.6613 & 0.1265 & 0.1913 & $-0.3437$ & - \\
		\midrule
		Sample-estimate covariance & - & 0.9034 & 0.1911 & 0.2116 & $-0.3538$ & 1.5748 \\
		Ledoit-Wolf Shrinkage covariance & - & 0.9119 & 0.1932 & 0.2118 & $-0.3551$ & 1.5656 \\
		\midrule
		\multirow{3}{*}{Factor-model-based covariance} 
		& 1 & 0.9838 & 0.1750 & 0.1779 & $-0.2407$ & 1.5069 \\
		& 3 & 1.0183 & 0.1937 & 0.1902 & $-0.2702$ & 1.6642 \\
		& 5 & 0.9975 & 0.1875 & 0.1880 & $-0.2698$ & 1.6458 \\
		\bottomrule
	\end{tabular}
	\caption{\small Performance of benchmark portfolios under daily rebalancing frequency. 
		The table compares portfolios based on different covariance matrix estimation methods including the sample-estimate covariance, Ledoit–Wolf shrinkage, and factor-model-based covariance using 1, 3, and 5 factors.
		SR, Return, Std, MDD and Turnover indicate the Sharpe ratio, annualized excess return, annualized standard deviation, maximum drawdown, and monthly turnover. }
\label{tbl:benchmark_portfolios_daily}
\end{table}
In comparison, portfolios based on the sample covariance matrix and Ledoit–Wolf shrinkage perform slightly worse but remain economically meaningful. These results imply that synthetic-data-based approaches should be evaluated relative to covariance estimation methods that already exhibit strong out-of-sample performance, rather than against weak or ad hoc benchmarks.

We next examine portfolios constructed from synthetic returns, $\boldsymbol{r}_{t+1}$, generated by MarketGAN and factor-model-based bootstrap methods, with future factor realizations forecast, $\boldsymbol{F}_{t+1}$, using standard time-series techniques. Two forecasting methods are considered: a one-year rolling average and a VAR. The resulting portfolio performance is reported in Tables \ref{tbl:port_rolling_avg} and \ref{tbl:port_var}, respectively.

\begin{table}[!htb]
\centering
\begin{tabular}{lcccccc}
	\toprule
	Model & Factors & SR & Return & Std & MDD & Turnover \\
	\midrule
	\multirow{4}{*}{MarketGAN} 
	& 1 & 0.7406 & 0.1357 & 0.1832 & $-0.3555$ & 5.2746 \\
	\cmidrule(lr){2-7}
	& 3 & 0.8952 & 0.1593 & 0.1780 & $-0.3312$ & 4.9584 \\
	\cmidrule(lr){2-7}
	& 5 & 0.8360 & 0.1426 & 0.1706 & $-0.3084$ & 4.1749 \\
	\midrule
	\multirow{4}{*}{\begin{tabular}{@{}l@{}} Factor-model-based \\ bootstrap \end{tabular}}
	& 1 & 0.8044 & 0.1399 & 0.1739 & $-0.3080$ & 4.7030 \\
	\cmidrule(lr){2-7}
	& 3 & 0.8194 & 0.1444 & 0.1763 & $-0.3091$ & 4.5435 \\
	\cmidrule(lr){2-7}
	& 5 & 0.8174 & 0.1449 & 0.1772 & $-0.3040$ & 4.5067 \\
	\bottomrule
\end{tabular}
\caption{\small Performance of long-only mean-variance portfolios using synthetic sample covariance matrices, with factors $\boldsymbol{F}_{t+1}$ predicted via 1-year rolling averages. SR, Return, Std, MDD, and Turnover indicate the Sharpe ratio, annualized excess return, annualized standard deviation, maximum drawdown, and monthly turnover. }
\label{tbl:port_rolling_avg}
\end{table}

\begin{table}[!htb]
\centering
\begin{tabular}{lcccccc}
	\toprule
	Model & Factors & SR & Return & Std & MDD & Turnover \\
	\midrule
	\multirow{4}{*}{MarketGAN} 
	& 1 & 1.0610 & 0.2110 & 0.1989 & $-0.3223$ & 13.4172 \\
	\cmidrule(lr){2-7}
	& 3 & 0.6849 & 0.1444 & 0.2109 & $-0.4108$ & 17.5554 \\
	\cmidrule(lr){2-7}
	& 5 & 0.6428 & 0.1388 & 0.2159 & $-0.3953$ & 18.0836 \\
	\midrule
	\multirow{4}{*}{\begin{tabular}{@{}l@{}} Factor-model-based \\ bootstrap \end{tabular}}
	& 1 & 0.8588 & 0.1979 & 0.2304 & $-0.4189$ & 12.8247 \\
	\cmidrule(lr){2-7}
	& 3 & 0.6328 & 0.1481 & 0.2340 & $-0.4506$ & 16.5555 \\
	\cmidrule(lr){2-7}
	& 5 & 0.8109 & 0.1926 & 0.2375 & $-0.4528$ & 17.9810 \\
	\bottomrule
\end{tabular}
\caption{Performance of mean-variance portfolios using synthetic sample covariance matrices, with factors $\boldsymbol{F}_{t+1}$ predicted via a VAR model. SR, Return, Std, MDD, and Turnover indicate the Sharpe ratio, annualized excess return, annualized standard deviation, maximum drawdown, and monthly turnover. }
\label{tbl:port_var}
\end{table}

Across both forecasting approaches, MarketGAN-based portfolios modestly outperform their corresponding bootstrap counterparts in some cases, but their performance is not consistently superior to that of benchmark portfolios based on traditional covariance estimators or factor-model-based covariance estimates. In particular, Sharpe ratios achieved by MarketGAN-based portfolios under forecasted factors remain comparable to, but not markedly higher than, those of the best-performing benchmark portfolios. These results imply that when factor forecasts are imprecise, improvements in modeling return distributions alone are insufficient to generate large economic gains. In such settings, factor predictability becomes the primary bottleneck, and even sophisticated generative models cannot fully compensate for inaccurate inputs.

A key insight underlying these results is that independent resampling errors propagate nonlinearly through the optimization process. Factor-model-based bootstrap methods generate synthetic returns by resampling idiosyncratic components independently across assets, implicitly assuming that residual dependence is either negligible or sufficiently captured by the factor structure. While this assumption may appear innocuous at the level of marginal distributions or covariance estimation, it becomes consequential once synthetic data are used as inputs to nonlinear decision problems such as portfolio optimization. Mean–variance portfolio choice is a nonlinear mapping from the joint return distribution to optimal allocations, and small distortions in dependence structures—particularly in low-variance directions or in the tails—can be amplified through covariance inversion and constraint interactions. This mechanism motivates the perturbed forecast simulation design that follows, which isolates the value of accurately modeling joint dependence independently of factor predictability.

To isolate the contribution of synthetic data generation from factor forecasting accuracy, we conduct a perturbed forecast simulation. Instead of relying on a specific factor forecasting model, we analyze the economic value of the generated joint distributions across a spectrum of factor information quality. This is achieved by taking the realized future factors $\boldsymbol{F}_{t+1}$ and injecting controlled levels of Gaussian noise $\boldsymbol{\eta}_{t+1}$ to simulate varying degrees of predictive reliability. Specifically, we construct noisy predictors $\tilde{\boldsymbol{F}}_{t+1} = \boldsymbol{F}_{t+1} + \boldsymbol{\eta}_{t+1}$, where the noise variance is calibrated to yield targeted predictive $R^2$ levels: 100\% (i.e., without any perturbation), 50\%, 10\%, 1\%, and 0.1\%. Given these simulated factor realizations, we generate the $N$-dimensional asset returns through the generative models and utilize them to construct portfolios. This simulation experiment enables us to assess the fidelity with which MarketGAN replicates high-dimensional dependence structures, independent of the inherent difficulty in forecasting the underlying factors.


The results, summarized in Table \ref{tbl:port_simulation_r2}, reveal a sharp contrast between MarketGAN-based portfolios and both bootstrap-based portfolios and traditional benchmarks. When factor information is highly informative, MarketGAN3 and MarketGAN5 achieve exceptionally high Sharpe ratios, substantially exceeding those of all benchmark portfolios. In particular, under ideal predictive conditions ($R^2=100\%$), mean-variance portfolio Sharpe ratios approach values around 5. Notably, even at low predictive accuracies for future factors (e.g., $R^2=0.1\%$), MarketGAN3 and MarketGAN5 achieve Sharpe ratios exceeding 3.3 and outperform their corresponding bootstrap counterparts by margins greater than 1. Their results also substantially surpass those of MarketGAN1, which delivers less pronounced improvements while still outperforming its corresponding bootstrap method. In contrast, bootstrap portfolios experience a pronounced deterioration in performance as factor predictability declines. This divergence underscores the importance of accurately modeling cross-sectional dependence structures, which MarketGAN captures through joint generation of asset returns. This superior performance can be attributable to the capability of MarketGAN3 and MarketGAN5 to accurately capture cross-sectional correlation structures as previously demonstrated in Table \ref{tbl:metric_full}.

The superior performance of MarketGAN-based portfolios, particularly under perturbed factor simulations, is accompanied by higher turnover due to daily rebalancing. To assess economic robustness, we evaluate performance under plausible assumptions with a transaction cost of 10 basis points and a monthly turnover of 3,400\%. Even after accounting for these assumptions, MarketGAN3 and MarketGAN5 retain economically meaningful Sharpe ratios and positive excess returns across a wide range of factor predictability levels. For example, even at a low predictive accuracy $R^2=0.1\%$, the annualized returns net of transaction costs are 38.74\% and 35.80\%, respectively, with Sharpe ratios of 1.62 in both cases. Importantly, benchmark portfolios with lower turnover do not exhibit comparable robustness when factor information is noisy. This suggests that the gains from MarketGAN are not solely artifacts of aggressive trading, but reflect genuine improvements in covariance estimation driven by higher-fidelity synthetic data.

\begin{table}[!p]
\centering
\footnotesize{
\begin{tabular}{llcccccc}
	\toprule
	$R^2$ & Model & Factors & Sharpe Ratio & Annualized Return & Annualized Std & MDD & Turnover \\
	\midrule
	\multirow{8}{*}{$100\%$}
	& \multirow{4}{*}{MarketGAN} 
	& 1 & 2.4339 & 0.5739 & 0.2358 & $-0.3310$ & 28.0504 \\
	\cmidrule(lr){3-8}
	&  & 3 & 5.3444 & 1.1332 & 0.2120 & $-0.2570$ & 31.9897 \\
	\cmidrule(lr){3-8}
	&  & 5 & 4.9003 & 1.0748 & 0.2193 & $-0.2402$ & 32.3922 \\
	\cmidrule(lr){2-8}
	& \multirow{4}{*}{\begin{tabular}{@{}l@{}} Factor-model-based \\ bootstrap \end{tabular}}
	& 1 & $-0.2663$ & $-0.0905$ & 0.3399 & $-0.7172$ & 26.7663 \\
	\cmidrule(lr){3-8}
	&  & 3 & 2.1898 & 0.6241 & 0.2850 & $-0.3763$ & 27.9007 \\
	\cmidrule(lr){3-8}
	&  & 5 & 3.3567 & 0.8860 & 0.2639 & $-0.3584$ & 27.9944 \\
	\midrule
	\multirow{8}{*}{$50\%$}
	& \multirow{4}{*}{MarketGAN} 
	& 1 & 2.0380 & 0.4853 & 0.2381 & $-0.3262$ & 28.4286 \\
	\cmidrule(lr){3-8}
	&  & 3 & 3.9247 & 0.9073 & 0.2312 & $-0.2766$ & 33.3657 \\
	\cmidrule(lr){3-8}
	&  & 5 & 3.5286 & 0.7978 & 0.2261 & $-0.2087$ & 33.6198 \\
	\cmidrule(lr){2-8}
	& \multirow{4}{*}{\begin{tabular}{@{}l@{}} Factor-model-based \\ bootstrap \end{tabular}}
	& 1 & 0.1598 & 0.0600 & 0.3753 & $-0.4712$ & 26.1087 \\
	\cmidrule(lr){3-8}
	&  & 3 & 2.1049 & 0.6413 & 0.3047 & $-0.4238$ & 28.8547 \\
	\cmidrule(lr){3-8}
	&  & 5 & 2.4166 & 0.6858 & 0.2838 & $-0.4113$ & 28.5285 \\
	\midrule
	\multirow{8}{*}{$10\%$}
	& \multirow{4}{*}{MarketGAN} 
	& 1 & 2.0055 & 0.4758 & 0.2372 & $-0.3267$ & 28.6623 \\
	\cmidrule(lr){3-8}
	&  & 3 & 3.3418 & 0.7978 & 0.2387 & $-0.3222$ & 33.7781 \\
	\cmidrule(lr){3-8}
	&  & 5 & 3.5675 & 0.7844 & 0.2199 & $-0.2666$ & 33.9254 \\
	\cmidrule(lr){2-8}
	& \multirow{4}{*}{\begin{tabular}{@{}l@{}} Factor-model-based \\ bootstrap \end{tabular}}
	& 1 & 0.5258 & 0.1986 & 0.3777 & $-0.4712$ & 25.7059 \\
	\cmidrule(lr){3-8}
	&  & 3 & 1.8999 & 0.6037 & 0.3178 & $-0.4705$ & 29.1764 \\
	\cmidrule(lr){3-8}
	&  & 5 & 2.4771 & 0.6706 & 0.2707 & $-0.4174$ & 28.6850 \\
	\midrule
	\multirow{8}{*}{$1\%$}
	& \multirow{4}{*}{MarketGAN} 
	& 1 & 2.0284 & 0.4829 & 0.2381 & $-0.3267$ & 28.6691 \\
	\cmidrule(lr){3-8}
	&  & 3 & 3.3033 & 0.7880 & 0.2386 & $-0.3148$ & 33.8637 \\
	\cmidrule(lr){3-8}
	&  & 5 & 3.4772 & 0.7668 & 0.2205 & $-0.2673$ & 33.9707 \\
	\cmidrule(lr){2-8}
	& \multirow{4}{*}{\begin{tabular}{@{}l@{}} Factor-model-based \\ bootstrap \end{tabular}}
	& 1 & 0.5201 & 0.1968 & 0.3785 & $-0.4712$ & 25.6837 \\
	\cmidrule(lr){3-8}
	&  & 3 & 2.0226 & 0.6345 & 0.3137 & $-0.4704$ & 29.1606 \\
	\cmidrule(lr){3-8}
	&  & 5 & 2.4797 & 0.6722 & 0.2711 & $-0.4197$ & 28.6375 \\
	\midrule
	\multirow{8}{*}{$0.1\%$}
	& \multirow{4}{*}{MarketGAN} 
	& 1 & 2.0334 & 0.4844 & 0.2382 & $-0.3267$ & 28.7304 \\
	\cmidrule(lr){3-8}
	&  & 3 & 3.3312 & 0.7954 & 0.2388 & $-0.3148$ & 33.8667 \\
	\cmidrule(lr){3-8}
	&  & 5 & 3.4729 & 0.7660 & 0.2206 & $-0.2673$ & 33.9751 \\
	\cmidrule(lr){2-8}
	& \multirow{4}{*}{\begin{tabular}{@{}l@{}} Factor-model-based \\ bootstrap \end{tabular}}
	& 1 & 0.5287 & 0.1996 & 0.3774 & $-0.4712$ & 25.7047 \\
	\cmidrule(lr){3-8}
	&  & 3 & 2.0253 & 0.6358 & 0.3139 & $-0.4704$ & 29.1663 \\
	\cmidrule(lr){3-8}
	&  & 5 & 2.4513 & 0.6659 & 0.2716 & $-0.4199$ & 28.6864 \\
	\bottomrule
	\end{tabular}}
	\caption{\footnotesize Performance of long-only mean-variance portfolios using synthetic sample covariance matrices under various levels of predictive $R^2$ for future factors $\boldsymbol{F}_{t+1}$. SR, Return, Std, MDD, and Turnover indicate the Sharpe ratio, annualized excess return, annualized standard deviation, maximum drawdown, and monthly turnover. }
	\label{tbl:port_simulation_r2}
\end{table}

In sum, when future factor realizations are poorly predicted, portfolio performance is largely constrained by factor predictability, and even sophisticated generative models cannot fully overcome this limitation. However, once factor information is even weakly informative, MarketGAN consistently achieves superior portfolio outcomes relative to factor-model-based bootstrap methods, while remaining competitive with strong covariance-based benchmarks. Crucially, these gains do not arise from improved factor forecasting or from marginal distributional accuracy alone. Instead, they stem from MarketGAN’s ability to generate coherent high-dimensional return distributions that preserve cross-sectional dependence and tail co-movement across assets. Traditional covariance estimators and bootstrap methods, while effective in estimating second moments, are inherently limited in capturing complex joint dependence structures in high-dimensional settings. By generating returns as a single joint vector and learning dependence patterns directly from the data, MarketGAN provides an additional layer of information that is particularly valuable for portfolio construction. These findings suggest that synthetic data-based approaches should be evaluated not only by their marginal or low-dimensional performance, but also by their capacity to generate economically meaningful high-dimensional structures. In this sense, MarketGAN complements existing covariance estimation techniques by extending their applicability to settings where accurate modeling of joint return distributions is essential for economic decision-making.

\section{Conclusion} \label{sec: conclusion}

This paper proposes MarketGAN, a factor-based generative framework for high-dimensional asset return generation under severe data scarcity. MarketGAN reframes the problem from point prediction to distributional learning, embedding a standard factor structure as an economic inductive bias while using conditional generative models with a temporal convolutional backbone to generate stochastic, time-varying factor loadings and idiosyncratic risks. By generating returns as a single joint vector, the framework is designed to learn and preserve cross-sectional dependence patterns that are central to portfolio analysis and risk management.

Empirically, using daily excess returns of 98 large U.S. equities from the S\&P 100 universe, we evaluate MarketGAN against transparent factor-model-based bootstrap baselines across one-, three-, and five-factor specifications. MarketGAN-generated data more closely match empirical marginal distributions. particularly in higher-order moments, while reproducing key inter-temporal stylized facts such as linear unpredictability, volatility clustering, and leverage effects. The most pronounced improvements arise in cross-sectional dependence, where MarketGAN more accurately replicates asset-level correlation structures, including dependence in extreme returns, as reflected in standard quantitative distributional metrics.

We further assess economic value through mean–variance portfolio optimization using MarketGAN-generated samples. Rather than relying on a specific factor forecasting model, we isolate the role of distributional fidelity through perturbed forecast simulations. In this setting, covariance matrices estimated from MarketGAN-generated data lead to substantially improved portfolio performance relative to bootstrap-based approaches, even when factor information is only weakly informative. These results indicate that the gains do not arise from improved forecasts or marginal accuracy alone, but from preserving joint dependence structures that are amplified through optimization.

From a broader decision-systems perspective, our findings underscore that the primary value of synthetic data lies in its role as input to downstream decision problems. In high-dimensional environments with limited and noisy data, decisions depend on nonlinear mappings of uncertainty, leading to small misspecifications in dependence being magnified. MarketGAN addresses this challenge by combining structural domain knowledge with generative distribution learning to produce joint uncertainty representations that remain informative under weak identification. While we illustrate the framework using financial portfolios, the underlying problem of learning and extending high-dimensional joint distributions for optimization and control under data scarcity is ubiquitous across operations, risk management, and data-driven decision systems. Viewed through this lens, generative models such as MarketGAN act as structural regularizers for decision-making, offering a principled approach to improving decision quality when historical data alone are insufficient.

\section*{Acknowledgement}
This work is also supported by the National Research Foundation of Korea(NRF) grant funded by the Korea government(MSIT) (RS-2025-00513038).

\section*{Declaration of competing interests}
The authors declare that they have no known competing financial interests or personal relationships that could have appeared to influence the work reported in this paper.

\bibliographystyle{apalike}
\bibliography{MarketGAN_bib}

\newpage
\appendix

\begin{center}
    \setstretch{1.2}
    \LARGE \textbf{Online Appendix for ``MarketGANs: Multivariate financial time-series data augmentation using generative adversarial networks"} \\[1.5em]
    
\end{center}





%

\tableofcontents

\section{Formal Introductions of GAN and TCN Models}\label{appendix: formal introduction of gan and tcn}

\subsection{Generative Adversarial Networks}

GANs \citep{goodfellow2014generative} are neural networks designed to generate synthetic samples analogously distributed to the target data samples.
GANs implicitly estimate complex target distributions through an adversarial training framework involving two distinct neural network components: a generator and a discriminator.
Formally, the GAN framework aims to model an unknown data distribution $p_{data}(x)$ based on a given set of samples $\{x_i\}_{i=1}^n$, where $x_i$ represents observed data such as handwritten digits, human face images, or asset returns. 
The generator $G$ takes as input a random noise vector $z\sim p_z(z)$ drawn from a simple prior distribution (e.g. Gaussian) and outputs synthetic samples $G(z)$ that resemble the real data. 
The discriminator $D$, on the other hand, acts as a binary classifier that assesses whether an input sample originates from the real data distribution or from the generator's synthetic data distribution.
Specifically, the discriminator is designed to output values in the interval $[0, 1]$, and it is trained to output 1 for real samples and 0 for generated samples. To this end, a sigmoid activation function is typically applied at the output layer of $D$.
These two neural networks compete against each other in a minimax game described by the following objective function:
\begin{equation*}
    \min_G \max_D V(D, G)=\mathbb{E}_{x\sim p_{data}(x)} \left[ \log D \left(x \right) \right] + \mathbb{E}_{z\sim p_z(z)} \left[ \log \left( 1-D\left(G(z)\right) \right) \right].
\end{equation*}
In this adversarial game, the discriminator $D$ seeks to maximize its ability to distinguish between real and synthetic samples by estimating the divergence between the two distributions. 
Simultaneously, the generator $G$ is trained to minimize this discrepancy and thereby generate increasingly realistic synthetic samples. 
This training process continues until convergence is achieved; i.e., the generator successfully captures the true data distribution $p_{data}(x)$, and the discriminator can no longer distinguish real from synthetic samples.
When the discriminator is optimal, the minimization of this objective corresponds to minimizing the Jensen-Shannon divergence between the real and generated distributions:
\begin{equation*}
    JSD(p_{data}, q_{G}) = \frac{1}{2} D_{KL} \left( p_{data} \Vert M  \right) + \frac{1}{2} D_{KL} \left( q_{G} \Vert M \right),
\end{equation*}
where $q_G$ denotes the model distribution induced by the generator, and $M=\frac{1}{2} \left(p_{data}+q_{G}\right)$ and $D_{KL}$ is the Kullback-Leibler divergence.
In essence, the generator is trained to approximate the real data distribution $p_{data}$.

When the goal is to generate data conditioned on auxiliary information (e.g., class labels or macroeconomic indicators), conditional GANs \citep{mirza2014conditional} can be employed.
In this setting, both the generator and discriminator take the conditioning variable $y$ as an additional input.
The generator is trained to learn the conditional distribution of the data given $y$, and the discriminator is trained to distinguish real from synthetic data conditioned on the same information.
The resulting objective function becomes:
\begin{equation*}
    \min_G \max_D V(D, G)=\mathbb{E}_{x\sim p_{data}(x|y), \ y\sim p_y(y)} \left[ \log D \left(x; y \right) \right] + \mathbb{E}_{z\sim p_z(z), \ y\sim p_y (y)} \left[ \log \left(1-D\left(G(z; y); y\right)  \right) \right].
\end{equation*}
This framework enables tasks such as generating specific digit classes in handwritten digit image datasets or synthesizing asset returns under specific macroeconomic conditions.

Despite their conceptual elegance, GANs often face practical challenges during training.
One major challenge is the inherent instability caused by the adversarial minimax optimization process. 
Another common issue is the vanishing gradient problem, which frequently arises from the use of a sigmoid activation function at the discriminator’s output layer. 
To mitigate such limitations, improved models such as WGAN \citep{arjovsky2017wasserstein} and WGAN-GP \citep{gulrajani2017improved} have been proposed.
These models introduce different optimization formulations that enhance training stability and convergence.

The WGAN replaces the Jensen-Shannon divergence used in the original GAN formulation with the Wasserstein-1 distance.
While the Wasserstein distance is generally computationally intractable, WGAN utilizes the Kantorovich-Rubinstein duality to reformulate the problem into a tractable optimization form. 
The WGAN objective is given by:
\begin{equation*}
    \min_G \max_{D\in \mathcal{D}} \mathbb{E}_{x\sim p_{data}(x)} \left[ D \left(x\right)\right]  - \mathbb{E}_{z\sim p_z(z)} \left[ D\left( G(z)\right) \right],
\end{equation*}
where $\mathcal{D}$ is the set of 1-Lipschitz functions. 
In this framework, $D$ is no longer a binary classifier but a \emph{critic} that estimates the Wasserstein distance between the real and generated distributions.
As a result, WGAN eliminates the need for a sigmoid activation function at the discriminator’s output layer and thereby avoids the vanishing gradient problem.

To enforce the 1-Lipschitz constraint in practice, the original WGAN employs weight clipping on the discriminator’s parameters.
However, this crude approach can restrict the model capacity and lead to suboptimal performance. To mitigate this, \citet{gulrajani2017improved} proposed the WGAN-GP, which introduces a soft gradient penalty based on the theoretical result that the optimal critic function has gradients of unit norm almost everywhere. The modified objective of the WGAN-GP becomes
\begin{equation} \label{eqn: wgan gp loss}
    \min_G \max_D\mathbb{E}_{x\sim p_{data}(x)}\left[D \left(x \right)\right] - \mathbb{E}_{z\sim p_z(z)} \left[ D \left(G(z)\right) \right] + \lambda \mathbb{E}_{\tilde{x} \sim p_{\tilde{x}} (\tilde{x})}\left[\left( \lVert\nabla_{\tilde{x}} D(\tilde{x})\rVert_2 - 1 \right)^2  \right],
\end{equation}
where $\tilde{x}$ denotes points sampled uniformly along straight lines between real and generated samples, and $\lambda$ is a penalty coefficient typically set to 10.
This modification has been shown to yield more stable training and reliable convergence, and is therefore widely adopted in applications involving continuous-valued data.


\subsection{Temporal Convolutional Networks (TCNs)}

To formalize TCN model, let $X\in\mathbb{R}^{C_\text{in}\times T}$ denote a one-dimensional input sequence of length $T$ with $C_\text{in}$ channels.
A dilated causal convolutional operator with kernel size $k$, dilation factor $D$, and output channels $C_\text{out}$ is defined as:
\begin{equation*}
\left( W * X \right)_{c,t} = \sum_{i=0}^{k-1} \sum_{j=1}^{C_{\text{in}}} W_{i,j,c} \cdot X_{j, t - D \cdot i} + b_c,
\end{equation*}
for $c=1,\cdots, C_\text{out}$ and $t\geq D(k-1)$, where $W\in \mathbb{R}^{k\times C_\text{in} \times C_\text{out}}$ is the weight tensor and $b\in\mathbb{R}^{C_\text{out}}$ is the bias vector.
When $D=1$, the operation reduces to a standard causal convolution.

\begin{figure}[!htb]
	\centering
	\includegraphics[scale=0.5]{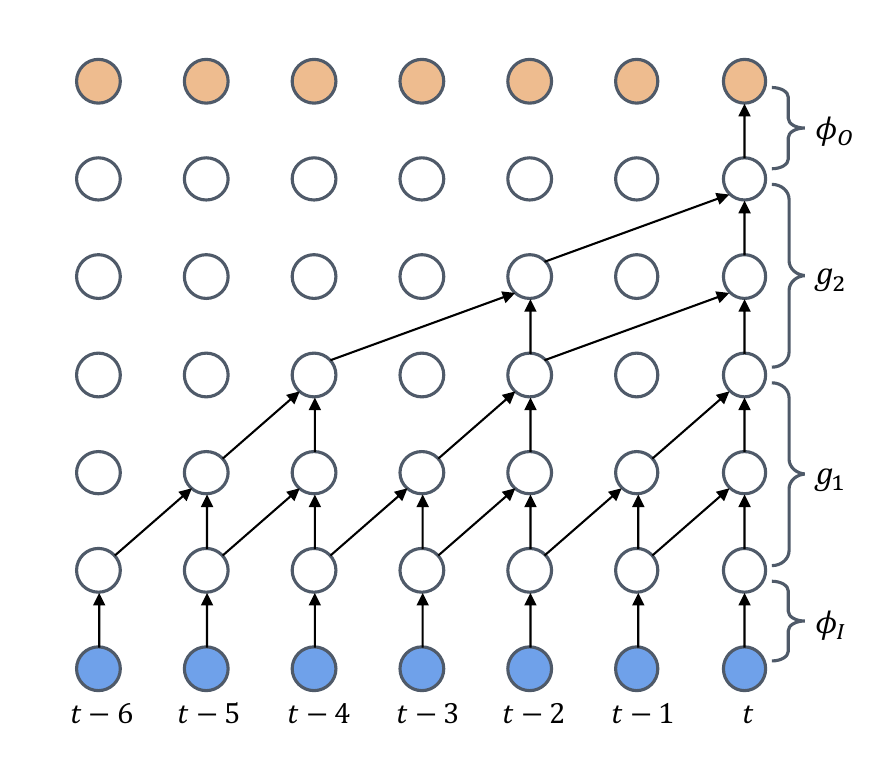}
	\caption{\centering TCN with the kernel size $k=2$, the dilation factor $D=2$, and the number of residual blocks $L=2$}
	\label{fig: TCN}
\end{figure}

A complete TCN consists of a stack of such convolutional blocks with increasing dilation factors at deeper layers.
Formally, a TCN $f:\mathbb{R}^{n_\text{in}\times T_0}\rightarrow \mathbb{R}^{n_\text{out}\times T_L}$ is composed of:
\begin{equation} \label{eqn: tcn composite functional form}
f(X) = \phi_O \circ g_L \circ g_{L-1} \circ \dots \circ g_1 \circ \phi_I (X),
\end{equation}
where $\phi_I:\mathbb{R}^{n_\text{in} \times T_0}\rightarrow \mathbb{R}^{C_0 \times T_0}$ and $\phi_O:\mathbb{R}^{C_L \times T_L}\rightarrow \mathbb{R}^{n_\text{out} \times T_L}$ are $1\times 1$ convolutional layers that project the input and output channels, respectively.
Each intermediate mapping $g_\ell: \mathbb{R}^{C_{\ell-1}\times T_{\ell-1}}\rightarrow \mathbb{R}^{C_\ell\times T_\ell}$ is implemented as a residual block composed of two dilated causal convolutions with dilation $d=D^{\ell-1}$, non-linear activation (e.g., ReLU), dropout, and a residual skip connection.
This composition can be written as
\begin{equation*}
g_\ell(X) = \mathrm{ReLU} \left( \mathrm{Conv}^{(2)}_\ell \left( \mathrm{Dropout} \left( \mathrm{ReLU} \left( \mathrm{Conv}^{(1)}_\ell (X) \right) \right) \right) \right) + \mathrm{Res}_\ell(X),
\end{equation*}
where $\mathrm{Conv}^{(1)}_\ell$ and $\mathrm{Conv}^{(2)}_\ell$ are dilated causal convolutional layers with dilation $D^{\ell-1}$, and $\mathrm{Res}_\ell(X)$ denote either the identity map or a $1\times 1$ convolution applied to $X$ to match dimensions when $C_{\ell-1}\neq C_{\ell}$. The overall structure of the TCN is illustrated in Figure \ref{fig: TCN}.

For computational convenience, we ensure that the output sequence length $T_L$ matches the input sequence length $T_0$ by applying zero padding to the input of each convolutional layer $\mathrm{Conv}_\ell^{(j)}$.
Each convolution extracts features from $k$ time points within a fixed window of size $D(k-1)+1$ and compresses them into a single output vector at the corresponding time step.
This operation reduces the length along the temporal axis.
To prevent this reduction, we pad zeros to the beginning of each input sequence, so that the output sequence at each layer remains the same length as the input.
However, since the padded values do not originate from actual data, they inevitably introduce some bias into the output.
To mitigate this effect, we generate an output sequence over a sufficiently long time horizon and discard the initial portion affected by zero padding, which ensures that the segment of interest is unaffected by its influence.

\subsection{GAN with TCN Backbones}

To enhance the generative performance of GANs, various neural network architectures have been employed as the backbone structure of GAN depending on the specific characteristics of the target data (e.g., \cite{radford2015unsupervised, mogren2016c, zhang2019self, wang2018high}).
In this work, to capture the temporal dependencies of financial data, we adopt TCNs as the backbone structure for both the generator and discriminator in GAN to exploit their ability to model long-range inter-temporal dependencies inherent in sequential data.

In this architecture, the generator $G$ takes as input a sequence of i.i.d. random noise vectors $\{z_t\}_{t=1}^T$ typically drawn from a standard normal distribution.
As this input sequence propagates through stacked layers of dilated convolutions in the TCN, initially independent noise variables become progressively intertwined.
This process induces intricate inter-temporal dependencies across time steps in the output sequence. 
Through adversarial training, the generator learns to map these latent noise sequences into synthetic data sequences that accurately reflect the temporal dynamics and distributional properties of the real data.
Simultaneously, the discriminator $D$ is trained to measure discrepancies between the true data distribution $p_{data}(x)$ and the model-induced distribution $p_{G}(x)$, explicitly taking into account the sequential structure of the data.

This framework can be naturally extended to the conditional GANs by concatenating additional covariates to the input of both the generator and the discriminator.
The generator then learns to generate condition-specific sequences while the discriminator evaluates distributional discrepancies conditioned on these covariates.

\section{Implementational Details and Hyperparameters of MarketGAN} \label{appendix: hyperparam}

The generator networks are implemented using a TCN with 80 hidden channels across all layers.
Each convolutional layer is followed by ReLU activation and weight normalization.
A dropout rate of 20\% is applied after each activation.
The weights of all convolutional layers are initialized from a normal distribution $\mathcal{N}(0, 0.5^2)$.
To ensure that the output sequence length $T_L$ matches the input sequence length $T_0$, we apply zero-padding to the input of each convolutional layer. 
In our implementation, we set $T_0 = T_L = \mathrm{RFS} + 252 \times 4$, where the relatively long $T_L$ helps mitigate the impact of zero padding during training.
The discriminator shares the same architecture as the generator except that it uses 160 hidden channels instead of 80.
We use a batch size of $B = 128$.
The Adam optimizer is employed with a learning rate of 0.001 for all neural networks. 
The gradient penalty coefficient $\lambda$ is set to 10.
The overall training procedure is summarized in Algorithm~\ref{alg:marketgan-training}.

\begin{algorithm}[!p]
\caption{Training Procedure for MarketGAN}
\label{alg:marketgan-training}
\small{
\For{epoch $= 1$ to $E$}{
  \For{each training mini-batch of time indices $\{t_1, \dots, t_B\}$}
    {
    \For{$k = 1$ to $n_D$} 
    {
      \textbf{(a) Sample latent inputs and prepare factor values, real excess return data and covariates} \\
      \quad For each $t_b$: \\
      \quad \quad Sample $\boldsymbol{z}_{t_b : t_b + T_0} \sim \mathcal{N}(0, I)$ \\
      \quad \quad Obtain covariates $\boldsymbol{y}_{t_b : t_b + T_0 - 1}$ \\
      \quad \quad Obtain factor values $\boldsymbol{F}_{t_b+1 : t_b + T_L}$ and real excess returns $\boldsymbol{r}^{\mathrm{real}}_{t_b+1 : t_b + T_L}$ \\
      \textbf{(b) Generate synthetic excess returns} \\
      \quad Generate coefficients 
      $\boldsymbol{\alpha}_{t_b : t_b + T_L - 1}$, 
      $\boldsymbol{\beta}_{t_b : t_b + T_L - 1}$, 
      $\boldsymbol{\sigma}_{t_b : t_b + T_L - 1}$ from 
      $\left( \boldsymbol{z}_{t_b : t_b + T_0 - 1}, \boldsymbol{y}_{t_b : t_b + T_0 - 1} \right)$ using~\eqref{eqn: alpha alphahat correction} \\
      \quad Generate residuals 
      $\boldsymbol{\epsilon}_{t_b+1 : t_b + T_L}$ 
      from $\left( \boldsymbol{z}_{t_b+1 : t_b + T_0}, \boldsymbol{y}_{t_b : t_b + T_0 - 1} \right)$ \\
      \quad Compute $\boldsymbol{r}^{\mathrm{synthetic}}_{t_b+1 : t_b + T_L}$ using~\eqref{eqn: marketgan} \\
      \textbf{(c) Update discriminator $D$ by maximizing the conditional WGAN-GP loss:}
      \begin{equation*}
      \max_D \frac{1}{B} \sum_{b=1}^B 
      \left[
      D\left(\boldsymbol{r}^{\mathrm{real}}_{t_b+1 : t_b + T_L}; \boldsymbol{y}_{t_b : t_b + T_0 - 1} \right)
      -    D\left(\boldsymbol{r}^{\mathrm{synthetic}}_{t_b+1 : t_b + T_L}; \boldsymbol{y}_{t_b : t_b + T_0 - 1} \right)
      \right.
      \end{equation*}
      \begin{equation*}
      \left.
      + \lambda \left( \left\| \nabla_{\tilde{\boldsymbol{r}}_{t_b+1 : t_b + T_L}} D(\tilde{\boldsymbol{r}}_{t_b+1 : t_b + T_L}; \boldsymbol{y}_{t_b : t_b + T_0 - 1}) \right\|_2 - 1 \right)^2
      \right],
      \end{equation*}
      where $\tilde{\boldsymbol{r}} = u \boldsymbol{r}^{\mathrm{real}} + (1 - u) \boldsymbol{r}^{\mathrm{synthetic}}$, and $u \sim \mathcal{U}(0,1)$
    }

    \For{$k = 1$ to $n_G$} 
    {
      \textbf{(a) Sample latent inputs and prepare factor values and covariates} \\
      \quad For each $t_b$: \\
      \quad \quad Sample $\boldsymbol{z}_{t_b : t_b + T_0} \sim \mathcal{N}(0, I)$ \\
      \quad \quad Obtain covariates $\boldsymbol{y}_{t_b : t_b + T_0 - 1}$
      \quad \quad Obtain factor values $\boldsymbol{F}_{t_b+1 : t_b + T_L}$

      \textbf{(b) Generate synthetic excess returns} \\
      \quad Generate coefficients 
      $\boldsymbol{\alpha}_{t_b : t_b + T_L - 1}$, 
      $\boldsymbol{\beta}_{t_b : t_b + T_L - 1}$, 
      $\boldsymbol{\sigma}_{t_b : t_b + T_L - 1}$ from 
      $\left( \boldsymbol{z}_{t_b : t_b + T_0 - 1}, \boldsymbol{y}_{t_b : t_b + T_0 - 1} \right)$ using~\eqref{eqn: alpha alphahat correction} \\
      \quad Generate residuals 
      $\boldsymbol{\epsilon}_{t_b+1 : t_b + T_L}$ from 
      $\left( \boldsymbol{z}_{t_b+1 : t_b + T_0}, \boldsymbol{y}_{t_b : t_b + T_0 - 1} \right)$ \\
      \quad Compute $\boldsymbol{r}^{\mathrm{synthetic}}_{t_b+1 : t_b + T_L}$ using~\eqref{eqn: marketgan} \\

      \textbf{(c) Update generator $G$ by minimizing the conditional WGAN-GP loss:}
      \begin{equation*}
      \min_G \frac{1}{B} \sum_{b=1}^B 
      \left[
      - D\left(\boldsymbol{r}^{\mathrm{synthetic}}_{t_b+1 : t_b + T_L}; \boldsymbol{y}_{t_b : t_b + T_0 - 1} \right)
      \right]
      \end{equation*}
    }
}
}

  \textbf{Validation check:} \\
  \quad Generate synthetic samples $\boldsymbol{r}^{\mathrm{synthetic}}$ over the validation period \\
  \quad Estimate correlation matrix $\hat{\Sigma}_{\text{syn}}$ from synthetic returns \\
  \quad Estimate correlation matrix $\hat{\Sigma}_{\text{real}}$ from real returns \\
  \quad Compute Frobenius norm $\|\hat{\Sigma}_{\text{real}} - \hat{\Sigma}_{\text{syn}}\|_F$ \\
  \lIf{the Frobenius norm is reduced}{Save model parameters}

\textbf{Fine-tuning:} \\
\quad Train generator for $E_{\text{fine}}$ epochs on the validation set with reduced learning rate
}
\end{algorithm}

\section{Histograms of Aggregated Excess Returns at Weekly and Monthly Frequencies} \label{appendix: weekly and monthly histogram}

\begin{figure}[!p]
\centering
\subfigure[One-factor model]{
\includegraphics[scale=0.4]{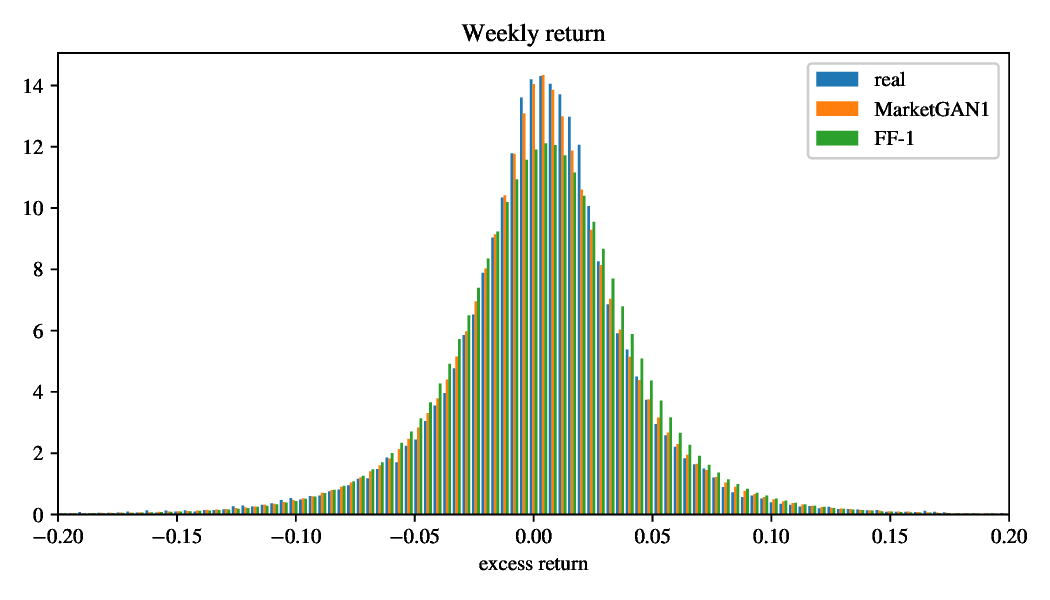}
\includegraphics[scale=0.4]{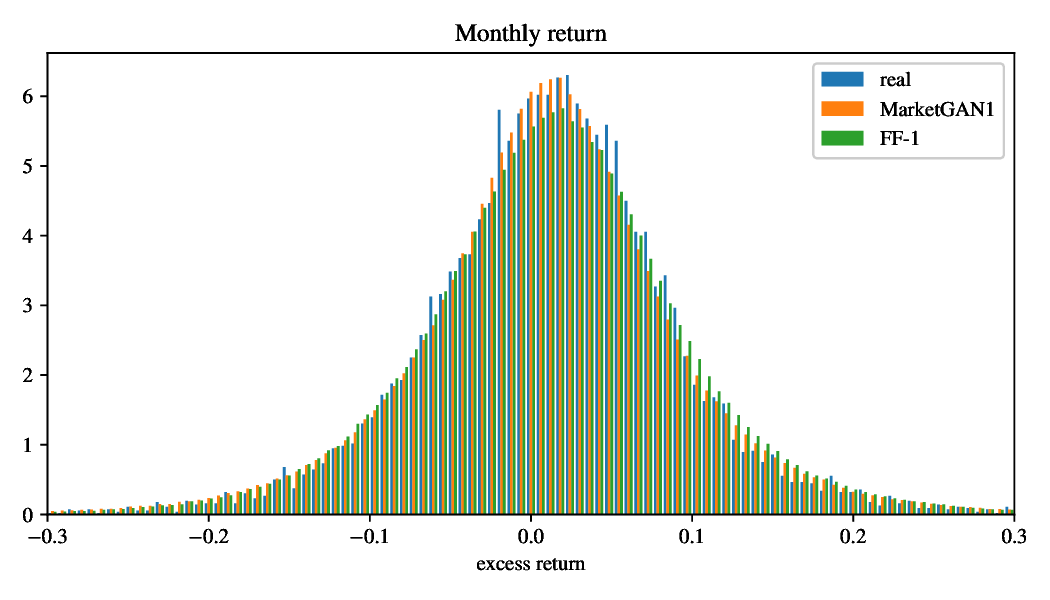}}

\subfigure[Three-factor model]{
\includegraphics[scale=0.4]{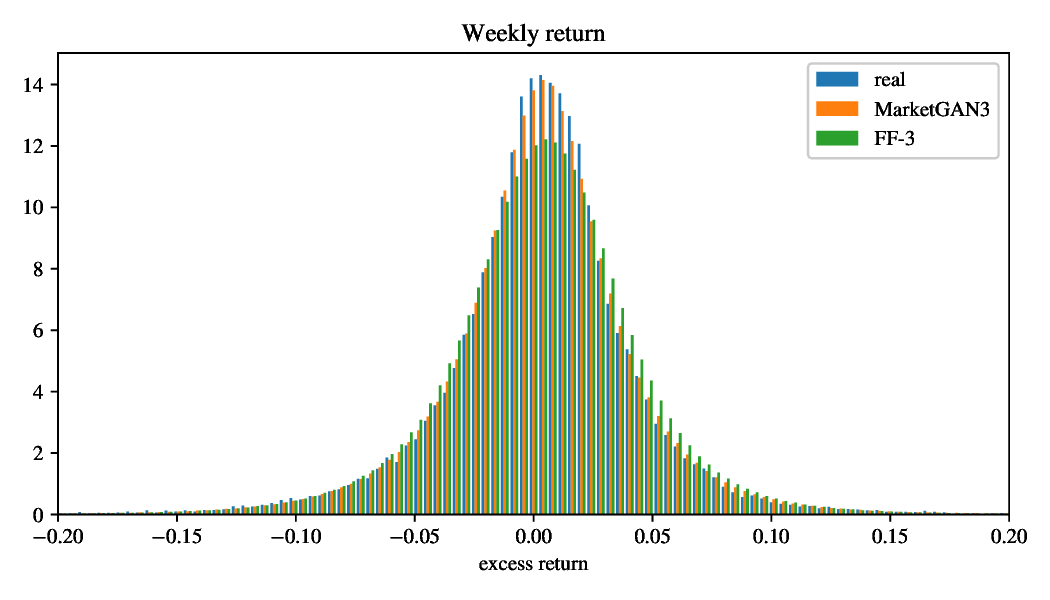}
\includegraphics[scale=0.4]{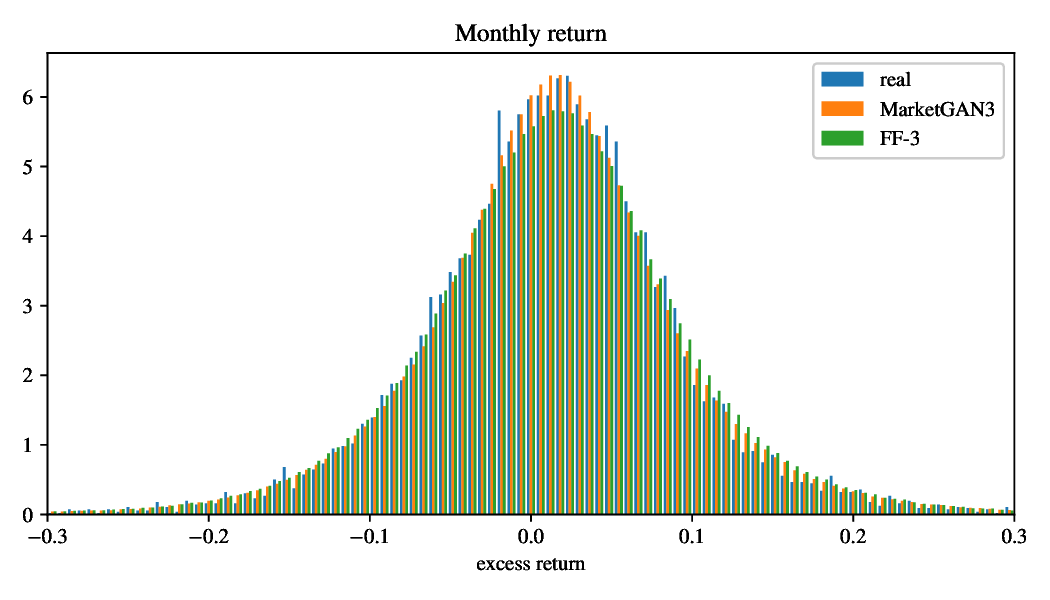}}

\subfigure[Five-factor model]{
\includegraphics[scale=0.4]{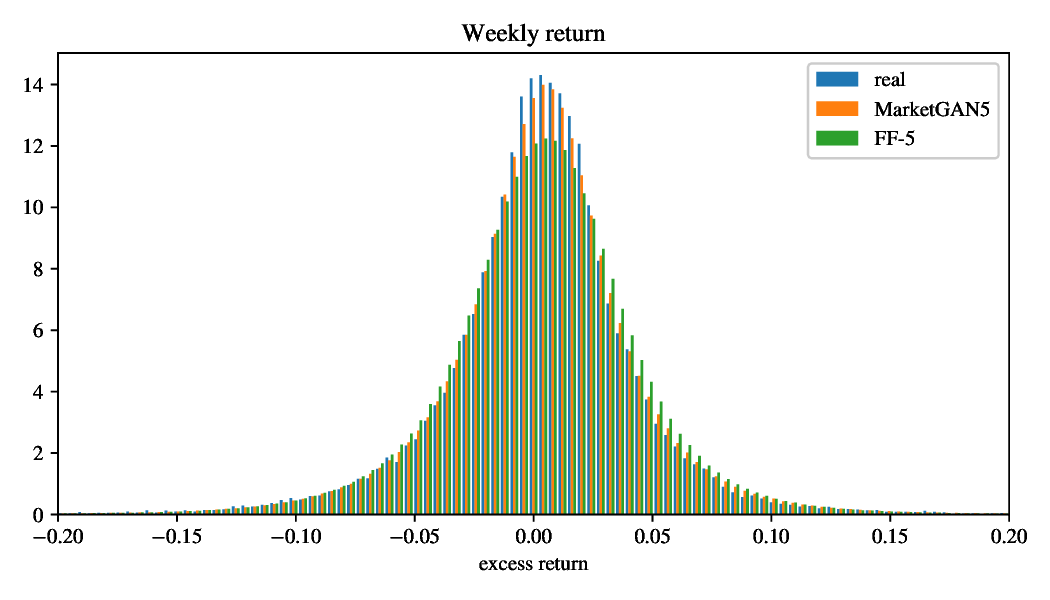}
\includegraphics[scale=0.4]{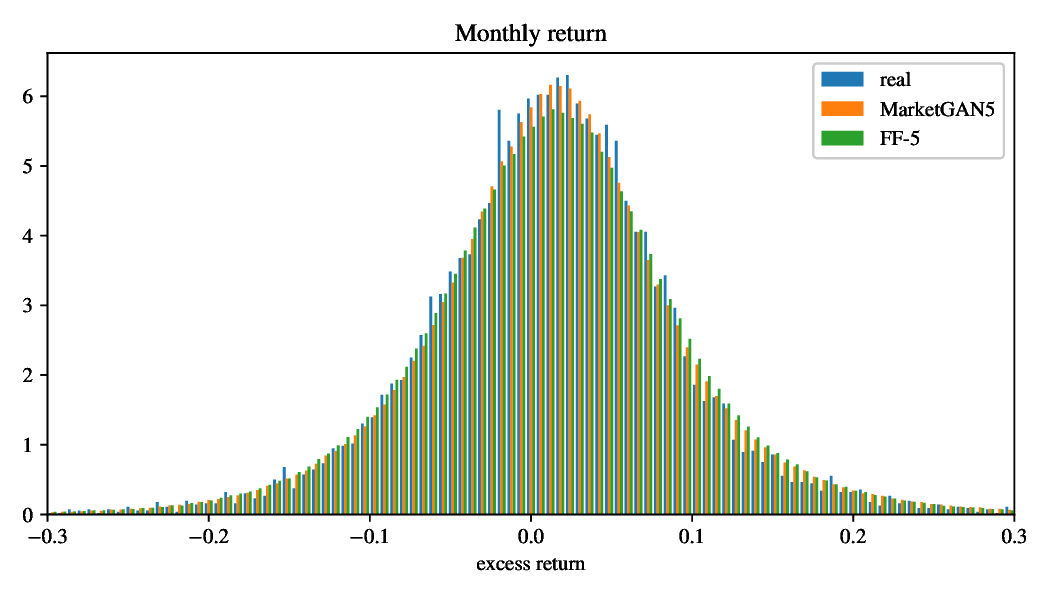}}

\caption{Marginal distributions of weekly and monthly aggregated excess returns. FF-1, FF-3, and FF-5 denote the respective factor-model-based bootstrap methods.}
\label{fig: weekly and monthly return histogram}
\end{figure}

To evaluate how well the generative models capture return distributional properties beyond the daily level, we analyze the marginal distributions of excess returns aggregated to weekly and monthly frequencies.
Figure \ref{fig: weekly and monthly return histogram} shows the corresponding histograms for both the real data and the synthetic data generated by the MarketGAN models and the factor-model-based bootstrap methods.
Compared to the daily returns shown in Figure \ref{fig: return histogram}, the aggregated return distributions exhibit lower kurtosis, which aligns with the well-known empirical observations that temporal aggregation tends to smooth extreme events.
At these lower frequencies, the differences between models become less pronounced, and increasing the number of factors does not lead to a clear improvement in the shape of the marginal distribution for either method.
Nonetheless, MarketGAN models consistently provide return distributions that more closely resemble the real data than those generated by the bootstrap methods. 
Across both weekly and monthly horizons, the MarketGAN-generated returns preserve higher peaks and better replicate the overall shape of the real return distributions.
In contrast, bootstrap-generated distributions tend to underestimate the kurtosis.

\section{Mathematical Formulation of Distance Metrics}
\label{appendix: distance metrics}

This appendix provides the mathematical definitions and underlying assumptions for the distance metrics used to quantitatively compare the distributions of synthetic and real data in Section \ref{sec:quantitative_evaluation_metrics}. 

\textbf{Fréchet Inception Distance (FID)}
Given two sets of data points or feature embeddings—one from real data and one from generated data—FID approximates the Wasserstein-2 distance between the two datasets by using only their empirical means $\mu_r$, $\mu_g$ and covariances $\Sigma_r$, $\Sigma_g$.
The distance is computed as:
\begin{equation*}
\mathrm{FID}^2 = \lVert\mu_r - \mu_g\rVert_2^2 + \mathrm{Tr}\left(\Sigma_r + \Sigma_g - 2(\Sigma_r \Sigma_g)^{1/2}\right).
\end{equation*}
This formulation captures the discrepancy in both the first and second moments. 
It implicitly assumes that the distributions are approximately Gaussian in the data or embedding space.
Higher-order differences or multimodality are not reflected.

\textbf{Sliced Wasserstein Distance (SWD)}
The SWD approximates the Wasserstein-1 distance between high-dimensional distributions $P$ and $Q$ by projecting them onto multiple random directions $\theta \in \mathbb{S}^{d-1}$ and computing the 1-dimensional Wasserstein distance for each projection:
\begin{equation*}
\mathrm{SWD}(P, Q) \approx \frac{1}{L} \sum_{\ell=1}^{L} W_1(P_{\theta_\ell}, Q_{\theta_\ell}),
\end{equation*}
where $L$ is the number of projection directions, and $P_{\theta_\ell}$ and $Q_{\theta_\ell}$ denote the one-dimensional marginal distributions obtained by projecting $P$ and $Q$ onto the direction $\theta_\ell$, respectively. In our experiments, we set $L = 100$. SWD does not rely on distributional assumptions such as Gaussianity. It is effective at detecting fine-grained differences, provided that the projection directions are sufficiently expressive.

\textbf{Mahalanobis Distance (MD)}
The MD between a sample $x$ and a distribution with mean $\mu$ and covariance matrix $\Sigma$ is defined as:
\begin{equation*}
\mathrm{MD}(x, \mu) = \sqrt{(x - \mu)^\top \Sigma^{-1} (x - \mu)}.
\end{equation*}
This metric takes into account the covariance structure by rescaling the distance along directions of varying variance. MD is particularly sensitive to deviations from the main distributional structure and therefore useful in detecting anomalies. If $\Sigma$ is ill-conditioned or contaminated by outliers, MD may yield misleading results.

\textbf{Dynamic Time Warping (DTW)}
DTW computes the optimal nonlinear alignment between two time series $X = (x_1, \dots, x_n)$ and $Y = (y_1, \dots, y_m)$ using dynamic programming. The recurrence relation for the cumulative cost matrix $\gamma(i,j)$ is:
\begin{equation}
\gamma(i,j) = D(i,j) + \min \left\{ \gamma(i-1,j),\; \gamma(i,j-1),\; \gamma(i-1,j-1) \right\},
\end{equation}
where $D(i,j) = \|x_i - y_j\|$ is the pointwise cost, and $\gamma(n,m)$ gives the final DTW distance. Unlike other metrics, DTW does not rely on distributional assumptions but assumes that meaningful similarity can be recovered via time-aligned warping. While effective for misaligned sequences, it may be sensitive to local noise and computationally expensive.

\section{Additional Figures} \label{appendix: acf corrmatrix figures}

\begin{figure}[!p]
	\centering
	\subfigure[MarketGAN1]{
		\includegraphics[scale=0.42]{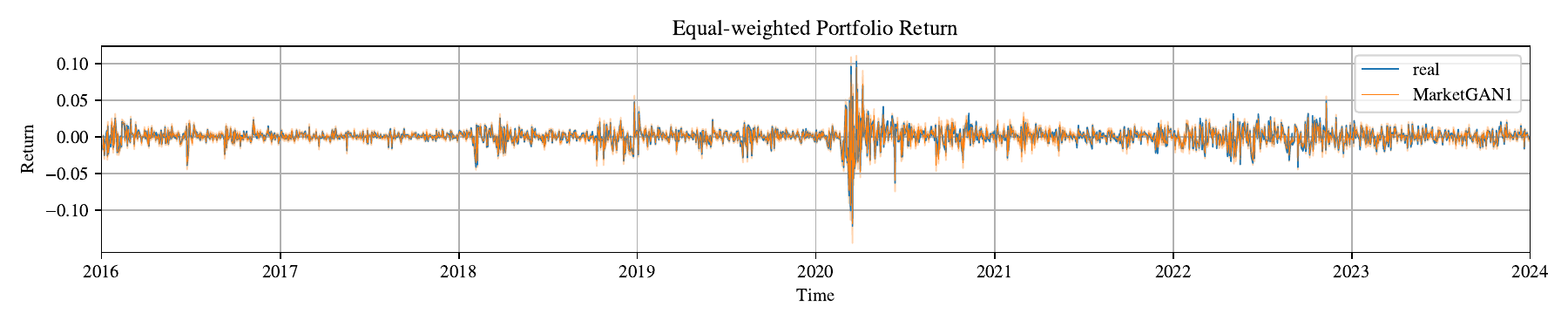}}
	\subfigure[FF-1 bootstrap]{
		\includegraphics[scale=0.42]{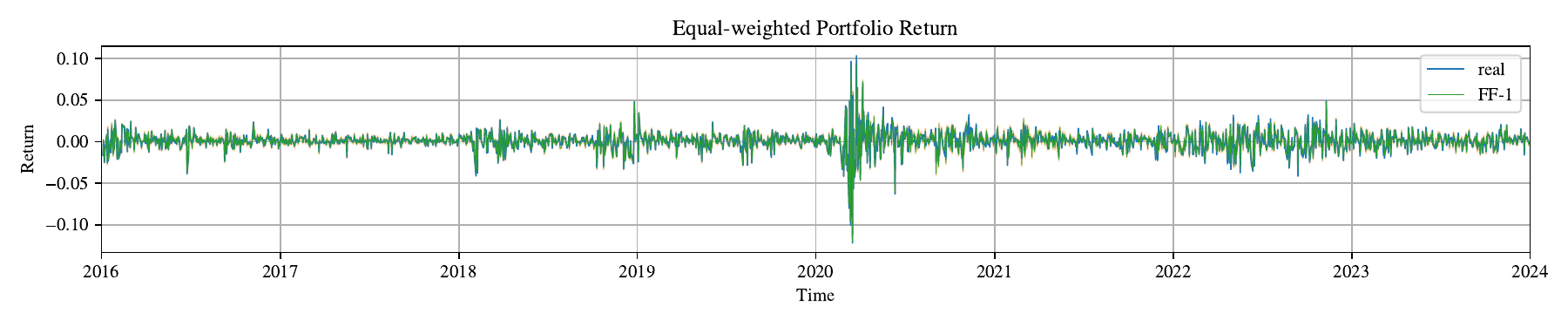}}
	\subfigure[MarketGAN3]{
		\includegraphics[scale=0.42]{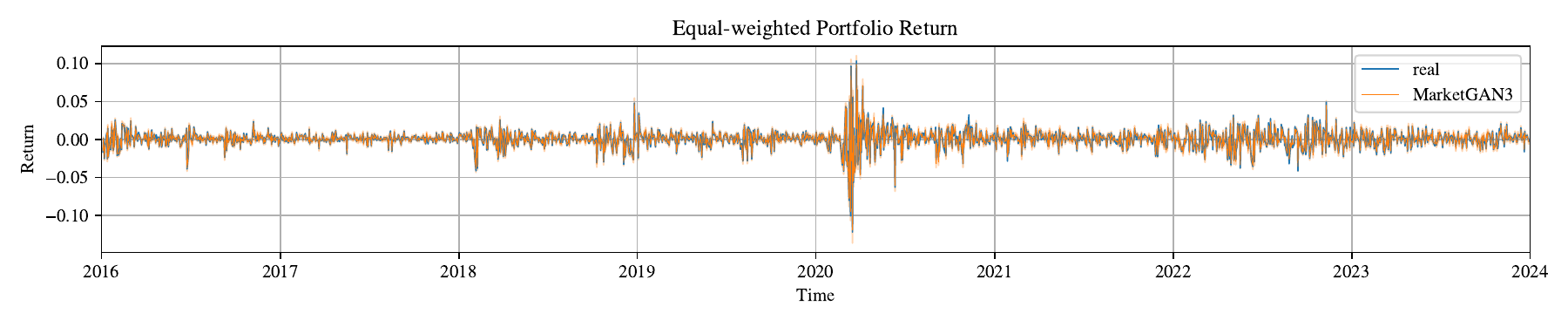}}
	\subfigure[FF-3 bootstrap]{
		\includegraphics[scale=0.42]{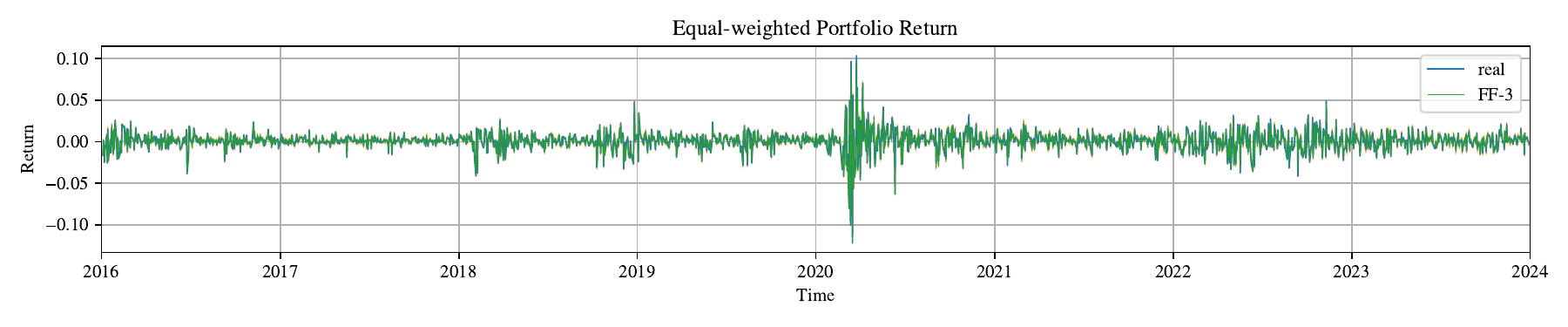}}
	\subfigure[MarketGAN5]{
		\includegraphics[scale=0.42]{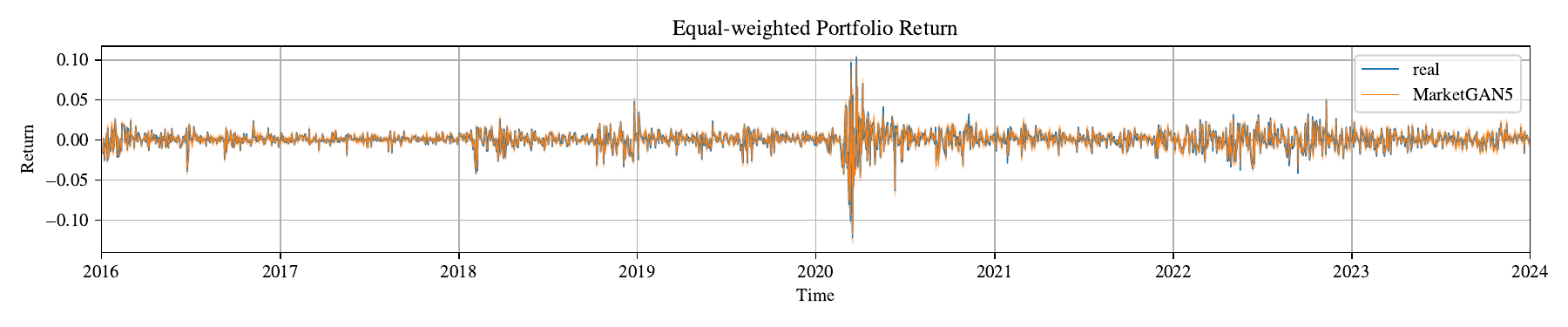}}
	\subfigure[FF-5 bootstrap]{
		\includegraphics[scale=0.42]{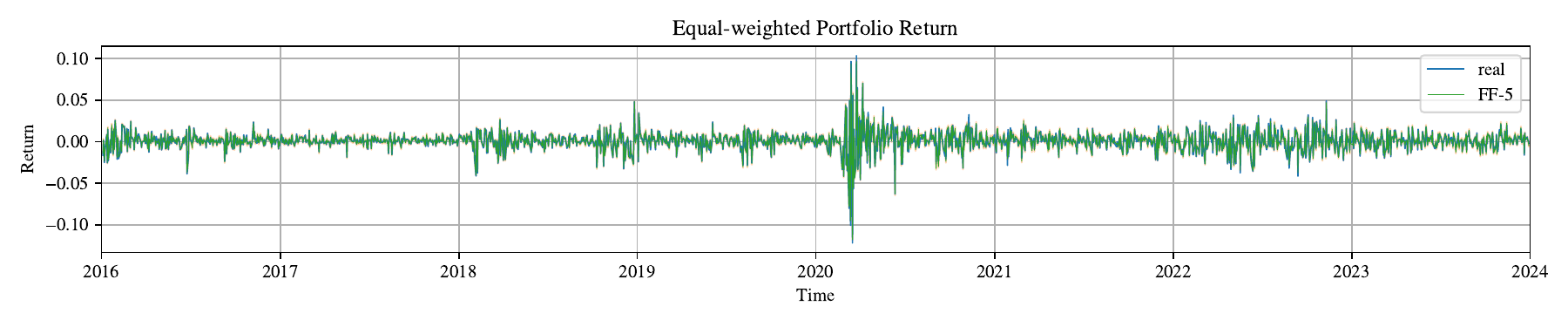}}
	\caption{Comparison of real and synthetic equal-weighted excess returns generated by the MarketGAN models and factor-model-based bootstrap methods.}
	\label{fig: ew return figure}
\end{figure}

\begin{figure}[!htbp]
  \centering
    \subfigure[MarketGAN1]{
        \includegraphics[scale=0.42]{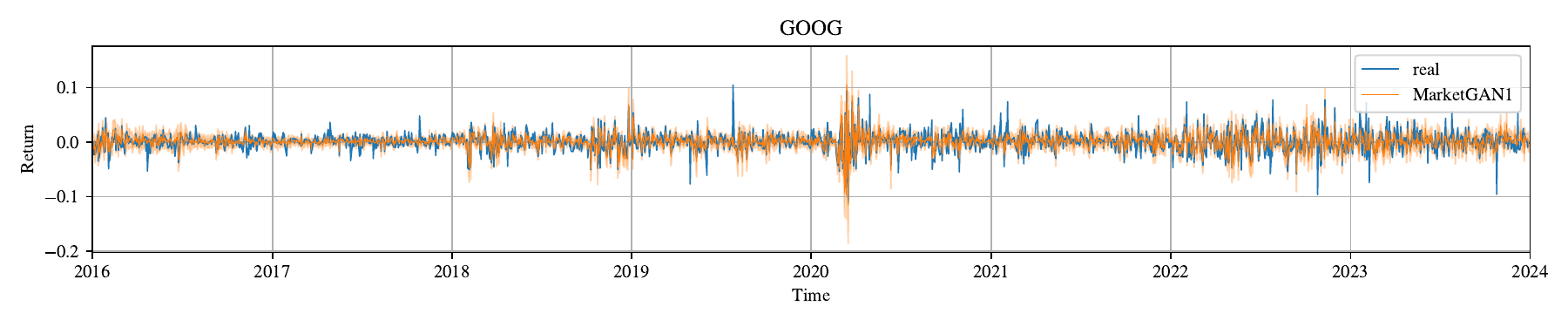}}
      \subfigure[FF-1 bootstrap]{
        \includegraphics[scale=0.42]{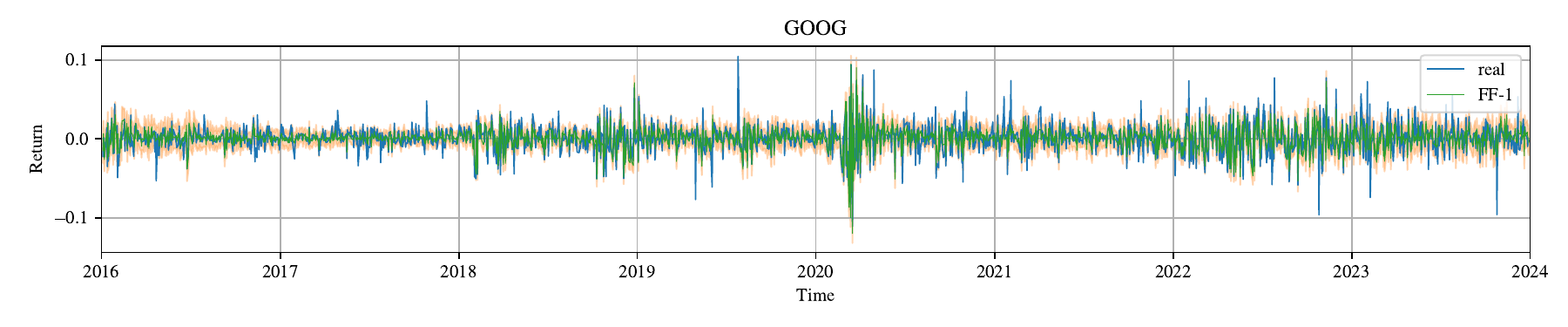}}
      \subfigure[MarketGAN3]{
        \includegraphics[scale=0.42]{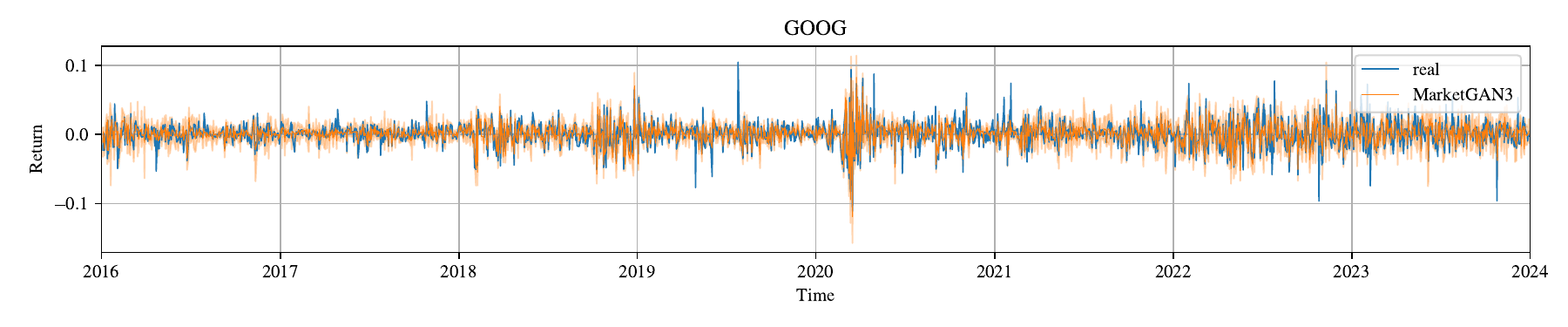}}
      \subfigure[FF-3 bootstrap]{
        \includegraphics[scale=0.42]{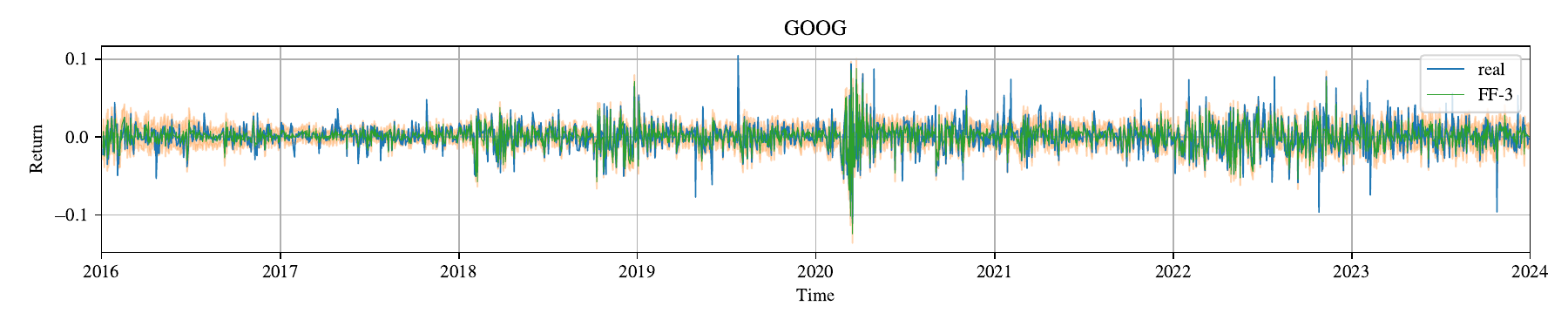}}
      \subfigure[MarketGAN5]{
        \includegraphics[scale=0.42]{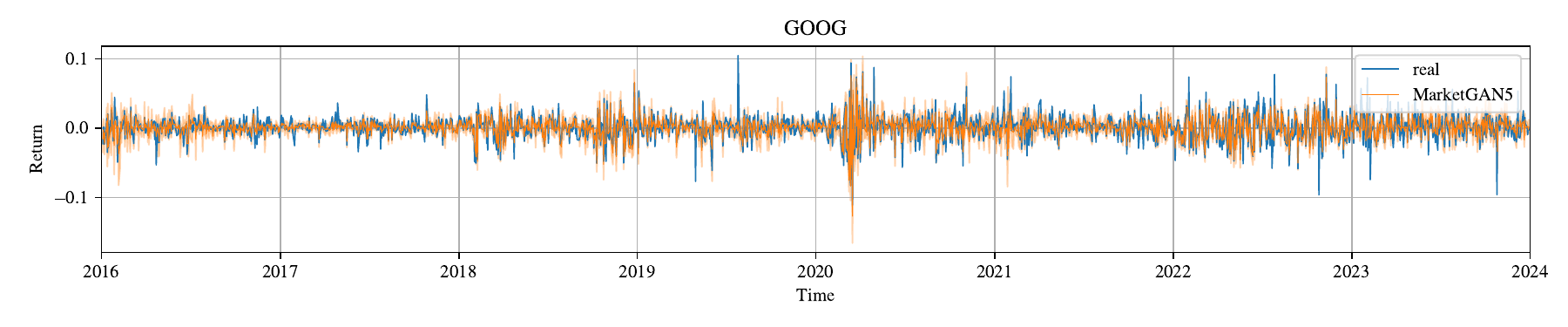}}
      \subfigure[FF-5 bootstrap]{
        \includegraphics[scale=0.42]{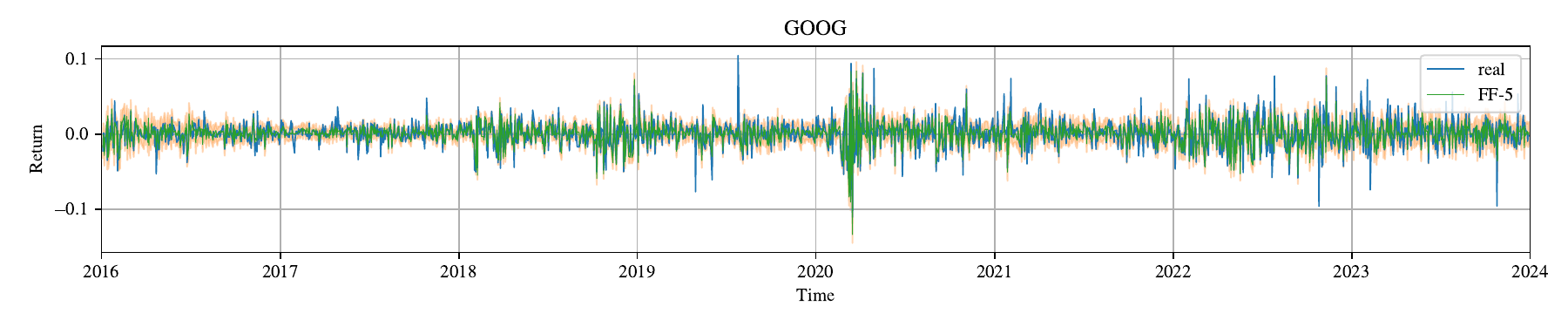}}
  \caption{Comparison of real and synthetic excess returns for GOOG generated by the MarketGAN models and factor-model-based bootstrap methods.}
  \label{fig: goog_return_comparison}
\end{figure}

\begin{figure}[!htbp]
  \centering
  \subfigure[ACF of MarketGAN1]{
    \includegraphics[scale=0.48]{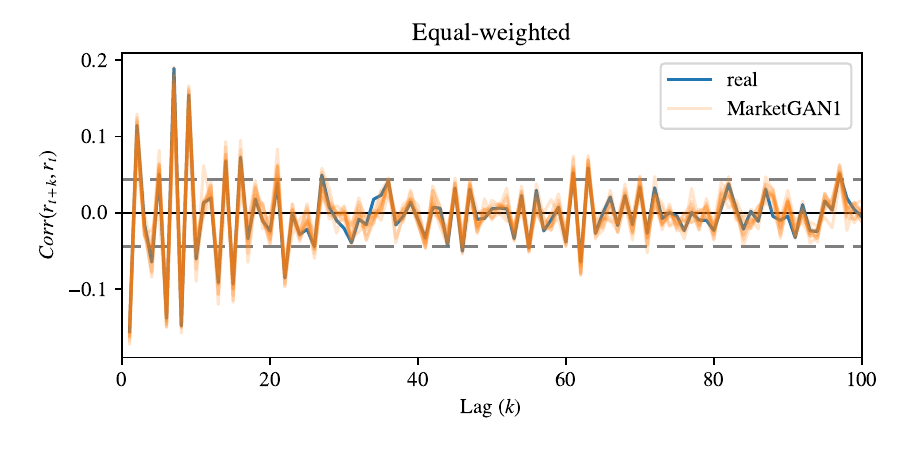}}
  \hfill
  \subfigure[ACF of FF-1 bootstrap]{
    \includegraphics[scale=0.48]{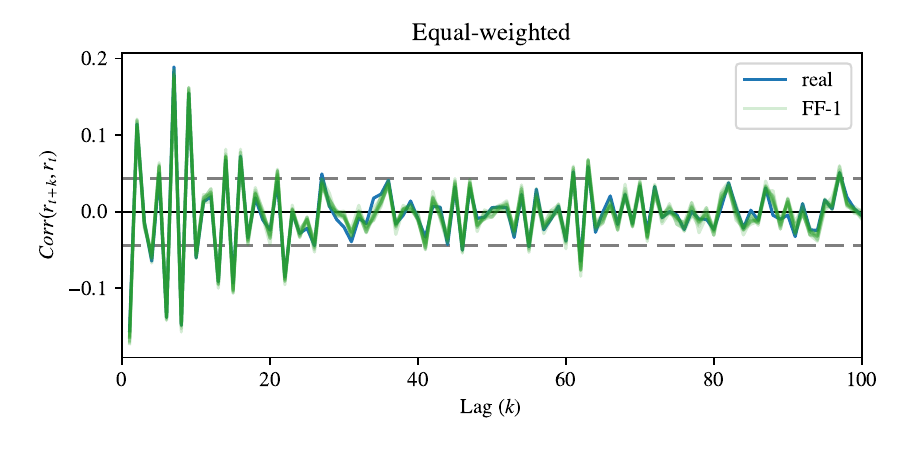}}
 \\
  \subfigure[VC of MarketGAN1]{
    \includegraphics[scale=0.48]{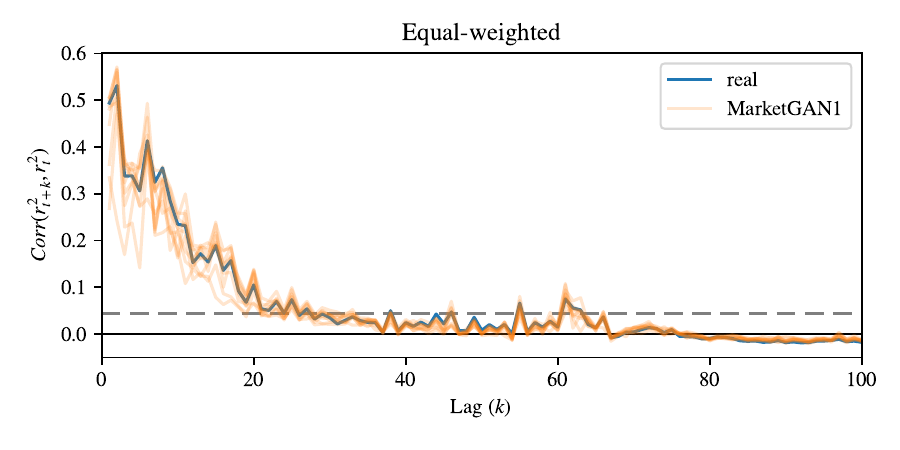}}
  \hfill
  \subfigure[VC of FF-1 bootstrap]{
    \includegraphics[scale=0.48]{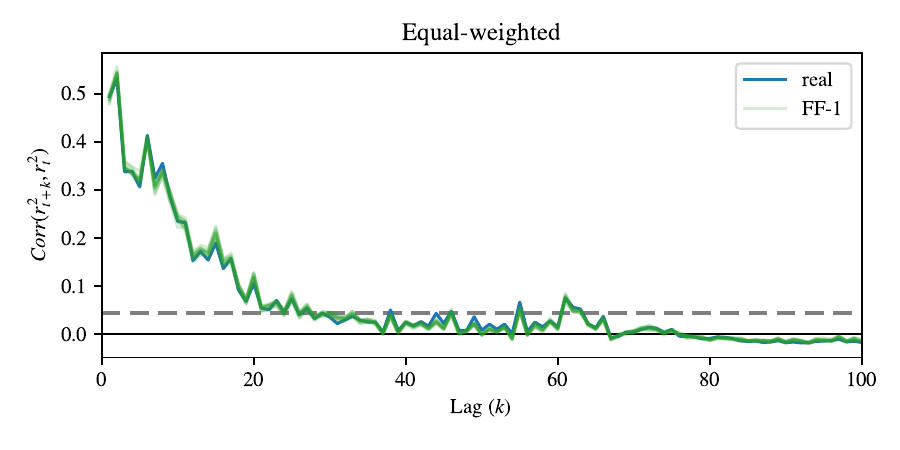}}
\\
  \subfigure[Lev of MarketGAN1]{
    \includegraphics[scale=0.48]{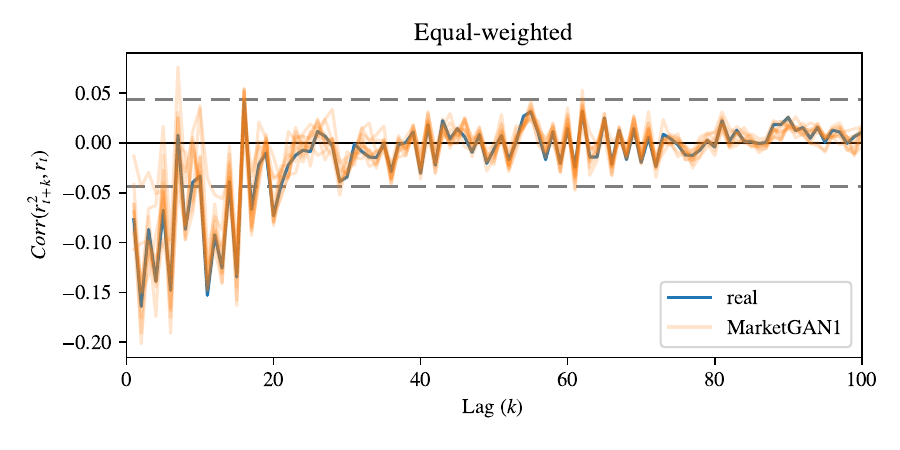}}
  \hfill
  \subfigure[Lev of FF-1 bootstrap]{
    \includegraphics[scale=0.48]{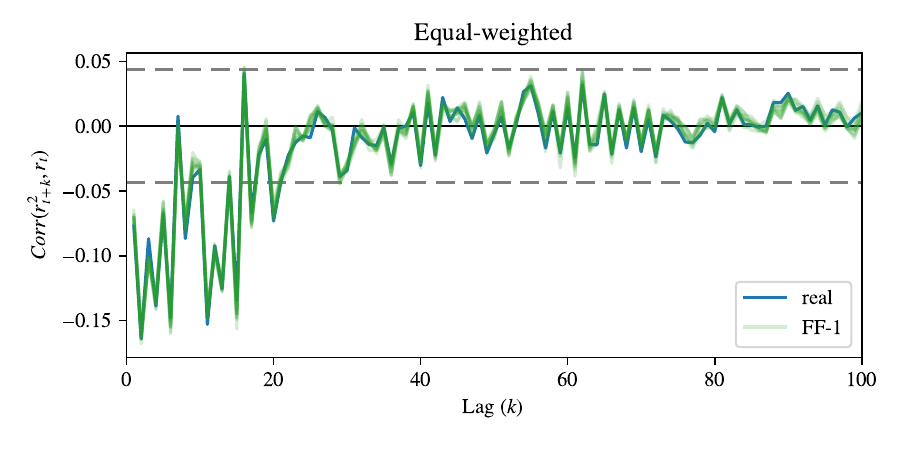}}
  \caption{ACF, VC, and Lev of real and synthetic equal-weighted excess return samples generated by the MarketGAN1 and FF-1 bootstrap.}
  \label{fig: autocorr_ew_1factor}
\end{figure}

\begin{figure}[!htbp]
  \centering
  \subfigure[ACF of MarketGAN3]{
    \includegraphics[scale=0.48]{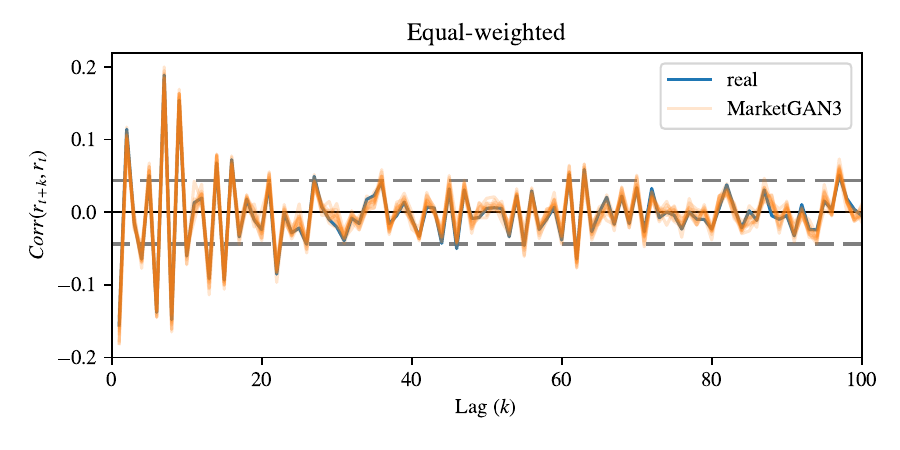}}
  \hfill
  \subfigure[ACF of FF-3 bootstrap]{
    \includegraphics[scale=0.48]{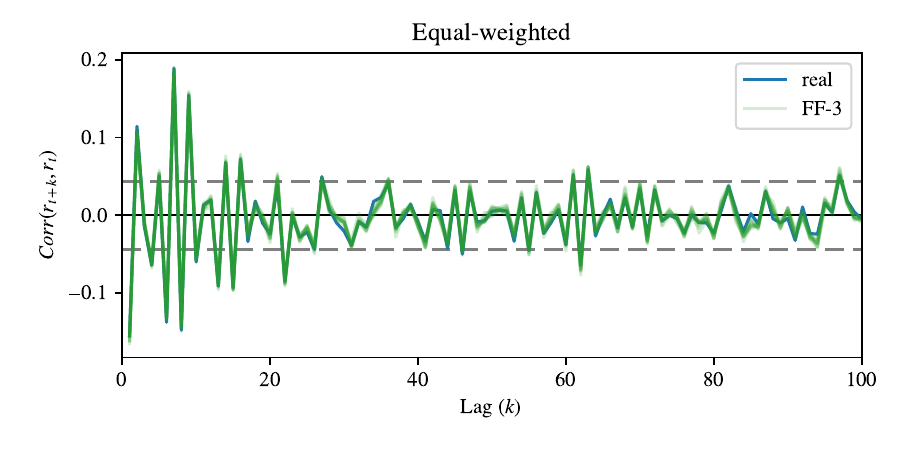}}
 \\
  \subfigure[VC of MarketGAN3]{
    \includegraphics[scale=0.48]{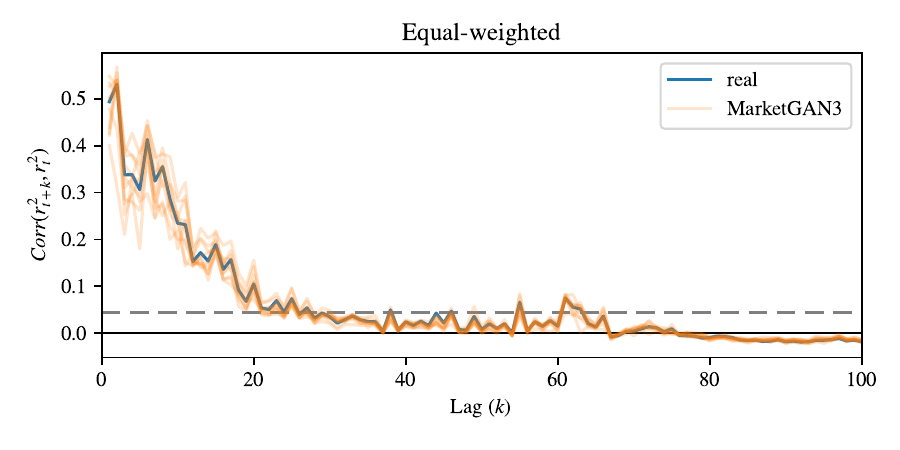}}
  \hfill
  \subfigure[VC of FF-3 bootstrap]{
    \includegraphics[scale=0.48]{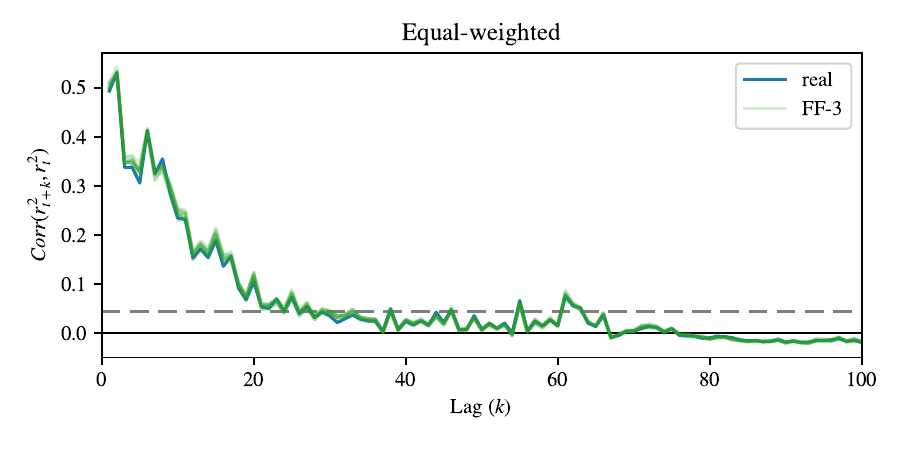}}
\\
  \subfigure[Lev of MarketGAN3]{
    \includegraphics[scale=0.48]{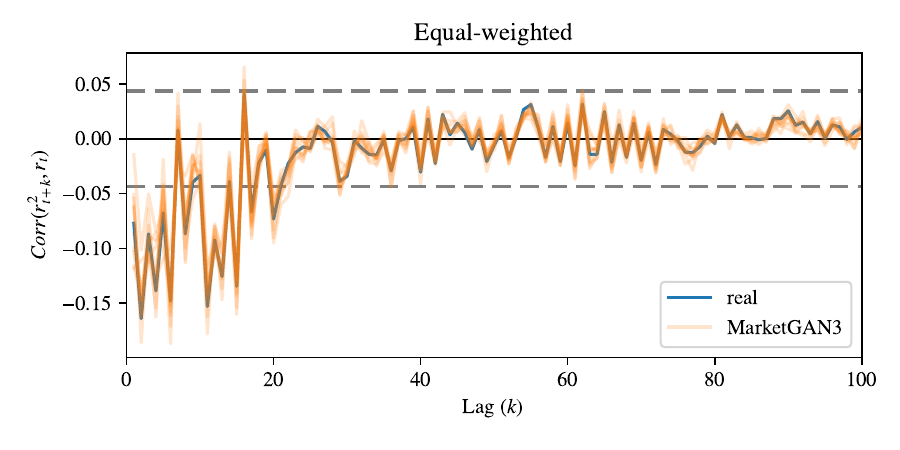}}
  \hfill
  \subfigure[Lev of FF-3 bootstrap]{
    \includegraphics[scale=0.48]{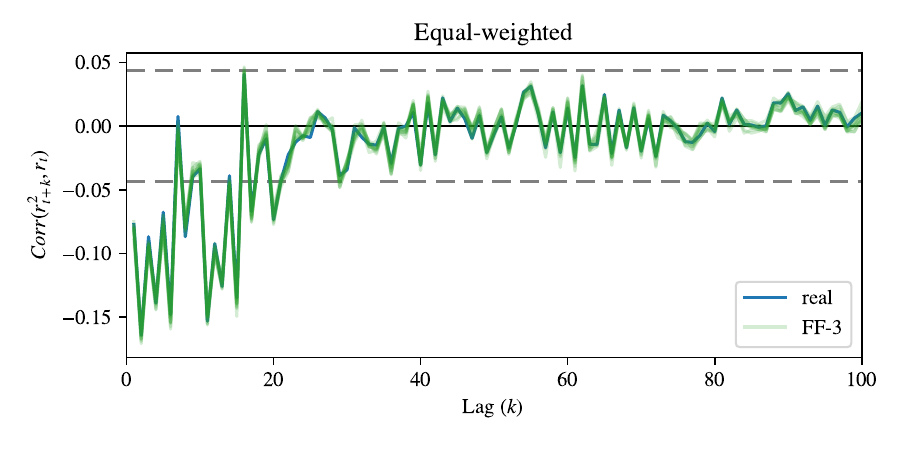}}
  \caption{ACF, VC, and Lev of real and synthetic equal-weighted excess return samples generated by the MarketGAN3 and FF-3 bootstrap.}
  \label{fig: autocorr_ew_3factor}
\end{figure}

\begin{figure}[!htbp]
  \centering

    \subfigure[ACF of MarketGAN1]{
    \includegraphics[scale=0.48]{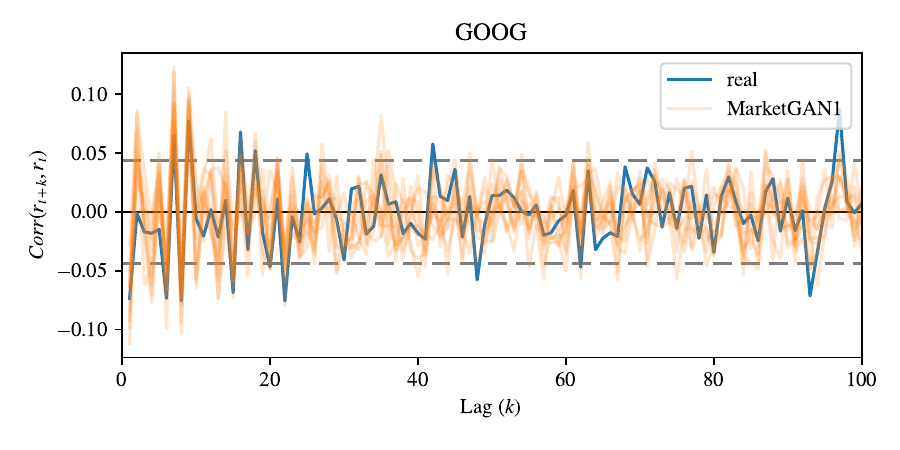}}
  \hfill
  \subfigure[ACF of FF-1 bootstrap]{
    \includegraphics[scale=0.48]{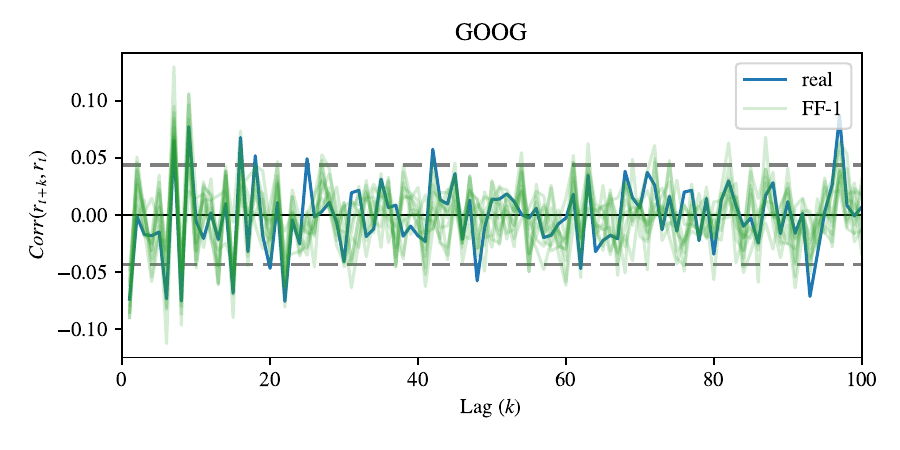}}
 \\
  \subfigure[VC of MarketGAN1]{
    \includegraphics[scale=0.48]{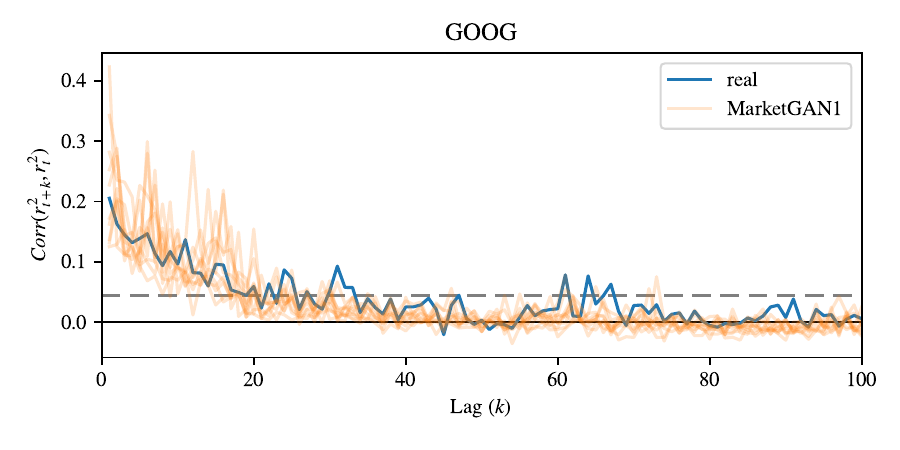}}
  \hfill
  \subfigure[VC of FF-1 bootstrap]{
    \includegraphics[scale=0.48]{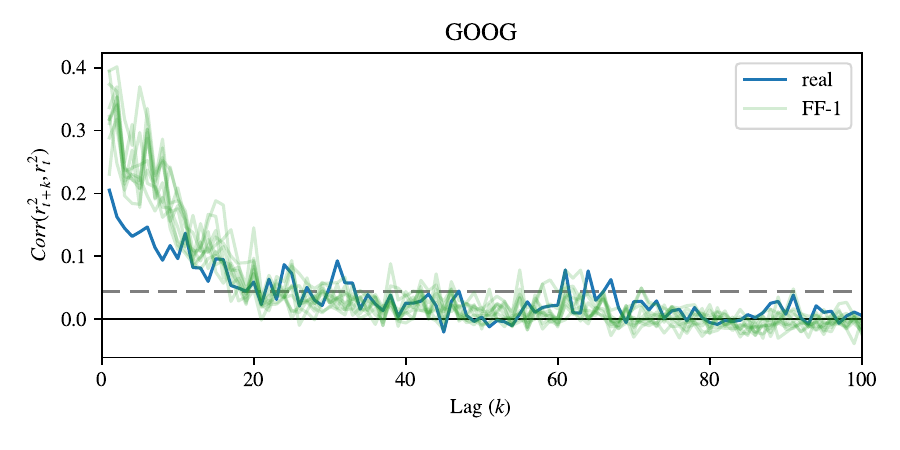}}
\\
  \subfigure[Lev of MarketGAN1]{
    \includegraphics[scale=0.48]{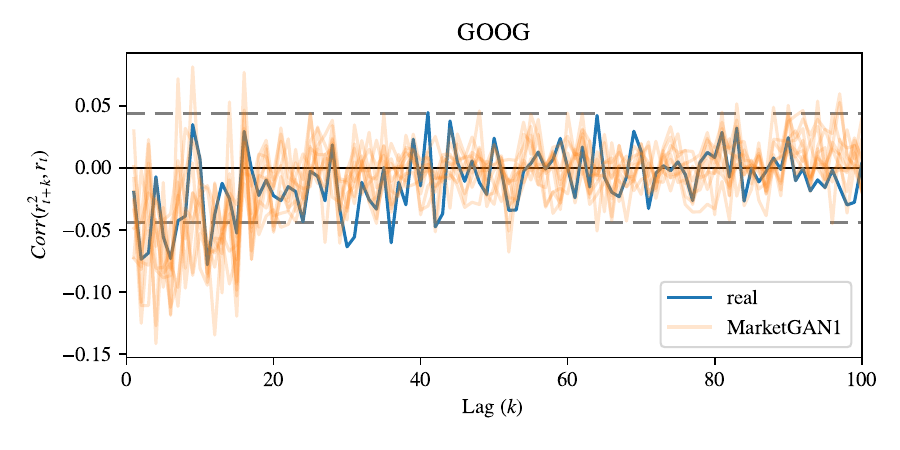}}
  \hfill
  \subfigure[Lev of FF-1 bootstrap]{
    \includegraphics[scale=0.48]{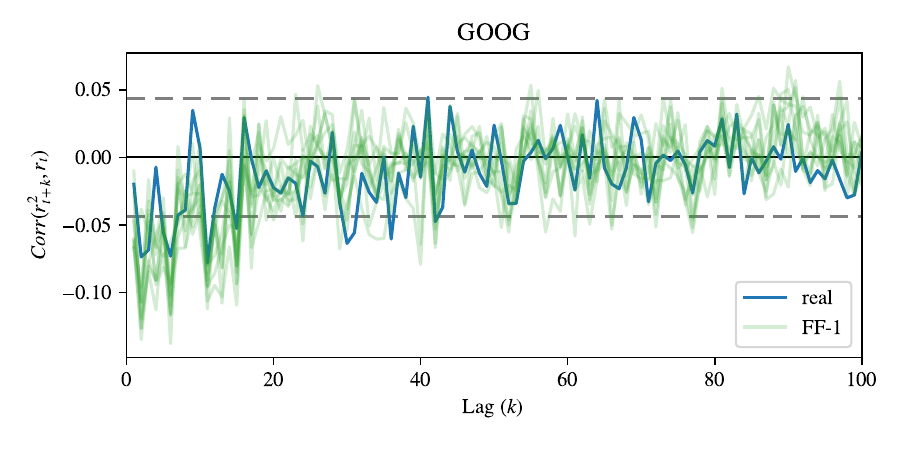}}
  \caption{ACF, VC, and Lev of real and synthetic excess return samples for GOOG generated by the MarketGAN1 and FF-1 bootstrap.}
  \label{fig: autocorr_goog_1}
\end{figure}

\begin{figure}[!htbp]
  \centering

    \subfigure[ACF of MarketGAN3]{
    \includegraphics[scale=0.48]{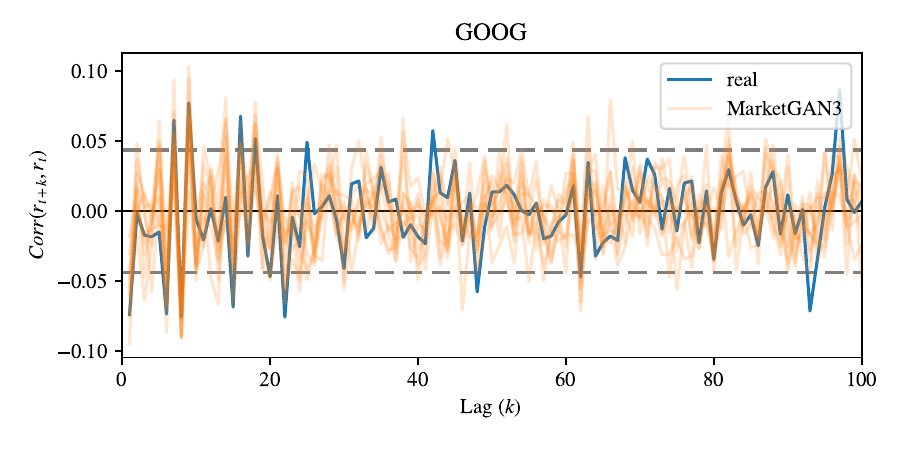}}
  \hfill
  \subfigure[ACF of FF-3 bootstrap]{
    \includegraphics[scale=0.48]{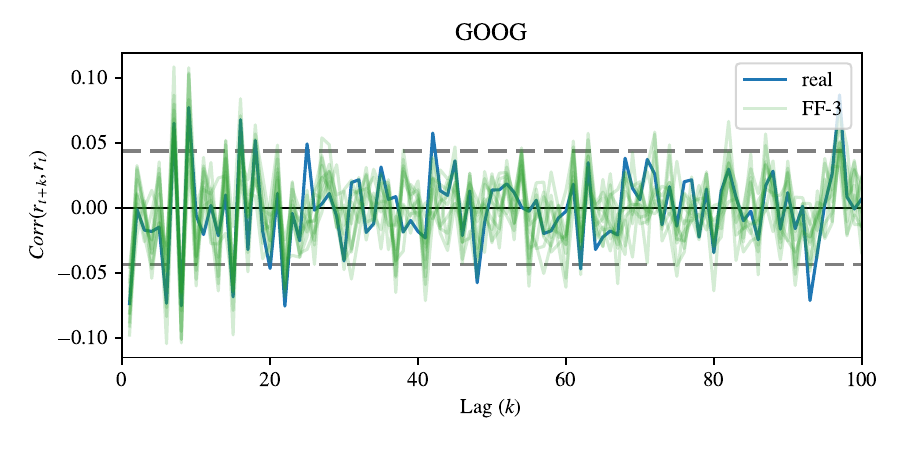}}
 \\
  \subfigure[VC of MarketGAN3]{
    \includegraphics[scale=0.48]{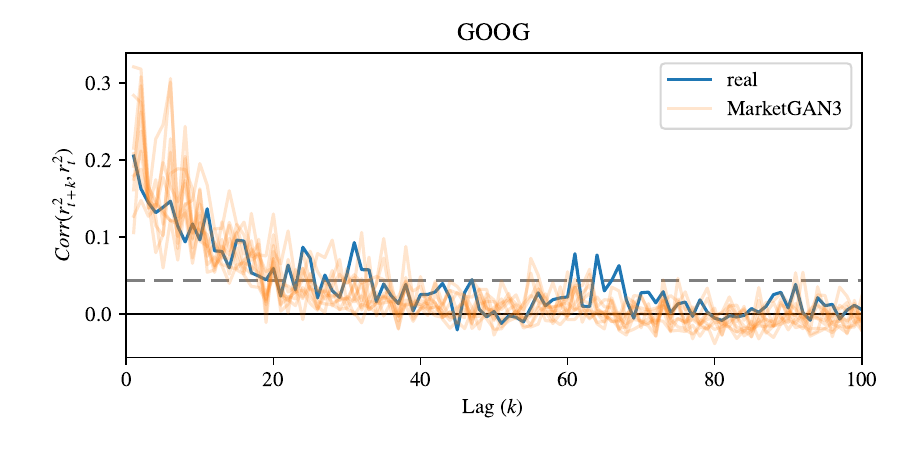}}
  \hfill
  \subfigure[VC of FF-3 bootstrap]{
    \includegraphics[scale=0.48]{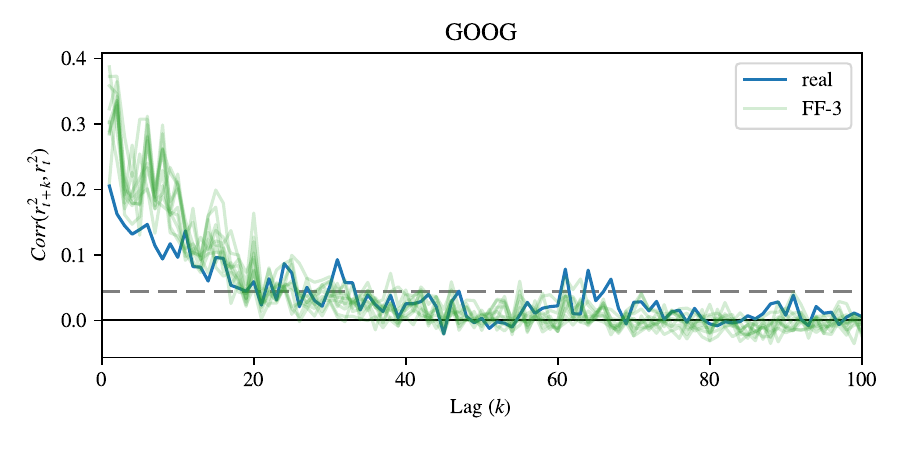}}
\\
  \subfigure[Lev of MarketGAN3]{
    \includegraphics[scale=0.48]{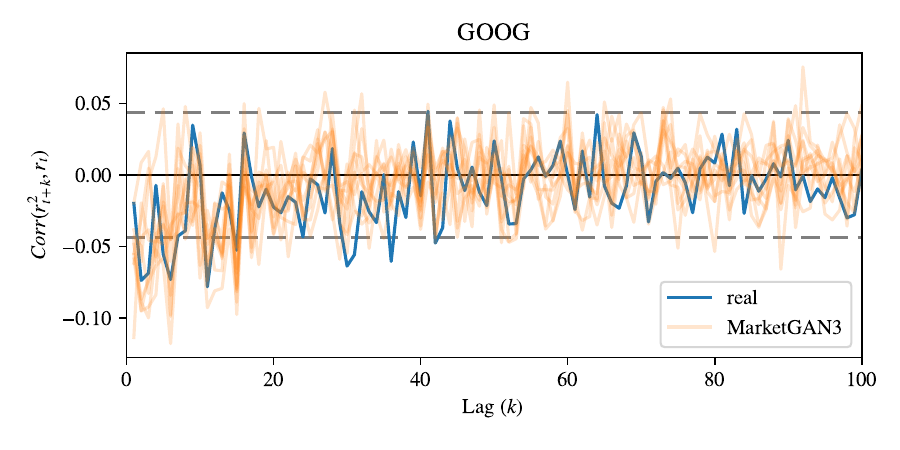}}
  \hfill
  \subfigure[Lev of FF-3 bootstrap]{
    \includegraphics[scale=0.48]{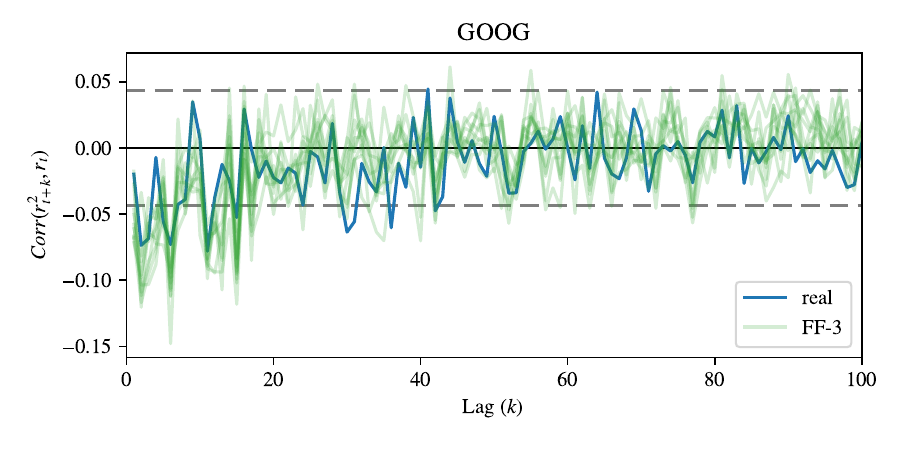}}
  \caption{ACF, VC, and Lev of real and synthetic excess return samples for GOOG generated by the MarketGAN3 and FF-3 bootstrap.}
  \label{fig: autocorr_goog_3}
\end{figure}

\begin{figure}[!htbp]
  \centering
        \subfigure[ACF of MarketGAN5]{
        \includegraphics[scale=0.48]{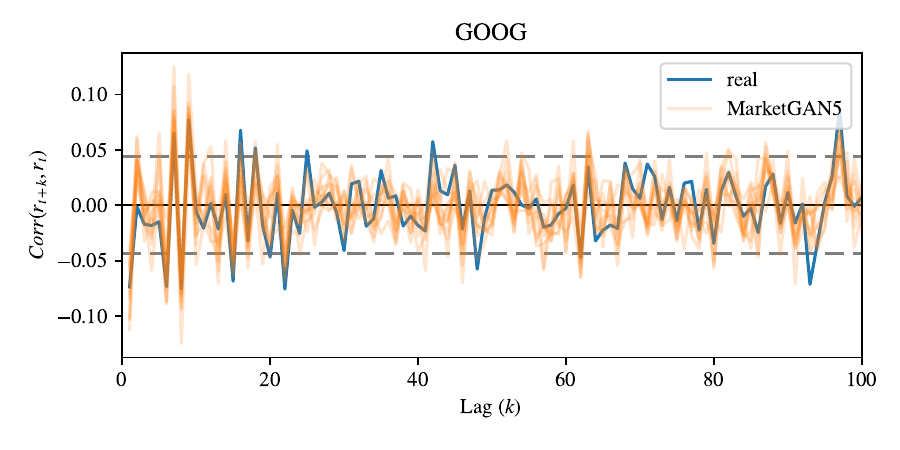}}
      \subfigure[ACF of FF-5 bootstrap]{
        \includegraphics[scale=0.48]{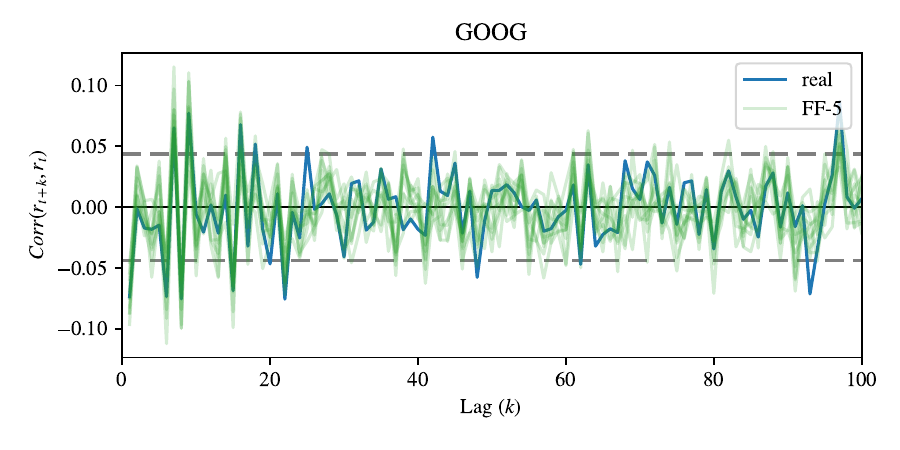}}
        \\
      \subfigure[VC of MarketGAN5]{
        \includegraphics[scale=0.48]{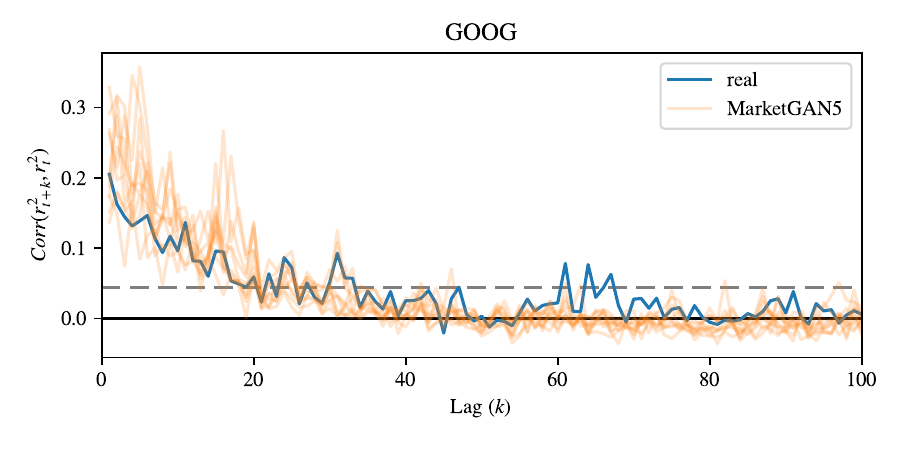}}
      \subfigure[VC of FF-5 bootstrap]{
        \includegraphics[scale=0.48]{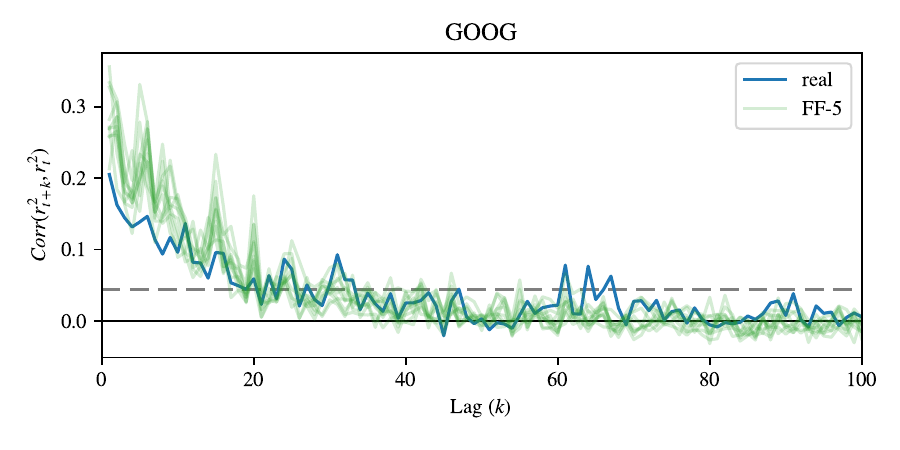}}
        \\
      \subfigure[Lev of MarketGAN5]{
        \includegraphics[scale=0.48]{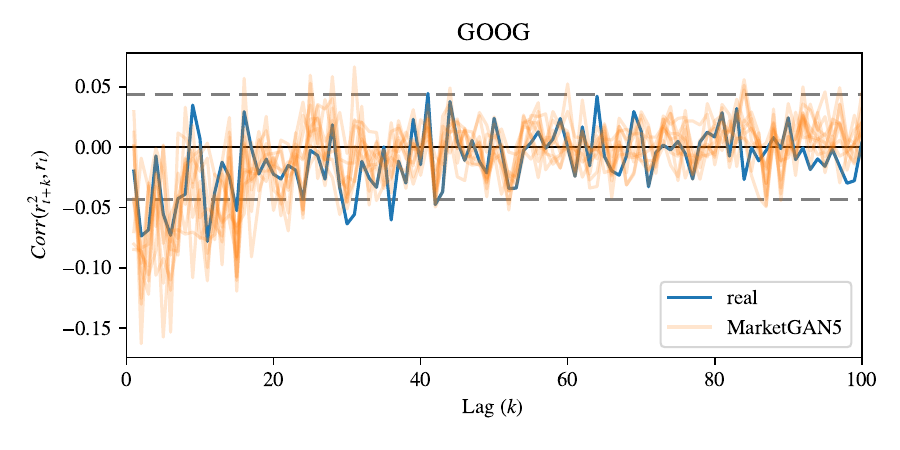}}
      \subfigure[Lev of FF-5 bootstrap]{
        \includegraphics[scale=0.48]{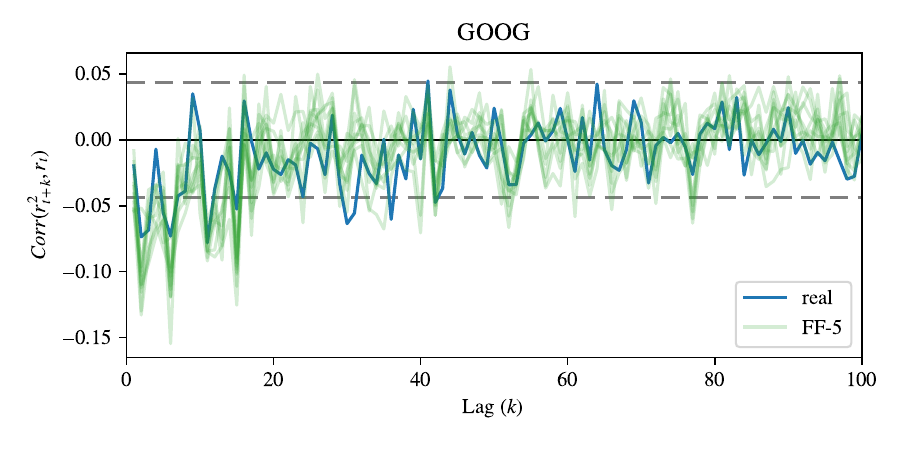}}
  \caption{ACF, VC, and Lev of synthetic excess return samples for GOOG generated by the MarketGAN5 and FF-5 bootstrap.}
  \label{fig: autocorr_goog_5}
\end{figure}

\begin{figure}[!htbp]
	\centering
	
	\subfigure[real]{
		\includegraphics[scale=0.4]{figs/real_cross_corr_MarketGAN5.eps}}
	
	\vspace{1em}  
	\hfill
	\subfigure[MarketGAN1]{
		\includegraphics[scale=0.4]{figs/fake_cross_corr_MarketGAN1.eps}}
	\hfill
	\subfigure[MarketGAN3]{
		\includegraphics[scale=0.4]{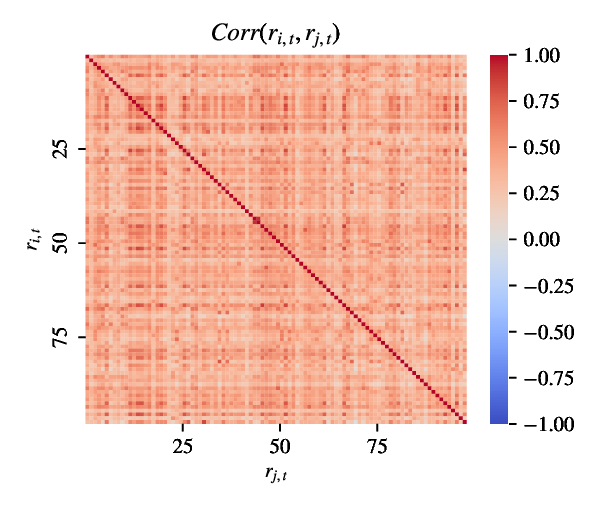}}
	\hfill
	\subfigure[MarketGAN5]{
		\includegraphics[scale=0.4]{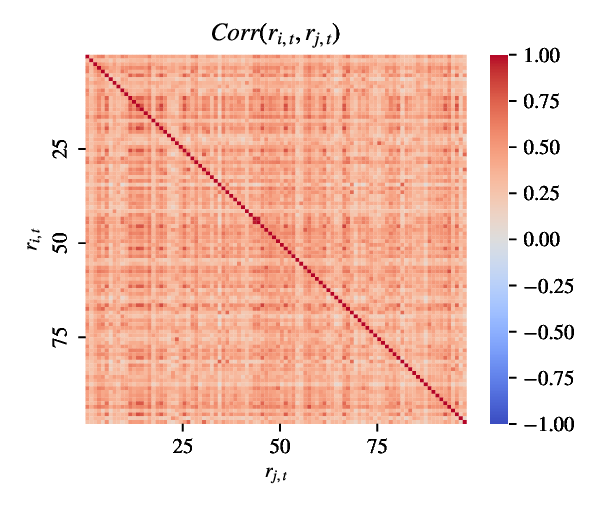}}
	\hfill
	
	\vspace{1em}
	\hfill
	\subfigure[FF-1 bootstrap]{
		\includegraphics[scale=0.4]{figs/FF_btstr_cross_corr_MarketGAN1.eps}}
	\hfill
	\subfigure[FF-3 bootstrap]{
		\includegraphics[scale=0.4]{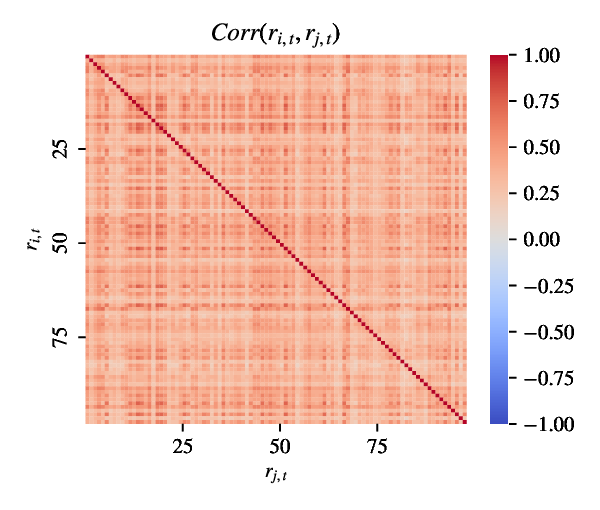}}
	\hfill
	\subfigure[FF-5 bootstrap]{
		\includegraphics[scale=0.4]{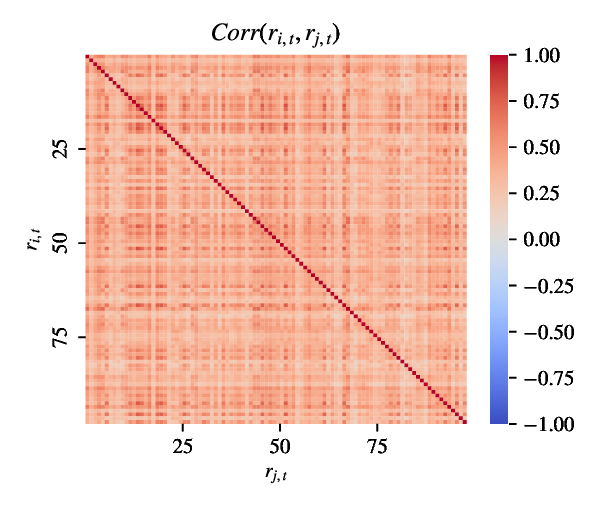}}
	\hfill
	\caption{Cross-correlation of real and synthetic excess return samples generated by MarketGANs and factor-model-based bootstrap methods.}
	\label{fig:crosscorr_comparison}
\end{figure}

\begin{figure}[!htbp]
  \centering
  \subfigure[real]{
    \includegraphics[scale=0.45]{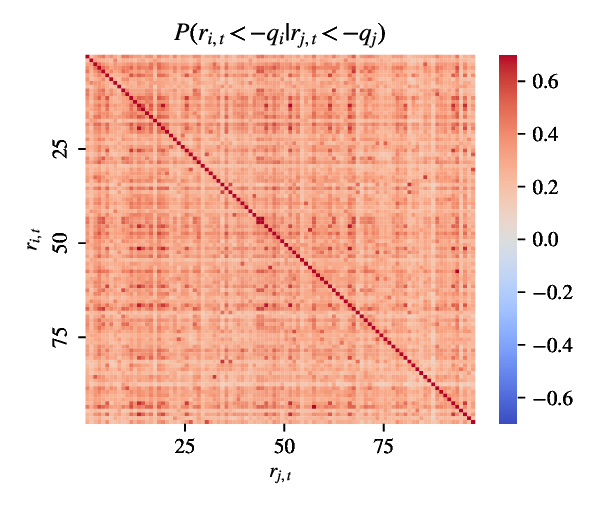}}

  \vspace{1em}
    \hfill
  \subfigure[MarketGAN1]{
    \includegraphics[scale=0.45]{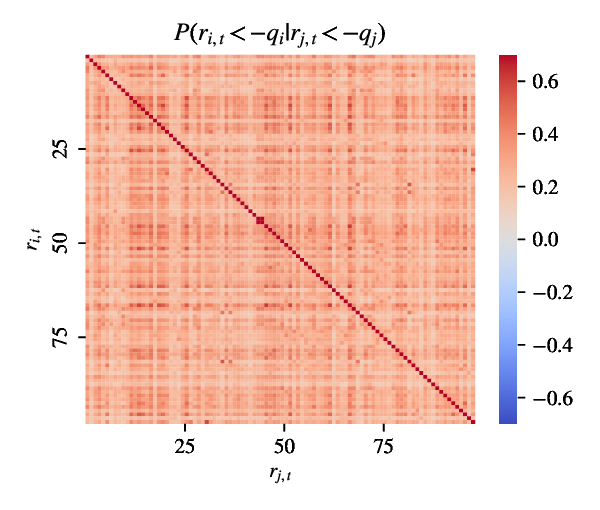}}
  \hfill
  \subfigure[MarketGAN3]{
    \includegraphics[scale=0.45]{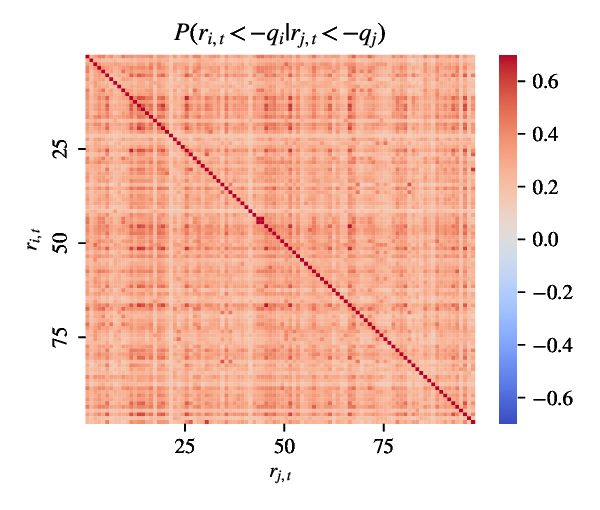}}
  \hfill
  \subfigure[MarketGAN5]{\includegraphics[scale=0.45]{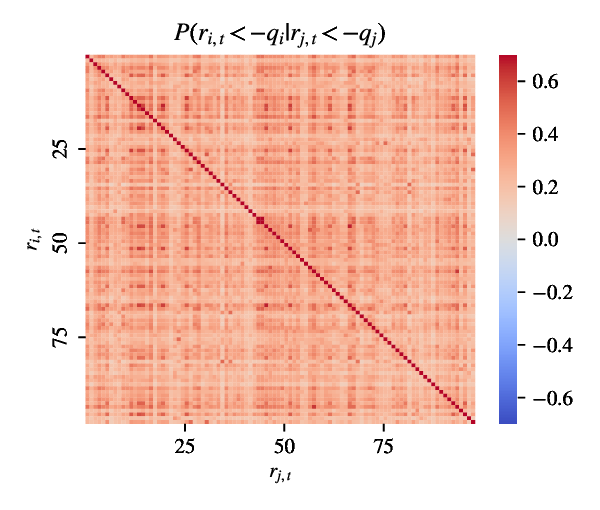}}
    \hfill

  \vspace{1em}
    \hfill
  \subfigure[FF-1 bootstrap]{
    \includegraphics[scale=0.45]{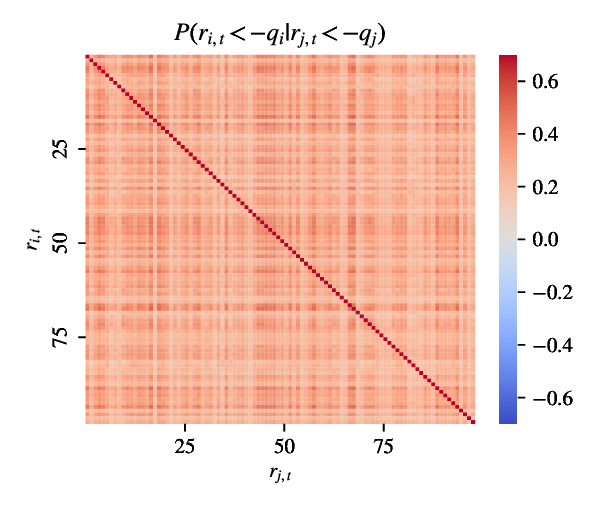}}
  \hfill
  \subfigure[FF-3 bootstrap]{\includegraphics[scale=0.45]{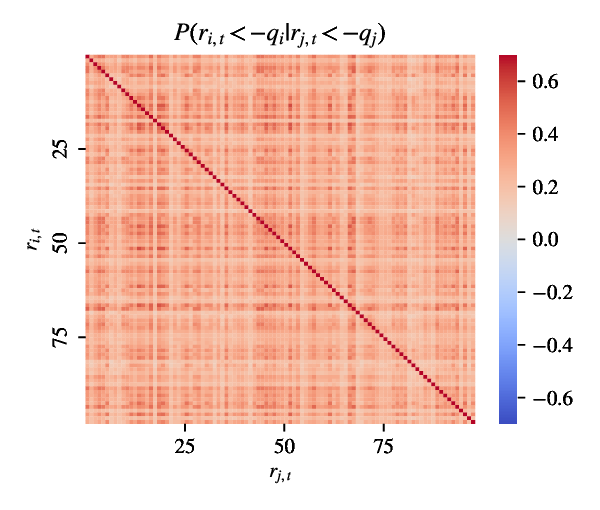}}
  \hfill
  \subfigure[FF-5 bootstrap]{\includegraphics[scale=0.45]{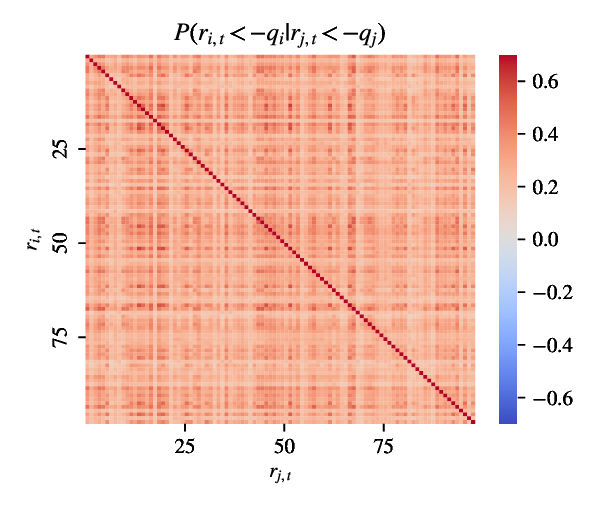}}
\hfill
  \caption{Extreme cross-correlation of real and synthetic excess return samples generated by the MarketGAN models and factor-model-based bootstrap methods.}
  \label{fig:extreme_crosscorr_comparison}
\end{figure}

\newpage



\end{document}